\documentclass[10pt,twocolumn,twoside]{IEEEtran}

\usepackage{color}
\usepackage{theorem}
\usepackage{times,amsmath,epsfig}
\usepackage{amssymb}
\usepackage{cite}
\usepackage{algorithmic}
\usepackage{algorithm}

\usepackage[font=small,format=plain,up,up]{caption}
\usepackage{subcaption}
\usepackage{needspace}

\usepackage{tikz}
\usetikzlibrary{shapes,arrows}

\input{mysymbol.sty}
\tikzstyle{phantom vertex} = [ ellipse, 
                               anchor = center, 
                               minimum height = 1*\unit, 
                               minimum width  = 1*\unit,
                               inner sep=0pt,
                               anchor=center]
\tikzstyle{red vertex}   = [black, fill = red!20,   phantom vertex, draw]
\tikzstyle{black vertex} = [black, fill = black!20, phantom vertex, draw]
\tikzstyle{blue vertex}  = [black, fill = blue!20,  phantom vertex, draw]
\tikzstyle{green vertex} = [black, fill = green!20,  phantom vertex, draw]
\tikzstyle{yellow vertex} = [black, fill = yellow!20,  phantom vertex, draw]
\tikzstyle{cyan vertex} = [black, fill = cyan!20,  phantom vertex, draw]
\tikzstyle{vertex}       = [draw, phantom vertex]

\tikzstyle{point} = [ellipse, inner sep=0pt, draw, fill=white, anchor = center,
                     minimum height = 0.05*\unit, minimum width  = 0.05*\unit]

\newcommand{\EE}{{\mathbb E}}

\def \MSE {\mathrm{MSE}}
\def \Sec {Sec.~}

\DeclareMathOperator{\Tr}{tr}

\DeclareMathOperator{\cg}{cg}
\DeclareMathOperator{\pg}{pg}

\theoremstyle{definition}

\theoremstyle{definition}
\newtheorem{mydefinition}{\bf Definition}

\newtheorem{myproposition}{\bf Proposition}
\newtheorem{remark}{\bf Remark}
\newtheorem{property}{\bf Property}

\title{Stationary Graph Processes and Spectral Estimation}
\author{\IEEEauthorblockN{Antonio G. Marques, Santiago Segarra, Geert Leus, and Alejandro Ribeiro}
	
\thanks{Part of the results in this paper were published in the Proceedings of the 2016 IEEE SAM Workshop \cite{marques2016stationarityRIO} (submitted February 2016, accepted May 2016)  and in the IEEE Transactions on Signal Processing (submitted March 2016, accepted August 2017). Personal use of this material is permitted. However, permission to use this material for any other purposes must be obtained from the IEEE. Work in this paper is supported by Spanish MINECO grants TEC2013-41604-R and TEC2016-75361-R, and USA NSF CCF-1217963. 
A. G. Marques is with the Dept. of Signal Theory and Comms., King Juan Carlos Univ. S. Segarra is with the Inst. for Data, Systems and Society, Massachusetts Inst. of Technology. G. Leus is with the Dept. of Electrical Eng., Mathematics
and Comp. Science, Delft Univ. of Technology. A. Ribeiro are with the Dept. of Electrical and Systems Eng., Univ. of Pennsylvania. Emails: antonio.garcia.marques@urjc.es, segarra@mit.edu, g.j.t.leus@tudelft.nl, aribeiro@seas.upenn.edu.
}}

\begin{document}
\maketitle

%
\begin{abstract}  Stationarity is a cornerstone property that facilitates the analysis and processing of random signals in the time domain. Although time-varying signals are abundant in nature, in many practical scenarios the information of interest resides in more irregular graph domains. This lack of regularity hampers the generalization of the classical notion of stationarity to graph signals. This paper proposes a definition of weak stationarity for random graph signals that takes into account the structure of the graph where the random process takes place, while inheriting many of the meaningful properties of the classical time domain definition. Provided that the topology of the graph can be described by a normal matrix, stationary graph processes can be modeled as the output of a linear graph filter applied to a white input. \black{This is shown equivalent to requiring} the correlation matrix to be diagonalized by the graph Fourier transform; a fact that is leveraged to define a notion of power spectral density (PSD). Properties of the graph PSD are analyzed and a number of methods for its estimation are proposed. This includes generalizations of nonparametric approaches such as periodograms, window-based average periodograms, and filter banks, as well as parametric approaches, using moving-average (MA), autoregressive (AR) and ARMA processes. \black{Graph stationarity and graph PSD estimation are investigated numerically for synthetic and real-world graph signals.}
\end{abstract}

%

%
\section{Introduction}\label{S:Introduction}

\black{Networks and graphs are used to represent pairwise relationships between elements of a set, and are objects of intrinsic interest. In graph signal processing (GSP), the object of study is not the network itself but a signal supported on the vertices of the graph \cite{EmergingFieldGSP,SandryMouraSPG_TSP13}. The graph is intended to represent a notion of proximity between the components of the signal and, as such, provides a structure that can be exploited in its processing. The use of a graph to process a signal is formalized by the definition of the graph shift operator (GSO), which is a matrix representation of the graph (\Sec \ref{S:Modeling}), and the graph Fourier transform (GFT), which projects signals in the eigenvector space of the GSO. The GSO and GFT have been proven useful in the processing of brain activity signals using brain networks, the study of epidemic spreading over social networks, and analysis of Markov random fields \cite{huang2016, BlindID_TSP, Egilmez_GMRF}.

When the GSO is particularized to a cyclic shift, the associated GFT turns out equivalent to the classical discrete Fourier transform (DFT) \cite{SandryMouraSPG_TSP14Freq}. Cyclic shifts and DFTs are central to the processing of time signals in general and to the analysis of (periodic) stationary stochastic processes in particular \cite{hayes2009statistical}. The latter is true for various reasons, one of the most important being that the DFT diagonalizes the covariance matrix of the stationary process and reduces its analysis to the study of its power spectral density (PSD) \cite{stoica2005spectral}. In this paper we consider analogous notions of stationary \textit{graph} processes, their concomitant PSDs and use those to study parametric and nonparametric \textit{graph} PSD estimation.

Stationary graph processes were first defined and analyzed in \cite{girault2015stationary}. The fundamental problem identified in that paper is that GSOs do not preserve energy in general and cannot therefore be isometric \cite{gavili2015ontheshiftoperator}. This problem is addressed with the definition of an isometric graph shift that preserves the eigenvector space of the Laplacian GSO but modifies its eigenvalues \cite{girault2015translation}. A stationary graph process is then defined as one whose probability distributions are invariant with respect to multiplications with the isometric shift. It is further shown that this definition requires the covariance matrix of the signal to be diagonalized by the eigenvectors of the GSO, which are also the eigenvectors of the isometric shift. Although not formally defined as such, this conclusion implies the existence of a graph PSD with components given by the covariance eigenvalues. The requirement of having a covariance matrix diagonalizable by the eigenvectors of the Laplacian GSO is itself adopted as a definition in \cite{EPFL16stationary}, where the requirement is shown to be equivalent to statistical invariance with respect to the translation operator introduced in \cite{shuman2016vertex}.  The PSD is explicitly defined in \cite{EPFL16stationary} and a generalization of the short time Fourier transform is proposed for its estimation.} 

\black{The main goal of this paper is to provide a comprehensive introduction to the spectral analysis and estimation of \textit{graph stationary processes}. We start with a review GSP concepts (\Sec \ref{S:Modeling}) and define weakly stationary graph processes (\Sec \ref{S:Def_GraphStationarity}), introduce useful properties (\Sec \ref{Ss:Properties}), and discuss some examples (\Sec \ref{S:DiffusionProcStationarity}). To emphasize connections with stationary processes in time we say a process is stationary with respect to a normal GSO if: (i) It can be modeled as the output of a linear shift invariant graph filter \cite{SandryMouraSPG_TSP14Freq, ssamar_distfilters_allerton15} applied to a white input. (ii) Its covariance matrix satisfies a form of invariance with respect to applications of the GSO. (iii) Its covariance matrix is diagonalized by the GSO's eigenvector matrix. The shift invariance in definition (ii) is more involved than its time-domain counterpart, but it does preserve the locality of the given GSO and sheds light on the local structure of graph stationary processes. These three definitions are equivalent under mild conditions (Prop. \ref{prop_all_defs_equivalent}) and, since (iii) is shared with \cite{EPFL16stationary} and a consequence of \cite{girault2015stationary}, they are also equivalent to the isometric invariance in \cite{girault2015stationary} and the translation invariance in \cite{EPFL16stationary}. Our discussion in this section differs from \cite{girault2015stationary, EPFL16stationary} in that we not only consider Laplacian GSOs but arbitrary normal GSOs. Although mathematically minor, this modification broadens applicability of graph stationarity, establishing connections with results in statistics and network topology inference \cite{segarra2016networktopologyID}. Normal GSOs that are not Laplacians covers all symmetric shifts and a subset of non-symmetric GSOs which is small but includes the directed cycle.}

\black{Since stationary processes are easier to understand in the frequency domain, we study different methods for \textit{spectral} (PSD) \textit{estimation}, which, if needed, can also be used to improve the estimate of the covariance matrix itself (\Sec \ref{S:nonparametric_PSD_est} and \Sec \ref{S:parametric_PSD_est}). We begin by looking at simple nonparametric methods for PSD estimation (\Sec \ref{S:nonparametric_PSD_est}). We first extend the periodogram and the correlogram to the graph domain, and analyze their estimation performance. We then generalize more advanced estimators such as window-based average periodograms and filter banks \cite[Chs. 2 and 5]{stoica2005spectral}. The estimation performance is evaluated analytically and differences relative to their time-domain counterparts are highlighted. After this, we shift attention to parametric estimation (\Sec \ref{S:parametric_PSD_est}). The focus is on estimating the PSD of autoregressive (AR), moving-average (MA), and ARMA graph processes, which are likely to arise in distributed setups driven by linear dynamics \cite{sandryhaila2014finite, segarra2015reconstruction, BlindID_TSP}. As in time, it turns out that the estimation of the ARMA parameters is a non-convex (quartic) problem, although for certain particular cases --including that of positive semidefinite shifts -- the optimization is tractable. Numerical results are presented to confirm theoretical claims and provide insights on the implications of graph stationarity and to discuss some potential applications (\Sec \ref{S:NumExper}). The latter include smoothing of face-image signals, characterization of brain networks and signals, and description of flow cytometry data.}

\medskip \noindent \textbf{Notation:} Entries of vector $\bbx$ are $[\bbx]_i=x_i$.  Entries of matrix $\bbX$ are $[\bbX]_{ij} = X_{ij}$. Conjugate, transpose, and transpose conjugate are $\bbX^*$, $\bbX^T$, and $\bbX^H$. For square matrix $\bbX$, we use $\Tr[\bbX]$ for its trace and $\diag(\bbX)$ for a vector with the diagonal elements of $\bbX$. For vector $\bbx$, $\diag(\bbx)$ denotes a diagonal matrix with diagonal elements $\bbx$. The elementwise product of $\bbx$ and $\bby$ is $\bbx\circ\bby$. The all-zero and all-one vectors are $\bbzero$ and $\bbone$. The $i$th element of the canonical basis of $\reals^N$ is $\bbe_i$.

%
\section{Graph Signals and Filters}\label{S:Modeling}

Let $\ccalG=(\ccalN, \ccalE)$ be a directed graph or network with a set of $N$ nodes $\ccalN$ and directed edges $\ccalE$ such that $(i,j)\in\ccalE$. We associate with $\ccalG$ the GSO $\bbS$, defined as an $N\times N$ matrix whose entry $S_{ji}\neq0$ only if $i=j$ or if $(i,j)\in\ccalE$ \cite{SandryMouraSPG_TSP13,SandryMouraSPG_TSP14Freq}. The sparsity pattern of $\bbS$ captures the local structure of $\ccalG$, but we make no specific assumptions on the values of the nonzero entries of $\bbS$. Frequent choices for $\bbS$ are the adjacency matrix \cite{SandryMouraSPG_TSP13,SandryMouraSPG_TSP14Freq}, the Laplacian \cite{EmergingFieldGSP}, and their respective generalizations \cite{godsil2001algebraic}. The intuition behind $\bbS$ is to represent a linear transformation that can be computed locally at the nodes of the graph. More rigorously, if the set $\ccalN_l(i)$ stands for the nodes within the $l$-hop neighborhood of node $i$ and the signal $\bby$ is defined as $\bby=\bbS^l\bbx$, then node $i$ can compute $y_i$ provided that it has access to the value of $x_j$ at $j\in \ccalN_l(i)$. We henceforth work with  \textit{normal} $\bbS$ to guarantee existence of a unitary matrix $\bbV$ and a diagonal matrix $\bbLambda$ such that $\bbS=\bbV\bbLambda\bbV^H$. 

A graph signal is a vector $\bbx=[x_1,...,x_N]^T \in  \mathbb{R}^N$ where the $i$-th element represents the value of the signal at node $i$ or, alternatively, as a function $f : \ccalN \to \mathbb{R}$, defined on the vertices of the graph. Given a graph signal $\bbx$, we refer to $\tilde{\bbx}:=\bbV^H\bbx$ as the frequency representation of $\bbx$, with $\bbV^H$ being the graph Fourier transform (GFT) \cite{SandryMouraSPG_TSP14Freq}. 

We further introduce the notion of a graph filter $\bbH:\;\mathbb{R}^N \to \mathbb{R}^N$, defined as a linear graph signal operator of the form
\begin{eqnarray}\label{E:def_graph_filters}
	&\bbH:=\sum_{l=0}^{L-1}h_l \bbS^l,&
\end{eqnarray}
where $\bbh=[h_0,\ldots,h_{L-1}]^T$ is a vector of $L\leq N$ scalar coefficients. According to \eqref{E:def_graph_filters}, graph filters are polynomials of degree $L-1$ in the GSO $\bbS$ \cite{SandryMouraSPG_TSP13}, which due to the local structure of the shift can be implemented locally too \cite{ssamar_distfilters_allerton15, loukas2015distributed}. It is easy to see that graph filters are invariant to applications of the shift in the sense that if $\bby=\bbH\bbx$, it must hold that $\bbS\bby=\bbH(\bbS\bbx)$. 
Using the factorization $\bbS=\bbV\bbLambda\bbV^H$ the filter in \eqref{E:def_graph_filters} can be rewritten as $\bbH=\bbV\big(\sum_{l=0}^{L-1}h_l \bbLam^l\big) \bbV^H$. The diagonal matrix $\sum_{l=0}^{L-1}h_l \bbLam^l$ is termed the frequency response of the filter and extracted in the vector $\tilde{\bbh}:=\diag(\sum_{l=0}^{L-1}h_l \bbLam^l)$.

To relate the frequency response $\tbh$ with the filter coefficients $\bbh$ let $\lam_k=[\bbLam]_{kk}$ be the $k$th eigenvalue of $\bbS$ and define the $N \times N$ Vandermonde matrix $\bbPsi$ with entries $\Psi_{kl} = \lam_k^{l-1}$. Further define $\bbPsi_L$ as a tall matrix containing the first $L$ columns of $\bbPsi$ to write $\tilde{\bbh}=\bbPsi_L\bbh$ and conclude that \eqref{E:def_graph_filters} can be written as
\begin{equation}\label{E:Filter_from_time_to_freq}
	\bbH \, =\, \sum_{l=0}^{L-1}h_l \bbS^l
	\, =\, \bbV\diag(\tilde{\bbh})\bbV^H
	\, =\, \bbV\diag\big(\bbPsi_L\bbh\big) \bbV^H .
\end{equation}
Equation \eqref{E:Filter_from_time_to_freq} implies that if $\bby$ is defined as $\bby=\bbH\bbx$, its frequency representation $\tilde{\bby} = \bbV^H \bby$ satisfies
\begin{equation}\label{E:Filter_input_output_freq}
	\tilde{\bby} \,=\, \diag\big(\bbPsi_L\bbh\big)\bbV^H\bbx
	\,=\, \diag\big(\tilde{\bbh}\big)\tilde{\bbx}
	\,=\, \tilde{\bbh}\circ\tilde{\bbx},
\end{equation}
which demonstrates that the output at a given frequency depends only on the value of the input and the filter response at that given frequency. Observe that when $L=N$ we have $\bbPsi_N=\bbPsi$ and that, if we are given a filter with order greater than $N-1$, we can rewrite it as a different filter of order not larger than $N-1$ due to the Cayley-Hamilton theorem.

%
%
\section{Weakly stationary random graph processes}\label{S:Def_GraphStationarity}

\black{This section leverages notions of weak stationarity in discrete time to define stationarity in the graph domain, discusses relations with existing definitions, and presents results that will be leveraged in the ensuing sections. The proposed definitions depend on the GSO $\bbS$, so that a process $\bbx$ can be stationary in $\bbS$ but not in $\bbS' \neq \bbS$. The shift does not need to be the adjacency or the Laplacian matrix, but it does need to be \textit{normal} and account for the topology of the graph. 

A standard zero-mean white random process $\bbw$ is one with mean $\E{\bbw}=\bbzero$ and covariance $\E{\bbw\bbw^H} = \bbI$. Our first definition, states that white signals processed by a linear shift-invariant \textit{graph} filter give rise to stationary \textit{graph} processes.}

%
\begin{mydefinition}\label{D:WeaklyStionaryGraphProcess_2}\normalfont 
	Given a normal shift operator $\bbS$, a zero-mean random process $\bbx$ is said to be weakly stationary with respect to $\bbS$ if it can be written as the response of a linear shift-invariant graph filter $\bbH\!=\!\sum_{l=0}^{N-1} h_l\bbS^l$ to a zero-mean white input $\bbw$.
\end{mydefinition}

%
\noindent The definition states that the process $\bbx$ is stationary if we can write $\bbx=\bbH\bbw$ for some filter $\bbH=\sum_{l=0}^{N-1}h_l \bbS^l$ that we excite with a white input $\bbw$. If we write $\bbx=\bbH\bbw$, the covariance matrix  $\bbC_x:=\E{\bbx\bbx^H}$ of the signal $\bbx$ can be written as
\begin{align}\label{E:cov_output_filter}
\bbC_x = \E{(\bbH\bbw)(\bbH\bbw)^H} 
= \bbH\E{\bbw\bbw^H}\bbH  
= \bbH\bbH^H,
\end{align}
which implies that the color of the process is determined by the filter $\bbH$. Def.~\ref{D:WeaklyStionaryGraphProcess_2} is constructive as it describes how a stationary graph signal can be generated. \black{Alternatively, one can define stationarity using requirements on the moments of  the random graph signal in either the vertex or the frequency domain.

%
\begin{mydefinition}\label{D:WeaklyStionaryGraphProcess_1}\normalfont 
	Given a normal shift operator $\bbS$, a zero-mean random process $\bbx$ is said to be weakly stationary with respect to $\bbS$ if the two equivalent properties hold
	\begin{itemize}
		\item[(a)]$\!$For any set of nonnegative integers $a$, $b$, and $c\!\leq\! b$ it holds
	\begin{align}\label{eqn:D:WeaklyStionaryGraphProcess_1}
\!\!\!	\!\!\!\mbE\Big[\big(\bbS^a\bbx\big)\big(\!\big(\bbS^H\big)^b\bbx\big)^H\Big]
	\!=\! \mbE\Big[\big(\bbS^{a+c}\bbx\big)\big(\!\big(\bbS^H\big)^{b-c}\bbx\big)^H\Big].
	\end{align} 
			\item[(b)]$\!$Matrices $\bbC_x$ and $\bbS$ are simultaneously diagonalizable.		
			
	\end{itemize}
	\end{mydefinition}
\noindent Before discussing its intuition, we start by showing that Defs.~\ref{D:WeaklyStionaryGraphProcess_1}.a and \ref{D:WeaklyStionaryGraphProcess_1}.b are indeed equivalent. To do this, suppose that Def.~\ref{D:WeaklyStionaryGraphProcess_1}.a holds and reorder terms in \eqref{eqn:D:WeaklyStionaryGraphProcess_1} to yield
\begin{align}\label{eqn:D:WeaklyStionaryGraphProcess_1_prop_proof}
\bbS^a \bbC_x \bbS^b  = \bbS^{a+c}\bbC_x\bbS^{b-c}.
\end{align} 
For \eqref{eqn:D:WeaklyStionaryGraphProcess_1_prop_proof} to be true, $\bbS^c$ and $\bbC_x$ must commute for all $c$. 
Since both matrices are diagonalizable, this will happen if and only if $\bbS$ and $\bbC_x$ have the same eigenvectors \cite{conrad1minimal},
which implies Def.~\ref{D:WeaklyStionaryGraphProcess_1}.b. Conversely, if Def.~\ref{D:WeaklyStionaryGraphProcess_1}.b. holds, then we can write $\bbC_x=\bbV \diag{(\bblam_c)} \bbV^H$. Substituting this expression into \eqref{eqn:D:WeaklyStionaryGraphProcess_1_prop_proof} the equality checks and Def.~\ref{D:WeaklyStionaryGraphProcess_1}.a follows.	
	
The idea under Def.~\ref{D:WeaklyStionaryGraphProcess_1}.a is that if the total number of times that we shift our signal is constant (regardless of how many of those times we shift the signal forward $\bbS \bbx$ or backward $\bbS^H\bbx$), then the correlation has to be the same. Indeed, in the two sides of \eqref{eqn:D:WeaklyStionaryGraphProcess_1} the total number of times the signal has been shifted is $a+b$. This resembles what happens for stationary signals in time, where correlation depends on the total number of shifts, but not the particular time instants. Formally, when $\bbS$ is a cyclic shift, $\bbS^H$ is a shift in the opposite direction with $\bbS^H\!=\!\bbS^{-1}$. Then, if we set $a\!=\!0$, $b\!=\!N$ and $c\!=\!l$ we recover $\E{\bbx\bbx^H} \!= \!\E{\bbS^l\bbx (\bbS^l\bbx)^H}$, which is the definition of a stationary signal in time. Intuitively, having the same number of shifts in \eqref{eqn:D:WeaklyStionaryGraphProcess_1} is necessary because the eigenvalues of the GSO $\bbS$ do not have unit magnitude and can change the energy of the signal.} \black{The vertex-based definition in \eqref{eqn:D:WeaklyStionaryGraphProcess_1} is based on the original GSO $\bbS$, which is local and real-valued. As a result, \eqref{eqn:D:WeaklyStionaryGraphProcess_1} provides intuition on the relations between stationarity and locality, which can be leveraged to develop stationarity tests or estimation schemes that work with local information; see \Sec \ref{S:nonparametric_PSD_est}. This is different from the vertex-based definitions in \cite{girault2015stationary} and \cite{EPFL16stationary}, which however are more advantageous than \eqref{eqn:D:WeaklyStionaryGraphProcess_1} from other perspectives; see Remark \ref{rmk_comparison_to_bananas}.}  

\black{While Def.~\ref{D:WeaklyStionaryGraphProcess_1}.a captures the implications of stationarity in the vertex domain, Def.~\ref{D:WeaklyStionaryGraphProcess_1}.b captures its implications in the graph frequency domain by requiring that the covariance $\bbC_x$ be diagonalized by the GFT matrix $\bbV$ -- this was deduced as a property in \cite{girault2015stationary} and proposed as a definition in \cite{EPFL16stationary}. When particularized to time, Def.~\ref{D:WeaklyStionaryGraphProcess_1}.b requires $\bbC_x$ to be circulant and therefore diagonalized by the Fourier matrix. This fact is exploited in (periodic) time stationary processes to define the PSD as the eigenvalues of the circulant covariance matrix, and motivates the following definition.} 
 
%
\begin{mydefinition}\label{def_psd}\normalfont 
The power spectral density (PSD) of a random process $\bbx$ that is stationary with respect to the normal graph shift $\bbS=\bbV \bbLambda \bbV^H$ is the nonnegative $N\times 1$ vector $\bbp$ 
		\begin{equation}\label{E:PSD_stat_graph_process}
		\bbp := \diag\left(\bbV^H \bbC_x \bbV\right).
		\end{equation}\end{mydefinition}
	
%
\black{Observe that since $\bbC_x$ is diagonalized by $\bbV$ (see Def.~\ref{D:WeaklyStionaryGraphProcess_1}.b) the matrix $\bbV^H \bbC_x \bbV$ is diagonal and it follows that the PSD in \eqref{E:PSD_stat_graph_process} corresponds to the eigenvalues of the positive semidefinite correlation matrix $\bbC_x$. Thus, \eqref{E:PSD_stat_graph_process} is equivalent to 
	\begin{equation}\label{eqn_covariance_from_psd}
	\bbC_x = \bbV \diag(\bbp) \bbV^H.
	\end{equation}
	%

	
%
Defs.~\ref{D:WeaklyStionaryGraphProcess_2} and \ref{D:WeaklyStionaryGraphProcess_1}, which are valid for any symmetric GSO $\bbS$ as well as for nonsymmetric but normal shifts, are equivalent for time signals by construction. They are also equivalent for an important class of graphs, as shown next.}
	
	%
	\begin{myproposition}\label{prop_all_defs_equivalent}
		\normalfont If $\bbS$ is normal and its eigenvalues are all distinct, Defs.~\ref{D:WeaklyStionaryGraphProcess_2} and \ref{D:WeaklyStionaryGraphProcess_1} are equivalent. 	
	\end{myproposition}
	
	%
	\begin{IEEEproof}  		
		Since Defs.~\ref{D:WeaklyStionaryGraphProcess_1}.b and \ref{D:WeaklyStionaryGraphProcess_1}.a are equivalent, we show that under the conditions in the proposition, Def.~\ref{D:WeaklyStionaryGraphProcess_2} is equivalent to Def.~\ref{D:WeaklyStionaryGraphProcess_1}.b. Assume first that Def.~\ref{D:WeaklyStionaryGraphProcess_2} holds. Since the graph filter $\bbH$ in \eqref{E:cov_output_filter} is linear shift invariant, it is completely characterized by its frequency response $\tbh=\bbPsi\bbh$ as stated in \eqref{E:Filter_from_time_to_freq}. Using this characterization we rewrite \eqref{E:cov_output_filter} as
		\begin{align}\label{E:cov_output_filter_freq}
			\!\!\bbC_x  \!= \!\bbV\diag(\tbh)\bbV^H       \!
			\big(\bbV\diag(\tbh)\bbV^H\big)\!^H  \! \!
			=\! \bbV\diag^2\big(|\tbh|\big)\bbV^H     \!.
		\end{align}
		This means that Def.~\ref{D:WeaklyStionaryGraphProcess_1}.b holds. Conversely, if Def.~\ref{D:WeaklyStionaryGraphProcess_1}.b holds it means that we can write $\bbC_x=\bbV \diag{(\bblambda_c)} \bbV^H$ for some component-wise \textit{nonnegative} vector $\bblambda_c$. Define now vector $\sqrt{\bblambda_c}$ where the square root is applied element-wise. Then, for Def.~\ref{D:WeaklyStionaryGraphProcess_2} to hold, there must exist a filter with coefficients $\bbh$ satisfying $\tbh=\bbPsi\bbh=\sqrt{\bblambda_c}$. Since $\bbPsi$ is Vandermonde, the system is guaranteed to have a solution with respect to $\bbh$ provided that all modes of $\bbPsi$ (the eigenvalues of $\bbS$) are distinct, as required in the proposition.  
	\end{IEEEproof}
	
	%
\black{For normal shifts, Def.~\ref{D:WeaklyStionaryGraphProcess_2} always implies Def.~\ref{D:WeaklyStionaryGraphProcess_1} as also shown in \cite{EPFL16stationary}. However, if $\bbS$ is not normal, as is the case for most directed graphs, the equivalence is lost and the question of which definition to use arises. Constructive approaches related to Def.~\ref{D:WeaklyStionaryGraphProcess_2} that use graph filters to model a set of observations have been successfully applied in the context of diffusions over \textit{directed} non-normal graphs \cite{SandryMouraSPG_TSP13,mei2015signal}.}

\begin{remark}\label{rmk_comparison_to_bananas}\normalfont
\black{When the shift is $\bbS=\bbL$ and the eigenvalues of $\bbS$ are all distinct, Defs.~\ref{D:WeaklyStionaryGraphProcess_2} and \ref{D:WeaklyStionaryGraphProcess_1} are equivalent to those in \cite{girault2015stationary} and \cite{EPFL16stationary}. The main differences between \cite{girault2015stationary}, \cite{EPFL16stationary}, and our frameworks are in the interpretation of stationarity in the vertex domain. To define stationarity, \cite{girault2015stationary} utilizes an isometric GSO $\tbS$ that has the same eigenvectors as $\bbS$ but whose eigenvalues are unitary complex exponentials \cite{gavili2015ontheshiftoperator}. Although the resultant operator is complex-valued and has a sparsity structure different from $\bbS$, it provides a more natural invariance. Indeed, since the resultant $\tbS$ preserves energy and gives rise to the same GFT, the counterpart to \eqref{eqn:D:WeaklyStionaryGraphProcess_1} can be simply written as $\EE[\bbx\bbx^H ]= \EE[\tbS^l\bbx (\tbS^l\bbx)^H]$. Differently, \cite{EPFL16stationary} proposes as starting point Def.~\ref{D:WeaklyStionaryGraphProcess_1}.b and analyzes the implications in the vertex domain using the so-called translation operator defined in \cite{shuman2016vertex}. The reason for having different possible definitions in the vertex domain is that the irregularity and finiteness of the graph support leads to non-isometric shifts. One can then modify the shift to make it isometric as in \cite{girault2015stationary}, define a different isometry as in \cite{EPFL16stationary}, or define a different form of invariance as in  \eqref{eqn:D:WeaklyStionaryGraphProcess_1}. We also point out that our work differs from \cite{EPFL16stationary} in that we consider normal shifts instead of Laplacians and that we see Def.~\ref{D:WeaklyStionaryGraphProcess_2} as a definition, not a property. These are mathematically minor differences that are important in practice. E.g., normal shifts that are not Laplacians include symmetric shifts with positive non-diagonal entries or the cyclic shift. Moreover, in cases where Defs.~\ref{D:WeaklyStionaryGraphProcess_2} and  \ref{D:WeaklyStionaryGraphProcess_1} differ, Def.~\ref{D:WeaklyStionaryGraphProcess_2} is useful because it models real-world processes (see \Sec \ref{S:DiffusionProcStationarity}).} 
\end{remark}

		%
		\subsection{Properties of Graph Stationary Processes}\label{Ss:Properties}
		
		\black{Three properties that will be leveraged in the ensuing sections are listed next. Since these properties are direct generalizations of their classical time counterparts, their proofs are omitted for conciseness. For Property \ref{P:locality_stat_MA_process}, recall that $\ccalN_l(i)$ denotes the $l$-hop neighborhood of node $i$.}
		
		%
		\vspace{-2mm}
		\begin{property}\label{P:PSD_output_based_PSD_input}\normalfont	
			Let $\bbx$ be a stationary process in $\bbS$ with covariance matrix $\bbC_x$ and PSD $\bbp_x$. Consider a filter $\bbH$ with coefficients $\bbh$ and frequency response $\tbh$ and define $\bby := \bbH\bbx$ as the response of $\bbH$ to input $\bbx$. Then, the process $\bby$:
			\begin{mylist}
				\item{(a)} Is stationary in $\bbS$ with covariance 
				$\bbC_y=\bbH \bbC_x \bbH^H$.
				\item{(b)} Has a PSD given by
				$ \bbp_y  = |\tbh|^2 \circ \bbp_x$.
			\end{mylist}
			\vspace{-5mm}
			\end{property} 
			
			\begin{property}\label{P:uncorrelated_freq_components}\normalfont
				Given a process $\bbx$ stationary in $\bbS=\bbV\bbLam\bbV^H$ with PSD $\bbp$, define the GFT process as $\tbx = \bbV^H\bbx$. \black{Then, it holds that $\tbx$ is uncorrelated and its covariance matrix is }
				\begin{equation}\label{eqn_uncorrelated_freq_components}
					\bbC_{\tdx} := \E{\tbx\tbx^H} 
					= \E{(\bbV^H\bbx) (\bbV^H\bbx)^H} 
					= \diag(\bbp) .
				\end{equation} 
				\vspace{-5mm}
				\end{property}
			
			\vspace{-4mm}
			\begin{property}\label{P:locality_stat_MA_process}
				\normalfont	Let $\bbx = \bbH\bbw$ be a process written as the response of a linear graph filter $\bbH=\sum_{l=0}^{L-1}h_l\bbS^l$ of degree $L-1$ to a white input. Then, $[\bbC_x]_{ij}=0$ for all $j\notin \ccalN_{2(L-1)}(i)$.
			\end{property}
			%
			
			%
	\black{Property \ref{P:PSD_output_based_PSD_input}, also identified in \cite{EPFL16stationary}, is a statement of the spectral convolution theorem for graph signals, which can also be viewed as a generalization of Def.~\ref{D:WeaklyStionaryGraphProcess_2} for inputs that are not white. Property \ref{P:uncorrelated_freq_components} (see also \cite{girault2015stationary}) is fundamental to motivate the analysis and modeling of stationary graph processes in the frequency domain, which we undertake in ensuing sections. It also shows that if a process $\bbx$ is stationary in the shift $\bbS=\bbV\bbLambda\bbV^H$, then the GFT $\bbV^H$ provides the Karhunen-Lo\`eve expansion of the process. The last property requires a bit more of a discussion. In time, the correlation matrix often conveys a notion of locality as the significant values of $\bbC_x$ accumulate close to the diagonal. 	Property \ref{P:locality_stat_MA_process}{, which follows from the locality of graph filters,} puts a limit on the spatial extent of the components $x_j$ of a graph signal that can be correlated with a given element $x_i$. Only those elements that are in the $2(L-1)$-hop neighborhood -- i.e., elements $x_j$ with indices $ j \in \ccalN_{2(L-1)}(i)$ -- can be correlated with $x_i$. This spatial limitation of correlations can be used to design windows for spectral estimation as we explain in \Sec \ref{sec_windowing}.}

\begin{remark}\label{R:mean_StationaryGraphProcess}\normalfont	
					\black{While Defs.~\ref{D:WeaklyStionaryGraphProcess_2} and \ref{D:WeaklyStionaryGraphProcess_1} assume that the random process $\bbx$ has mean $\bar{\bbx}:=\E{\bbx}=\bbzero$, traditional stationary time processes are allowed to have a (non-zero) constant mean $\bar{\bbx} = \alpha\bbone$, with $\alpha$ being an arbitrary scalar. Our proposal here is for stationary graph processes to be required to have a first-order moment of the form $\bar{\bbx} = \alpha \bbv_k$, i.e., a scaled version of an eigenvector of $\bbS$. This choice: i) takes into account the structure of the underlying graph; ii) maintains the validity of Property \ref{P:PSD_output_based_PSD_input}; and iii) allows to set $\bbv_k = \bbone$ for the cases where $\bbS$ is either the adjacency matrix of the directed cycle or the Laplacian of any graph, recovering the classical definition. Note that \cite{girault2015stationary,girault2015translation} require $\bar{\bbx}\!=\!\bbzero$, while \cite{EPFL16stationary} requires $\bar{\bbx}\!=\!\alpha\bbone$.} 
\end{remark}

				%
				\subsection{Examples of Stationary Graph Processes}\label{S:DiffusionProcStationarity}

				\black{We close this section providing some representative examples of stationary graph processes.}

					%
					\medskip
					\noindent \textit{Ex. 1: White noise.}	
					Zero-mean white noise is stationary in any graph shift $\bbS$. The PSD of white noise with covariance $\EE[\bbw\bbw^H] = \sigma^2\bbI$ is $\bbp = \sigma^2\bbone$. 
					
					%
					\medskip
					\noindent \textit{Ex. 2: Covariance matrix graph.}	
					Any random process $\bbx$ is stationary with respect to the shift $\bbS=\bbC_x$ defined by its covariance matrix. Since in this case the eigenvalues of $\bbS$ and $\bbC_x$ are the same, it holds that the PSD is $\bbp=\diag(\bbLam)$. This can be exploited, for example, in the context of network topology inference. Given a set of (graph) signal observations $\{\bbx_r\}_{r=1}^R$ it is common to infer the underlying topology by building a graph $\ccalG_{corr}$ whose weight links correspond to cross-correlations among the entries of the observations. \black{In that case, the process generating those signals is stationary in the} shift given by the adjacency of $\ccalG_{corr}$; see  \cite{segarra2016networktopologyID} for details. 

						\medskip
						\noindent \textit{Ex. 3: Precision matrix graph.}
						Let $\bbTheta$ denote the precision matrix of $\bbx$, which is defined as the (pseudo-)inverse $\bbTheta=\bbC_x^\dagger$. Per Def.~\ref{D:WeaklyStionaryGraphProcess_1}.b, it holds then than the process $\bbx$ is stationary in $\bbTheta$. The PSD in this case is $\bbp=\diag(\bbLam)^{^\dagger}$. This is particularly important when $\bbx$ is a Gaussian Markov Random Field (GMRF) whose Markovian dependence is captured by the unweighted graph $\ccalG_{MF}$. It is well-known \cite[Ch. 19]{murphy2012machine} that if $\bbTheta$ is the precision matrix of a GMRF, then $\Theta_{ij}$ can be non-zero only if $(i,j)$ is either a link of $\ccalG_{MF}$, or an element in the diagonal  $i\!=\!j$. Then, it holds that any GMRF is stationary with respect to the \emph{sparse} shift $\bbS\!=\!\bbTheta$, which captures the conditional independence of the elements of $\bbx$. 

				\medskip
				\noindent \textit{Ex. 4: Network Diffusion Processes.}		
				\black{Many graph processes are characterized by local interactions between nodes of a (sparse) graph that can be approximated as linear \cite{sandryhaila2014finite,segarra2015reconstruction, BlindID_TSP}. This implies that we can track the evolution of the signal at the $i$th node through a local recursion of the form $x_i^{(l+1)}=x_{i}^{(l)}  - \gamma^{(l)} \textstyle\sum_{j} s_{ij} x_{j}^{(l)}$ where $x_i^{(0)}$ is the initial condition of the system, $s_{ij}$ are elements of the GSO, and $\gamma^{(l)}$ are time varying diffusion coefficients. The diffusion halts after $L$ iterations -- which can be possibly infinite -- to produce the output $y_i = x_i^{(L)}$. Diffusion dynamics appear in, e.g., network control systems \cite{Pasqualetticontrollability}, opinion formation \cite{segarra2015reconstruction}, brain networks \cite{Friston2003dynamic}, and molecular communications \cite{Nakano_etal12}. Utilizing the GSO, diffusion dynamics can be written as $\bbx^{(l+1)}=(\bbI-\gamma^{(l)}\bbS)\bbx^{(l)}$, with the output $\bby = \bbx^{(L)}$ being
				\begin{equation}\label{E:local_diffusion_singlenode}
					\bby = \bbx^{(L)} = \textstyle \prod_{l=0}^L(\bbI-\gamma^{(l)}\bbS)\bbx.
				\end{equation}
				The Cayley-Hamilton theorem guarantees that any matrix polynomial can be expressed as a graph filter with degree less than $N$. If we now think of the initial condition $\bbx$ as a process stationary in $\bbS$, it follows that $\bby$, which is a filtered version of $\bbx$, is also stationary (cf. Property \ref{P:PSD_output_based_PSD_input}).}
				%
				%
				%
				
				%

%
%
\section{Nonparametric PSD Estimation}\label{S:nonparametric_PSD_est}

The interest in this and the following section is in estimating the PSD of a random process $\bbx$ that is stationary with respect to $\bbS$ using as input either \textit{one} or a few realizations $\{\bbx_r\}_{r=1}^R$ of $\bbx$. In this section we consider nonparametric methods that do not assume a particular model for $\bbx$. We generalize to graph signals the periodogram, correlogram, windowing, and filter bank techniques that are used for PSD estimation in the time domain. Parametric methods are analyzed in \Sec \ref{S:parametric_PSD_est}.

%
\subsection{Periodogram and Correlogram}\label{Ss:nonparametric_PSD_est}

Since Property \ref{P:uncorrelated_freq_components} implies that $\bbC_{\tdx}$ is diagonal, we can rewrite \eqref{eqn_uncorrelated_freq_components} to conclude that the PSD can be written as $\bbp = \E{|\bbV^H\bbx |^2}$. This yields a natural approach to estimate $\bbp$ with the GFT of realizations of $\bbx$. Thus, compute the GFTs $\tbx_r = \bbV^H\bbx_r$ of each of the samples $\bbx_r$ in the training set and estimate $\bbp$ as
\begin{equation}\label{E:nonpar_PSD_periodogram}
\hbp_{\pg}\ :=\ \frac{1}{R}\sum_{r=1}^R \left|\tbx_r \right|^2
\  =\ \frac{1}{R}\sum_{r=1}^R \left|\bbV^H\bbx_r \right|^2.
\end{equation}
The estimator in \eqref{E:nonpar_PSD_periodogram} is the analogous of the \textit{periodogram} of time signals and is referred to as such from now on. Its intuitive appeal is that it writes the PSD of the process $\bbx$ as the average of the squared magnitudes of the GFTs of realizations of $\bbx$.

Alternatively, one can replace $\bbC_x$ in \eqref{E:PSD_stat_graph_process} by \black{its sample-based version} $\hbC_x\!=\!(1/R)\sum_{r=1}^R\bbx_r\bbx_r^H$ \black{and estimate the PSD as}
\begin{equation}\label{E:nonpar_PSD_correlogram}
\hbp_{\cg} \!:=\! \diag\left(\bbV^H
\hbC_x
\bbV\right) 
\!:=\! \diag\bigg[\bbV^H 
\bigg[\frac{1}{R}\!\sum_{r=1}^R\bbx_r\bbx_r^H\bigg] 
\bbV\bigg].\!
\end{equation}
An important observation in the correlogram definition in \eqref{E:nonpar_PSD_correlogram} is that the empirical covariance $\hbC_x$ is not necessarily diagonalized by $\bbV$. However, since we know that the \black{actual covariance $\bbC_{\tilde{x}}$} is diagonal, we retain only the diagonal elements of $\bbV^H\hbC_x\bbV$ to estimate the PSD of $\bbx$. The expression in \eqref{E:nonpar_PSD_correlogram} is the analogous of the time
\textit{correlogram}. Its intuitive appeal is that it estimates the PSD with a double GFT transformation of the empirical covariance matrix. 

\black{Although different in genesis, the periodogram in \eqref{E:nonpar_PSD_periodogram} and the correlogram in \eqref{E:nonpar_PSD_correlogram} are identical estimates. To see this, consider the last equality in \eqref{E:nonpar_PSD_correlogram}, and move matrices $\bbV^H$ and $\bbV$ into the empirical covariance sum. Observe then that the summands end up being of the form $(\bbV^H\bbx_r )(\bbV^H\bbx_r )^H$. The diagonal elements of these outer products are $|\bbV^H\bbx_r |^2$, which are the summands in the last equality in \eqref{E:nonpar_PSD_periodogram}. This is consistent with the equivalence of correlograms and periodograms in time signals.} Henceforth, we choose to call $\hbp_{\pg} = \hbp_{\cg}$ the periodogram estimate of $\bbp$. 

To evaluate the performance of the periodogram estimator in \eqref{E:nonpar_PSD_periodogram} we assess its mean and variance. The estimator is unbiased by design and, as we shall prove next, this is easy to establish formally. To study the estimator's variance we need the additional hypothesis of the process having a Gaussian distribution. Expressions for means and variances of periodogram estimators are given in the following proposition.

%
\begin{myproposition}\label{P:CovariancePeriodogram}\normalfont
	Let $\bbp$ be the PSD of a process $\bbx$ that is stationary with respect to the shift $\bbS=\bbV\bbLam\bbV^H$. Independent samples $\{\bbx_r\}_{r=1}^R$ are drawn from the distribution of the process $\bbx$ and the periodogram $\hbp_{\pg}$ is computed as in \eqref{E:nonpar_PSD_periodogram}. The bias $\bbb_{\pg}$ of the estimator is zero,
	\begin{align}\label{E:BiasPeriodogram}
	\bbb_{\pg}\ :=\ \E{\hbp_{\pg}} - \bbp  
	\  =\ \bbzero. 
	\end{align} 
	Further define the covariance matrix of the periodogram estimator as $\bbSigma_{\pg} := \E{(\hbp_{\pg}-\bbp) (\hbp_{\pg}-\bbp)^H}$. If the process $\bbx$ is assumed Gaussian and $\bbS$ is symmetric, the covariance matrix can be written as
	\begin{align}\label{E:CovariancePeriodogram}
	\bbSigma_{\pg} := \E{(\hbp_{\pg}-\bbp) (\hbp_{\pg}-\bbp)^H} = (2/R) \diag^2(\bbp).
	\end{align} \end{myproposition}

%
\begin{IEEEproof} See Appendix A. \end{IEEEproof}

%
The expression in \eqref{E:BiasPeriodogram} states that the bias of the periodogram is $\bbb_{\pg}=\bbzero$, or, equivalently, that the expectation of the periodogram estimator is the PSD itself, i.e., $\E{\hbp_{\pg}} = \bbp$. The expression for the covariance matrix of $\bbSigma_{\pg}$ holds true only when the process $\bbx$ has a Gaussian distribution. The reason for this limitation is that the determination of this covariance involves operations with the fourth order moments of the process $\bbx$. This necessity and associated limitation also arise in time signals \cite[\Sec \ 8.2]{hayes2009statistical}. To help readability and intuition, the covariance $\bbSigma_{\pg}$ in \eqref{E:CovariancePeriodogram} is stated for a symmetric $\bbS$, however, in Appendix A the proof is done for a generic normal, not necessarily symmetric, $\bbS$.

The variance expression in \eqref{E:CovariancePeriodogram} is analogous to the periodogram variances of PSDs of time domain signals in that: (i) Estimates of different frequencies are uncorrelated -- because $\bbSigma_{\pg}$ is diagonal. (ii) The variance of the periodogram is proportional to the square of the PSD. The latter fact is more often expressed in terms of the mean squared error (MSE), which we define as $\MSE(\hbp_{\pg}):=\E{\|(\hbp_{\pg}-\bbp)\|_2^2}$ and write as [cf. \eqref{E:BiasPeriodogram} and \eqref{E:CovariancePeriodogram}]
\begin{equation}\label{E:mse_period}
\MSE(\hbp_{\pg})  = \|\bbb_{\pg}\|_2^2+\Tr[\bbSigma_{\pg}]
= (2/R)\|\bbp\|_2^2  .
\end{equation}
As it happens in time signals, the MSE in \eqref{E:mse_period} is large and results in estimation errors that are on the order of the magnitude of the frequency component itself. Two of the workarounds to reduce the MSE in \eqref{E:mse_period} are the use of windows and filter banks. Both tradeoff bias for variance as we explain in the following sections.

%
\subsection{Windowed Average Periodogram}\label{sec_windowing}

The Bartlett and Welch methods for PSD estimation of time signals utilize windows to, in effect, generate multiple samples of the process even if only a single realization is given \cite[\Sec \ 2.7]{stoica2005spectral}. These methods reduce variances of PSD estimates, but introduce some distortion (bias). The purpose of this section is to define counterparts of windowing methods for PSD estimation of graph signals.

To understand the use of windows in estimating a PSD let us begin by defining windows for graph signals and understanding their effect on the graph frequency domain. We say that a signal $\bbw$ is a window if its energy is $\|\bbw\|_2^2=\|\bbone\|_2^2=N$. Applying the window $\bbw$ to a signal\footnote{To keep notation simple, we use $\bbx$ to denote a realization of process $\bbx$.} $\bbx$ entails componentwise multiplication to produce the signal $\bbx_{\bbw} = \diag(\bbw) \bbx$. In the graph frequency domain we can use the definition of the GFT $\tbx_{\bbw} = \bbV^H \bbx_{\bbw}$, the definition of the windowed signal $\bbx_{\bbw} $ $= \diag(\bbw) \bbx$, and the definition of the inverse GFT to write
\begin{align}\label{eqn_windows}
\! \! \tbx_{\bbw} \!=\! \bbV^H \bbx_{\bbw}
\!=\! \bbV^H \diag(\bbw) \bbx
\!=\! \bbV^H \diag(\bbw) \bbV \tbx
\!:=\! \tbW \tbx,\!
\end{align}
where in the last equality we defined the dual of the windowing operator $\diag(\bbw)$ in the frequency domain as the matrix $\tbW:=\bbV^H\diag(\bbw)\bbV$. In time signals the frequency representation of a window is its Fourier transform and the dual operator of windowing is the convolution between the spectra of the window and the signal. This decoupled explanation is lost in graph signals\footnote{\black{See  \cite{shuman2016vertex} for a different definition of the windowing operation in the graph domain. While the definition in \cite{shuman2016vertex} does not amount to multiplication in the vertex domain, it  exhibits a number of convenient properties.}}. Nonetheless, \eqref{eqn_windows} can be used to design windows with small spectral distortion. Ideal windows are such that $\tbW=\bbI$ which can be achieved by setting $\bbw=\bbone$, although this is unlikely to be of any use. More interestingly, \eqref{eqn_windows} implies that good windows for spectral estimation must have $\tbW \approx \bbI$ or can allow nonzero values in columns $k$ where the components $\tdx_k$ of $\tbx$ are known or expected to be small.

Turning now to the problem of PSD estimation, consider the estimate $\hbp$ obtained after computing the periodogram in \eqref{E:nonpar_PSD_periodogram} using a \textit{single} realization $\bbx$, so that we have that $\hbp=|\bbV^H\bbx|^2$. Suppose now that we window the realization $\bbx$ to produce $\bbx_{\bbw} = \diag(\bbw) \bbx$ and compute the \textit{windowed periodogram} $\hbp_{\bbw} :=   \left| \bbV^H\bbx_\bbw \right|^2$. Utilizing the definition of the window's frequency representation, we can write this windowed periodogram as
\begin{equation}\label{E:nonpar_PSD_windowed_periodogram}
\hbp_{\bbw} :=   \big| \bbV^H \bbx_\bbw       \big|^2 
=   \big| \bbV^H \diag(\bbw)\bbx \big|^2
=   \big| \tbW \bbV^H\bbx        \big|^2 .                
\end{equation}
The expression in \eqref{E:nonpar_PSD_windowed_periodogram} can be used to compute the expectation of the windowed periodogram that we report in the following proposition.
%
\begin{myproposition}\label{prop_bias_windowed_periodogram}\normalfont 
	Let $\bbp$ be the PSD of a process $\bbx$ that is stationary with respect to the shift $\bbS=\bbV\bbLam\bbV^H$. The expectation of the windowed periodogram $\hbp_{\bbw}$ in \eqref{E:nonpar_PSD_windowed_periodogram} is,
	\begin{align}\label{eqn_windowed_periodogram}
	\E{\hbp_{\bbw}} = (\tbW\circ \tbW^*) \bbp .
	\end{align} \end{myproposition}

%
\begin{IEEEproof} Write $\hbp_{\bbw} = \big| \tbW \bbV^H\bbx\big|^2 = \diag(\tbW \bbV^H\bbx\bbx^H\bbV\tbW)$. Take the expectation and use $\E{\bbx\bbx^H}=\bbC_x$ to write $\E{\hbp_{\bbw}} = \diag(\tbW \bbV^H\bbC_x\bbV\tbW^H)$. Further observe that $\bbV^H\bbC_x\bbV = \diag(\bbp)$ to conclude that  $\hbp_{\bbw} = \diag(\tbW\diag(\bbp)\tbW^H)$. This latter expression is identical to \eqref{eqn_windowed_periodogram}.\end{IEEEproof}

%
Prop. \ref{prop_bias_windowed_periodogram} implies that the windowed periodogram in \eqref{E:nonpar_PSD_avw_periodogram} \black{is a biased estimate of $\bbp$, with the bias being} determined by the dual of the windowing operator in the frequency domain $\tbW$. 

Multiple windows yield, in general, better estimates than single windows. Consider then a bank of $M$ windows $\ccalW=\{\bbw_m\}_{m=1}^M$ and use each of the windows $\bbw_m$ to construct the windowed signal $\bbx_m := \diag(\bbw_m) \bbx$. We estimate the PSD $\bbp$ with the \textit{windowed average periodogram}
\begin{equation}\label{E:nonpar_PSD_avw_periodogram}
\hbp_{\ccalW} \!:=\frac{1}{M}\!\sum_{m=1}^M \left| \bbV^H\bbx_m \right|^2
=\!\frac{1}{M}\!\sum_{m=1}^M \left| \bbV^H\diag(\bbw_m)\bbx \right|^2\!.
\end{equation}
The estimator $\hbp_{\ccalW}$ is an average of the windowed periodograms in \eqref{E:nonpar_PSD_windowed_periodogram} but is also reminiscent of the periodogram in \eqref{E:nonpar_PSD_periodogram}. The difference is that in \eqref{E:nonpar_PSD_periodogram} the samples $\{\bbx_r\}_{r=1}^R$ are independent observations whereas in \eqref{E:nonpar_PSD_avw_periodogram} the samples $\{\bbx_m\}_{m=1}^M$ are all generated through multiplications with the window bank $\ccalW$. This means that: (i) There is some distortion in the windowed periodogram estimate because the windowed signals $\bbx_m$ are used in lieu of $\bbx$. (ii) The different signals $\bbx_m$ are correlated with each other and the reduction in variance resulting from the averaging operation in \eqref{E:nonpar_PSD_avw_periodogram} is less significant than the reduction of variance observed in Prop. \ref{P:CovariancePeriodogram}.

To study these effects, given the windowing operation $\diag(\bbw_m)$, we obtain its dual in the frequency domain as $\tbW_m:=\bbV^H\diag(\bbw_m)\bbV$ [cf. \eqref{eqn_windows}], and use those to define the power \textit{spectrum mixing} matrix of windows $m$ and $m'$ as the componentwise product
\begin{equation}\label{E:Spectrum_mixing_matrices}
\tbW_{mm'}:=\tbW_m\circ \tbW_{m'}^*.
\end{equation} 
We use these matrices to give expressions for the bias and covariance of the estimator in \eqref{E:nonpar_PSD_avw_periodogram} in the following proposition.

%
\begin{myproposition}\label{P:BiasCovMSE_Av_W_Period}\normalfont 
	Let $\bbp$ be the PSD of a process $\bbx$ that is stationary with respect to the shift $\bbS=\bbV\bbLam\bbV^H$. A single observation $\bbx$ is given along with the window bank $\ccalW=\{\bbw_m\}_{m=1}^M$ and the windowed average periodogram $\hbp_{\ccalW}$ is computed as in \eqref{E:nonpar_PSD_avw_periodogram}. The expectation of the estimator $\hbp_{\ccalW}$ is
	\begin{align}\label{E:bias_avw_period_PSD}
	\E{\hbp_{\ccalW}} = \frac{1}{M}   \sum_{m=1}^M \tbW_{mm} \bbp.
	\end{align}
	Equivalently, $\hbp_{\ccalW}$ is biased with bias $\bbb_{\ccalW}:=\E{\hbp_{\ccalW}}-\bbp$. Further define the covariance matrix of the windowed periodogram as $\bbSigma_{\ccalW}:=
	\E{(\hbp_{\ccalW} - \E{\hbp_{\ccalW}})(\hbp_{\ccalW} - \E{\hbp_{\ccalW}})^H}$. If the process $\bbx$ is assumed Gaussian and $\bbS$ is symmetric, the trace of the covariance matrix can be written as
	\begin{align}\label{E:cov_avw_period_PSD}
	\Tr[\bbSigma_{\ccalW}]
	= \frac{2}{M^2} \sum_{m=1,m'=1}^M
	\Tr\left[\big(\tbW_{mm'}\bbp\big)\big(\tbW_{mm'}\bbp\big)^H\right].
	\end{align} \end{myproposition}

%
\begin{IEEEproof} See Appendix B, where the expression of $\Tr[\bbSigma_{\ccalW}]$ for nonsymmetric normal shifts is given too [cf. \eqref{E:diag_elem_cov_avw_v2_2}].
\end{IEEEproof}

%
The first claim of Prop. \ref{P:BiasCovMSE_Av_W_Period} is a generalization of the bias expression for the bias of windowed periodograms in \eqref{eqn_windowed_periodogram}. It states that the estimator $\hbp_{\ccalW}$ is biased with a bias determined by the average of the spectrum mixing matrices $\tbW_{mm}=\tbW_m\circ \tbW_{m}^*$. The form of these mixing matrices depends on the design of the window bank $\ccalW=\{\bbw_m\}_{m=1}^{M}$ and on the topology of the graph $\mathcal{G}$. Observe that even if the individual spectrum mixing matrices $\tbW_{mm}$ are not close to identity we can still have small bias by making their weighted sum $M^{-1} \sum_{m} \tbW_{mm} \approx \bbI$.

The second claim in Prop. \ref{P:BiasCovMSE_Av_W_Period} is a characterization of the trace of the covariance matrix $\bbSigma_{\ccalW}$. To interpret the expression in \eqref{E:cov_avw_period_PSD} it is convenient to separate the summands with $m=m'$ from the rest to write
\begin{align}\label{E:cov_avw_period_PSD_term1}
\Tr[\bbSigma_{\ccalW}]
= &\frac{2}{M^2} \sum_{m=1}^M
\Tr\left[\big(\tbW_{mm}\bbp\big)\big(\tbW_{mm}\bbp\big)^H\right] \\ \nonumber
&+ \frac{2}{M^2} \sum_{m=1,m'\neq m}^M
\Tr\left[\big(\tbW_{mm'}\bbp\big)\big(\tbW_{mm'}\bbp\big)^H\right]
\end{align}
Comparing \eqref{E:cov_avw_period_PSD_term1} with \eqref{eqn_windowed_periodogram}, we see that the terms in the first summand are $\Tr[(\tbW_{mm}\bbp)(\tbW_{mm}\bbp)^H] = \|\mbE[\hbp_{\bbw_m}]\|_2^2$. Given that the windows have energy $\|\bbw\|_2^2=\|\bbone\|_2^2=N$ we expect $\|\mbE[\hbp_{\bbw_m}]\| \approx \|\E{\hbp_{\bbx}}\|_2=\|\bbp\|_2$. This implies that the first sum in \eqref{E:cov_avw_period_PSD_term1} is approximately proportional to $2\|\bbp\|_2^2/M$. Thus, this first term behaves as if the different windows in the bank $\ccalW$ were generating independent samples [cf. \eqref{E:mse_period}]. 

The effect of the correlation between different windowed signals appears in the cross terms of the second sum, which can be viewed as the price to pay for the signals $\{\bbx_m\}_{m=1}^M$ being generated from the same realization $\bbx$ instead of actually being independent. We explain next that the price of this second sum is smaller than the gain we get in the first term.

%
\medskip \noindent {\bf Window design. } The overall MSE is given by the squared bias norm summed to the trace of the covariance matrix,
\begin{align}\label{E:mse_avw_period_PSD}
\MSE(\hbp_{\ccalW}) = \|\bbb_{\ccalW}\|_2^2+\Tr[\bbSigma_{\ccalW}].
\end{align}
The expression in \eqref{E:mse_avw_period_PSD} can be used to design (optimal) windows with minimum MSE. Do notice that the bias in \eqref{E:mse_avw_period_PSD} depends on the unknown PSD $\bbp$. This problem can be circumvented by making $\bbp=\bbone$ in the bias and trace expressions in Prop. \ref{P:BiasCovMSE_Av_W_Period}. This choice implies that the PSD is assumed white a priori. If some other knowledge of the PSD is available, it can be used as an alternative prior. Irrespectively of the choice of prior, finding windows that minimize the MSE is computationally challenging.

An alternative approach to design the window bank $\ccalW=\{\bbw_m\}_{m=1}^M$ is to exploit the local properties of the random process $\bbx$. As stated in Property \ref{P:locality_stat_MA_process}, some stationary processes are expected to have correlations with local structure. It is then reasonable to expect that windows without overlap capture independent information that results in a reduction of the cross-terms in \eqref{E:cov_avw_period_PSD_term1}. This intuition is formalized next.

%
\begin{myproposition}\label{prop_local_windows}\normalfont
	Consider two windows $\bbw_m$ and $\bbw_{m'}$ and assume that the distance between any node in $\bbw_m$ and any node in $\bbw_{m'}$ is larger than $2L$ hops. If the process $\bbx$ can be modeled as the output of an $L$-degree filter, then it holds that $\Tr[(\tbW_{mm'}\bbp)(\tbW_{mm'}\bbp)^H] = 0$. 
\end{myproposition}

%
\begin{IEEEproof} Let $\ccalN(\bbw_m)=\{i:\;[\bbw_m]_i\neq 0\}$ be the set of nodes in the support of $\bbw_m$, and $\ccalN(\bbw_{m'})$ the ones in the support of $\bbw_{m'}$. We know that the distance between any $i\in\ccalN(\bbw_m)$ and any $i'\in\ccalN(\bbw_{m'})$ is greater than $2L$
	. Invoking Property \ref{P:locality_stat_MA_process}, this implies that $[\bbC_x]_{ii'}=0$, which allows us to write $\diag(\bbe_i)\bbC_x\diag(\bbe_{i'})=\bbzero$. Since this is true for any pair $(i,i')$ with $i\in\ccalN(\bbw_m)$ and $i'\in\ccalN(\bbw_{m'})$, we have that: \textit{f1}) $\diag(\bbw_m)\bbC_x\diag(\bbw_{m'})=\bbzero$. Note now that $\tbW_{mm'}\bbp =  \tbW_m\circ \tbW_{m'}^* \bbp= \tbW_{m}\diag(\bbp) \tbW_{m'}^*$, which using the definitions $\tbW_{m}=\bbV^H \diag(\bbw_m) \bbV$ and $\bbC_x = \bbV  \diag(\bbp) \bbV^H$ can be written as: \textit{f2}) $\tbW_{mm'}\bbp=\bbV^H \diag(\bbw_m) \bbC_x \diag(\bbw_{m'}^*)\bbV$. Substituting \textit{f1}) into \textit{f2}) yields $\tbW_{mm'}\bbp = \bbV^H \bbzero \bbV =\bbzero$.   \end{IEEEproof}

%
The result in Prop. \ref{prop_local_windows} implies that if we use windows without overlap, the second sum in \eqref{E:cov_avw_period_PSD_term1} is null. This means that the covariance matrix $\bbSigma_{\ccalW}$ of the PSD estimator in \eqref{E:nonpar_PSD_avw_periodogram} behaves as if the separate windowed samples were independent samples. We emphasize that Prop. \ref{prop_local_windows} provides a lax bound for a rather stringent correlation model. The purpose of the result is to illustrate the reasons why local windows are expected to reduce estimation MSE. In practice, we expect local windows to reduce MSE as long as the correlation decreases with the hop distance between nodes. Windowing design can then be related to clustering (if the windows are non-overlapping) and covering (if they overlap) with the goal of keeping the diameter of the window large enough so that most of the correlation of the process is preserved. \black{Recent results in the context of designing sampling schemes for covariance estimation of graph processes corroborate this point \cite{Sundeep2016CovarianceSampling}.} \Sec \ref{S:NumExper} evaluates the estimation performance when estimating the PSD of a graph process for different types of windows.

%
\subsection{Filter Banks}

Windows reduce the MSE of periodograms by exploiting the locality of correlations. Filter banks reduce MSE by exploiting the locality of the PSD \cite[\Sec \ 5]{stoica2005spectral}. To define filter bank PSD estimators for graph processes, suppose that we are given a filter bank $\ccalQ:=\{\bbQ_k\}_{k=1}^{N}$ with $N$ filters (as many as frequencies). The filters $\bbQ_k$ are assumed linear shift invariant with frequency responses $\tbq_k$, so that we can write $\bbQ_k=\bbV \diag(\tbq_k)\bbV^H$. We further assume that their energies are normalized $\|\tbq_k\|_2^2 = 1$. The $k$th (bandpass) filter is intended to estimate the $k$th component of the PSD $\bbp = [p_1,\ldots,p_N]^T$ through the energy of the output signal $\bbx_k := \bbQ_k \bbx$, i.e., 
\begin{align}\label{eqn_psd_estimate_filter_bank_time}
\hhatp_{\tbq_k} := \|\bbx_k\|_2^2 = \| \bbQ_k \bbx \|_2^2 .
\end{align}
The filter bank PSD estimate is given by the concatenation of the individual estimates in \eqref{eqn_psd_estimate_filter_bank_time} into the vector $\hbp_{\ccalQ}:=[\hhatp_{\tbq_1},\ldots,\hhatp_{\tbq_N}]^T$. We emphasize that we can think of filter banks as an alternative approach to generate multiple virtual realizations $\bbx_k$ from a single actual realization $\bbx$. In the case of windowing, realizations $\bbx_m$ correspond to different pieces of $\bbx$. In the case of filter banks, realizations $\bbx_k$ correspond to different filtered versions\footnote{\black{A related approach that filters $\bbx$ using a set of $N'\neq N$ graph filters whose frequency responses are designed by shifting a prespecified kernel and then obtains the PSD estimate via interpolation is presented in \cite{EPFL16stationary}.}} of $\bbx$. 

Using Parseval's theorem and the frequency representations $\tbq_k$ and $\tbx$ of the filter $\bbQ_k$ and the realization $\bbx$, respectively, the estimate in \eqref{eqn_psd_estimate_filter_bank_time} is equivalent to 
\begin{align}\label{E:filter_bank_window_freq}
\hhatp_{\tbq_k}\ =\ \| \tbx_k\|_2^2 
\ =\ \|\diag(\tbq_k)\tilde{\bbx}\|_2^2 .
\end{align}
The expression in \eqref{E:filter_bank_window_freq} guides the selection of the response $\tbq_k$. E.g., if the PSD values at frequencies $k$ and $k'$ are expected to be similar, i.e., if $p_k\approx p_{k'}$ we can make $[\tbq_k]_k = [\tbq_k]_{k'} = (1/\sqrt{2})$ so that the PSD components $[|\bbV^H\bbx|^2]_k$ and $[|\bbV^H\bbx|^2]_{k'}$ are averaged. More generically, the design of the filter $\tbq_k$ can be guided by the bias and variance expressions that we present in the following proposition.

%
\begin{myproposition}\label{P:BiasCovMSE_FilterBank}\normalfont 
	
	Let $\bbp = [p_1,\ldots,p_N]^T$ be the PSD of a process $\bbx$ that is stationary with respect to the shift $\bbS=\bbV\bbLam\bbV^H$. A single observation $\bbx$ is given along with the filter bank $\ccalQ=\{\tbq_k\}_{k=1}^N$ and the filter bank PSD estimates $\hbp_{\ccalQ}$ are computed as in \eqref{E:filter_bank_window_freq}. The expectation of the $k$ entry of $\hbp_{\ccalQ}$ is,
	\begin{align}\label{E:mean_filter_bank_period_PSD}
	\E{\hhatp_{\tbq_k}} :=  (|\tbq_k|^2)^T \bbp .
	\end{align}
	Equivalently, $\hhatp_{\tbq_k}$ is biased with bias $b_{\tbq_k}:=\E{\hhatp_{\tbq_k}}-p_k$. Further define the variance of the $k$th entry of the filter bank estimate $\hbp_{\ccalQ}$ as $\var{\hhatp_{\tbq_k}}:= \E{(\hhatp_{\tbq_k} - \E{\hhatp_{\tbq_k}})^2}$. If the process $\bbx$ is assumed Gaussian and $\bbS$ is symmetric, the variance can be written as
	\begin{align}\label{E:cov_filter_bank_PSD}
	\!\var{\hhatp_{\tbq_k}} \!:= \E{ (\hhatp_{\tbq_k}\!-\!\E{\hhatp_{\tbq_k}})^2}
	\!= 2 \left\|\diag\left(|\tbq_k|^2\right) \bbp \right\|_2^2.\!
	\end{align}\end{myproposition}

%
\begin{IEEEproof}	See Appendix C, where the expression of $\var{\hhatp_{\tbq_k}}$ for nonsymmetric normal shifts is given too [cf. \eqref{E:second_order_moment_fb_v2}]. 
\end{IEEEproof}

%
The variance expression in \eqref{E:mean_filter_bank_period_PSD} shows that the estimation accuracy of a filter bank benefits from an averaging effect -- recall that the filter is normalized to have unit energy $\|\tbq_k\|_2^2 = 1$. This averaging advantage manifests only if the bias in \eqref{E:mean_filter_bank_period_PSD} is made small so that the overall MSE, given by $\MSE(\hbp_{\ccalQ}) = \sum_{k=1}^N b_{\tbq_k}^2 + \var{\hhatp_{\tbq_k}}$, decreases. In time signals the bias is made small by exploiting the fact that the PSD of nearby frequencies are similar. In graph signals, some extra information is necessary to identify frequency components with similar PSD values. If, e.g., the process $\bbx$ is a diffusion process as the one in Example 4, the PSD components $p_k$ and $p_{k'}$ are similar -- irrespectively of $\alpha$ -- if the eigenvalues $\lam_k$ and $\lam_{k'}$ of the shift $\bbS$ are similar. If the eigenvalues of the Laplacian are further ordered, averaging of nearby PSD estimates can be interpreted as the use of a bandpass filter.

A generic approach to designing bandpass filters for PSD estimation is to exploit (FIR) filters, which are attractive due their ability to be implemented distributedly \cite{ssamar_distfilters_allerton15}. Formally, write $\bbQ_k = \sum_{l=1}^L q_k^l \bbS^l$, denote as $\bbq_k=[q_k^1,...,q_k^L]^T$ the vector of filter coefficients, and as $\tbq_k=\bbPsi_L\bbq_k$ its frequency response, where we recall that $\bbPsi_L$ stands for the first $L$ columns of $\bbPsi$. The coefficients $\bbq_k$ could be designed upon substituting $\tbq_k=\bbPsi_L\bbq_k$ into both \eqref{E:mean_filter_bank_period_PSD} and \eqref{E:cov_filter_bank_PSD} and minimizing the resultant MSE. This can be challenging because it involves fourth-order polynomials and requires some prior knowledge on $\bbp$. For the purpose of PSD estimation in time, a traditional approach for suboptimal FIR design is to select coefficients guaranteeing that $[\tbq_k]_k=1$ while minimizing the out-of-band power \cite{hayes2009statistical}. Defining $\bbpsi_{k,L}^T$ as the $k$th row of $\bbPsi_L$, this can be formalized as
\begin{align}\label{E:filter_bank_fir}
\bbq_k := \ \argmin_{\bbq_L}\ \|\bbPsi_L\bbq_L\|_2^2, \qquad
\ \st\              \bbpsi_{k,L}^T\bbq_L=1,
\end{align}
with the constraint forcing $[\tbq_k]_k\!=\!1$ and the objective attempting to minimize the contribution to $\hhatp_{\tbq_k}$ from frequencies other than $k$. If we make $L\geq N$ the solution to \eqref{E:filter_bank_fir} is $[\tbq_k]_{k'}=0$ for all $k'\neq k$. For $L<N$, the filter $\bbq_k$ has a response in which nonzero coefficients $[\tbq_k]_{k'}$ are clustered at frequencies $k'$ that are similar to $k$ -- in the sense of being associated with multipliers $\lam_{k'}\! \approx \!\lam_k $. We further observe that \eqref{E:filter_bank_fir} has the additional advantage of being solved in closed form as
\begin{align}\label{E:filter_bank_fir_sol}
\bbq_k = \big( \bbpsi_{k,L}^T \bbPsi_L^H\bbPsi_L \bbpsi_{k,L}^* \big)^{-1} \big(\bbPsi_L^H\bbPsi_L\big)^{-1}\bbpsi_{k,L},
\end{align}
which does not have unit energy but can be normalized.

%

%

%
\section{Parametric PSD Estimation}\label{S:parametric_PSD_est}

We address PSD estimation assuming that the graph process $\bbx$ can be well approximated by a parametric model in which $\bbx$ is the response of a graph filter $\bbH$ to a white input. As per Def.~\ref{D:WeaklyStionaryGraphProcess_2}, this is always possible if the filter's order is sufficiently large. The goal here is to devise filter representations of an order (much) smaller than the number of signal elements $N$. Mimicking time processes, we devise moving average (MA), autoregressive (AR), and ARMA models. \black{Due to space constraints the focus here is on the modeling of graph processes and the parametric estimation of the generating filters, but not on the design of those filters. Details on this latter topic can be found in, e.g., \cite{SandryMouraSPG_TSP14Freq,loukas2015distributed,segarra2015distributed}.}

%
\subsection{Moving Average Graph Processes}\label{Ss:parametric_PSD_est}

Consider a vector of coefficients $\bbbeta=[\beta_0,...,\beta_{L-1}]^T$ and assume that $\bbx$ is stationary in the graph $\bbS$ and generated by the FIR filter $\bbH(\bbbeta) = \sum_{l=0}^{L-1} \beta_l \bbS^l$. The degree of the filter is less that $N-1$ although we want in practice to have $L\ll N$. If process $\bbx$ is indeed generated as the response of $\bbH(\bbbeta)$ to a white input, the covariance of $\bbx$ can be written as
\begin{align}\label{E:cov_output_filter_FIR_eq1}
\bbC_x(\bbbeta) = \bbH(\bbbeta)\bbH^H(\bbbeta)
= \sum_{l=0,l'=0}^{L-1} (\beta_l \bbS^l)(\beta_{l'}\bbS^{H})^{l'}.
\end{align}
The PSD corresponding to the covariance in \eqref{E:cov_output_filter_FIR_eq1} is the magnitude squared of the frequency representation of the filter. To see this formally, notice that it follows from the definition in \eqref{E:PSD_stat_graph_process} that $\bbp(\bbbeta) = \diag\left(\bbV^H \bbC_x(\bbbeta) \bbV\right)$. Writing the covariance matrix as $\bbC_x(\bbbeta) = \bbH(\bbbeta)\bbH^H(\bbbeta)$ and the frequency representation of the filter as $\tbh(\bbbeta) = \diag(\bbV\bbH(\bbbeta)\bbV^H)$, it follows readily that $\bbp(\bbbeta) = |\tbh(\bbbeta)|^2 $. For the purposes of this section the latter will be written explicitly in terms of $\bbbeta$ as [cf. \eqref{E:Filter_input_output_freq}]
\begin{equation}\label{E:PSD_via_filter_coeff}
\bbp (\bbbeta)\ =\ |\tbh(\bbbeta)|^2
\ =\ |\bbPsi_L \bbbeta|^2.
\end{equation}
The covariance and PSD expressions in \eqref{E:cov_output_filter_FIR_eq1} and \eqref{E:PSD_via_filter_coeff} are graph counterparts of MA time processes generated by FIR filters -- see \Sec \ref{S:DiffusionProcStationarity} for a discussion on their practical relevance.

The estimation of the coefficients $\bbbeta$ can be addressed in either the graph or graph frequency domain. In the graph domain we compute the sample covariance $\hbC_x = \bbx\bbx^H$ and introduce a distortion function $D_\bbC(\hbC_x, \bbC_x(\bbbeta))$ to measure the similarity of $\hbC_x$ and $\bbC_x(\bbbeta)$. The filter coefficients $\bbbeta$ are then selected as the ones with minimal distortion,
\begin{align}\label{E:opt_hFIR_coef_space}
\hat{\bbbeta}=\argmin_{\bbbeta}\ D_\bbC(\hbC_x, \bbC_x(\bbbeta)).  
\end{align}
The expression for $\bbC_x(\bbbeta)$ in \eqref{E:cov_output_filter_FIR_eq1} is a quadratic function of $\bbbeta$ that is generally indefinite. The optimization problem in \eqref{E:opt_hFIR_coef_space} will therefore be not convex in general. 

To perform estimation in the frequency domain we first compute the periodogram $\hbp_{\pg}$ defined in \eqref{E:nonpar_PSD_periodogram}. We then introduce a distortion measure $D_\bbp (\hbp_{\pg}, |\bbPsi_L \bbbeta|^2)$ to compare the periodogram $\hbp_{\pg}$ with the PSD $|\bbPsi_L \bbbeta|^2$ and select the coefficients $\bbbeta$ that solve the following optimization
\begin{align}\label{E:opt_hFIR_coef_freq}
\hat{\bbbeta}:=\argmin_{\bbbeta} \ D_\bbp (\hbp_{\pg}, |\bbPsi_L \bbbeta|^2) .
\end{align}
We observe that although we use the periodogram in \eqref{E:opt_hFIR_coef_freq}, any of the nonparametric methods of \Sec \ref{Ss:nonparametric_PSD_est} can be used instead. Since the quadratic form $|\bbPsi_L \bbbeta|^2$ in \eqref{E:opt_hFIR_coef_freq} is also indefinite, the optimization problem in \eqref{E:opt_hFIR_coef_freq} is not necessarily convex. In the particular case when the distortion $D_\bbp (\hbp_{\pg}, |\bbPsi_L \bbbeta|^2) = \| \hbp_{\pg} - |\bbPsi_L \bbbeta|^2\|_2^2$ is the Euclidean 2-norm, efficient (phase-retrieval) solvers with probabilistic guarantees are available \cite{fienup1982phase, candes2015wirtingerflow}. Alternative tractable formulations of \eqref{E:opt_hFIR_coef_space} and \eqref{E:opt_hFIR_coef_freq} when the shifts are symmetric, or, when the shifts are positive semidefinite and the filter coefficients are nonnegative are discussed below.

%
\medskip\noindent {\bf Symmetric shifts. }\label{Sss:FIRsymmetricshifts} If the shift $\bbS$ is symmetric, the expression for the covariance matrix in \eqref{E:cov_output_filter_FIR_eq1} can be simplified to a polynomial of degree $2(L-1)$ in $\bbS$,
\begin{align}\label{E:cov_output_filter_FIR_eq3}
\bbC_x(\bbbeta) =\sum_{l=0,l'=0}^{L-1}\beta_l\beta_{l'}\bbS^{l+l'}
:= \sum_{l=0}^{2(L-1)}\gamma_{l}\bbS^{l}
:= \bbC_x(\bbgamma)
\end{align}
In the second equality in \eqref{E:cov_output_filter_FIR_eq3} we have defined the coefficients $\gamma_l := \sum_{l'+l''=l} \beta_{l'}\beta_{l''}$ summing all the coefficient crossproducts that multiply $\bbS^l$ and introduced $\bbC_x(\bbgamma)$ to denote the covariance matrix written in terms of the $\bbgamma$ coefficients. We propose now a relaxation of \eqref{E:opt_hFIR_coef_space} in which $\bbC_x(\bbgamma)$ is used in lieu of $\bbC_x(\bbbeta)$ to yield the optimization problem
\begin{align}\label{E:opt_hFIR_coef_space_reformulated}
\hat{\bbbeta} = \argmin_{\bbgamma}\ D_\bbC(\hbC_x, \bbC_x(\bbgamma)).
\end{align}
If we add the constraints $\gamma_l = \sum_{l'+l''=l} \beta_{l'}\beta_{l''}$, the problem in \eqref{E:opt_hFIR_coef_space_reformulated} is equivalent to \eqref{E:opt_hFIR_coef_space}. By dropping these constraints we end up with a tractable relaxation because \eqref{E:opt_hFIR_coef_space_reformulated} is convex for all convex distortion metrics $D_\bbC(\hbC_x, \bbC_x(\bbgamma))$. A tractable relaxation of \eqref{E:opt_hFIR_coef_freq} can be derived analogously.

%
\medskip\noindent {\bf Nonnegative filter coefficients. }\label{Sss:FIRSDPshifts} When the shift $\bbS$ is positive semidefinite, the elements of the matrix $\bbPsi$ are all nonnegative. If we further restrict the coefficients $\bbbeta$ to be nonnegative, all the elements in the product $\bbPsi_L \bbbeta$ are also nonnegative. This means that in \eqref{E:opt_hFIR_coef_freq} we can replace the comparison $D_\bbp (\hbp_{\pg}, |\bbPsi_L \bbbeta|^2)$ by the comparison $D_\bbp (\sqrt{\hbp_{\pg}}, \bbPsi_L \bbbeta)$
. We can then replace \eqref{E:opt_hFIR_coef_freq} by
\begin{align}\label{E:par_PSD_est_phase_retr_positive}
\hat{\bbbeta}:=\argmin_{\bbbeta\geq\bbzero} \ 
D_\bbp \big(\sqrt{\hbp_{\pg}}, \bbPsi_L \bbbeta\big) .
\end{align}
The optimization in \eqref{E:par_PSD_est_phase_retr_positive} is convex, therefore tractable, for all convex distortion metrics $D_\bbp (\sqrt{\hbp_{\pg}}, \bbPsi_L \bbbeta)$. Do notice that the objective costs in \eqref{E:par_PSD_est_phase_retr_positive} and \eqref{E:opt_hFIR_coef_freq} are {\it not} equivalent and that \eqref{E:par_PSD_est_phase_retr_positive} requires positive semidefinite shifts --such as the Laplacian-- and restricts coefficients to satisfy $\bbbeta\geq\bbzero$. A tractable restriction of \eqref{E:opt_hFIR_coef_space} can be derived analogously.

%
\subsection{Autoregressive Graph Processes}\label{Ss:parametric_PSD_est_AR}

For some processes it is more convenient to use a parametric model that generates an infinite impulse response through an autoregressive filter. \black{As a simple example, consider the diffusion process driven by the graph filter $\bbH=\alpha_0 \sum_{l=0}^{\infty}\alpha^l \bbS^l$, where $\alpha$ represents the diffusion rate and $\alpha_0$ a scaling coefficient.} If the series is summable, the filter can be rewritten as $\bbH=\alpha_0 (\bbI - \alpha \bbS)^{-1}$ from where we can conclude that its frequency response is $\tbh = \diag(\bbV^H\bbH\bbV) = \alpha_0\diag(\bbI - \alpha \bbLambda^{-1})$. This demonstrates that $\bbH$ can be viewed as a single pole AR filter -- see also \cite{loukas2015distributed}. 

Suppose now that $\bbx$ is a random graph process whose realizations are generated by applying $\bbH=\alpha_0 (\bbI - \alpha \bbS)^{-1}$ to a white input $\bbw$. Then, it readily holds that its covariance $\bbC_x$ is [cf. \eqref{E:cov_output_filter}]
\begin{align}\label{E:cov_output_filter_IIR_eq1}
\bbC_x(\alpha_0,\alpha)
= \bbH\bbH^H=\alpha_0^2 (\bbI - \alpha \bbS)^{-1} (\bbI - \alpha \bbS)^{-H},
\end{align}
which implies that the PSD of $\bbx$ is
\begin{equation}\label{E:ClosedForm_PSD_AR1Filter}
\bbp(\alpha_0,\alpha) = \diag\left[\alpha_0^2|\bbI-\alpha\bbLam|^{-2}\right],
\end{equation}
confirming the fact that the expression for the PSD of $\bbx$ is similar to that of a first-order AR time-varying process. We can now proceed to estimate the PSD utilizing the AR parametric models in \eqref{E:cov_output_filter_IIR_eq1} and \eqref{E:ClosedForm_PSD_AR1Filter} as we did in \Sec \ref{Ss:parametric_PSD_est} for MA models. Substituting $\bbC_x(\alpha_0,\alpha)$ for $\bbC_x(\bbbeta)$ in \eqref{E:opt_hFIR_coef_space} yields a graph domain formulation and substituting $\bbp(\alpha_0,\alpha)$ for $|\bbPsi_L \bbbeta|^2$ in \eqref{E:opt_hFIR_coef_freq} yields a graph frequency domain formulation. Since only two parameters must be estimated the corresponding optimization problems are tractable.

If the filter $\bbH=\alpha_0 (\bbI - \alpha \bbS)^{-1}$ is the equivalent of an AR process of order one, an AR process of order $M$ can be written as $\bbH= \alpha_0\prod_{m=1}^M  (\bbI - \alpha_m \bbS)^{-1}$ for some set of diffusion rates $\bbalpha=[\alpha_0,\ldots,\alpha_M]^T$. The frequency response $\tbh=\diag(\bbV^H\bbH\bbV)$ of this filter is 
\begin{equation}\label{E:ClosedForm_freqresp_ARMFilter}
\tbh = \alpha_0\,\diag\bigg[ \prod_{m=1}^M  (\bbI-\alpha_m\bbLam)^{-1}\bigg].
\end{equation}
If we define the graph process $\bbx=\bbH\bbw$ with $\bbw$ white and unitary energy, the covariance matrix $\bbC_x$ can be written as
\begin{align}\label{E:ClosedForm_cov_ARMFilter}
\bbC_x(\bbalpha) = \alpha_0^2\,\prod_{m=1}^M
(\bbI - \alpha_m \bbS)^{-1}(\bbI - \alpha_m \bbS)^{-H}.
\end{align}
The process $\bbx$ is stationary with respect to $\bbS$, because of, e.g., Def.~\ref{D:WeaklyStionaryGraphProcess_2}. The PSD of $\bbx$ can be written as
\begin{align}\label{E:ClosedForm_PSD_ARMFilter}
\bbp(\bbalpha) = \alpha_0^2\,\diag\bigg[\prod_{m=1}^M 
|\bbI-\alpha_m\bbLam|^{-2}\bigg].
\end{align}
As before, we substitute $\bbC_x(\bbalpha)$ for $\bbC_x(\bbbeta)$ in \eqref{E:opt_hFIR_coef_space} to obtain a graph domain formulation and substitute $\bbp(\bbalpha)$ for $|\bbPsi_L \bbbeta|^2$ in \eqref{E:opt_hFIR_coef_freq} to obtain a graph frequency domain formulation. \black{A related approach in the context of identifying the coefficients of a linear predictor using graph filters was presented in \cite{SandryMouraSPG_TSP13}.} For large degree $M$ the problems can become intractable. Yule-Walker schemes \cite[\Sec \ 3.4]{stoica2005spectral} tailored to graph signals may be of help. Their derivation and analysis are left as future work.

%
\begin{remark}\label{R:GMRF}\normalfont 
	To further motivate AR processes, consider the example of $\bbx=\bbH\bbw$ with $\bbH$ being a single-pole filter and $\bbw$ a white and Gaussian input, so that $\bbx$ is Gaussian too. The covariance of this first-order AR process is given by \eqref{E:cov_output_filter_IIR_eq1}, with its inverse covariance (precision matrix) being simply $\bbTheta := \bbC_x^{-1}=(\rho)^{-2} (\bbI - \alpha \bbS)^H(\bbI - \alpha \bbS)$
	. Since $\bbS$ is sparse, the precision matrix $\bbTheta$ is sparse too. Specifically, $\Theta_{i,j}\neq 0$ only if $j$ is in the two-hop neighborhood of $i$. Then, it follows that a Gaussian AR process of order one on the shift $\bbS$ is a GMRF, with the Markov blanket \cite[Ch. 19]{murphy2012machine} of a node $i$ being given by $\ccalN_2(i)$, i.e., the nodes that are within the two-hop neighborhood of $i$. As explained in Example 2, such a process is stationary both on $\bbS$ and on $\bbTheta$. The same holds true for an AR process of order $M$, which in this case will give rise to a GMRF whose Markov blankets are given by the $2M$-hop neighborhoods of the original graph. 
\end{remark}

%
\subsection{Autoregressive Moving Average Graph Processes}\label{Ss:parametric_PSD_est_ARMA}

The techniques in Secs. \ref{Ss:parametric_PSD_est} and \ref{Ss:parametric_PSD_est_AR} can be combined to form ARMA models for PSD estimation. However, as is also done for time signals, we formulate ARMA filters directly in the frequency domain as a ratio of polynomials in the graph eigenvalues. We then define coefficients $\bba:=[a_1,...,a_M]^T$ and $\bbb:=[b_0,...,b_{L-1}]^T$ and postulate filters with frequency response
%
%
%
\begin{equation}\label{E:FreqResponseARMAFilter}
\tbh = \textstyle \diag\Big[\big(\sum_{l=0}^{L-1}b_l\bbLam^l\big) 
\big(1-\sum_{m=1}^{M}a_m\bbLam^m\big)^{-1}\Big].
\end{equation}
To find the counterpart of \eqref{E:FreqResponseARMAFilter} in the graph domain define the matrices $\bbB:=\sum_{l=0}^{L-1} b_l \bbS^l$ and $\bbA:=\sum_{m=1}^M a_m \bbS^m$. It then follows readily that the filter whose frequency response is in \eqref{E:FreqResponseARMAFilter} is $\bbH=(\bbI-\bbA)^{-1}\bbB = \bbB(\bbI-\bbA)^{-1}$. These expressions confirm that we can interpret the filter as the sequential application of finite and infinite response filters.

If we now define the graph process $\bbx=\bbH\bbw$, its covariance matrix follows readily as 
\begin{equation}\label{E:cov_ARMA_param}
\bbC_x(\bba,\bbb)=(\bbI-\bbA)^{-1}\bbB\bbB^H(\bbI-\bbA)^{-H}.
\end{equation}
Since $\bbC_x(\bba,\bbb)$ is diagonalized by the GFT $\bbV$, the process $\bbx$ is stationary with PSD [cf. \eqref{E:FreqResponseARMAFilter}] 
\begin{equation}\label{E:FreqPSD_ARMAFilter}
\bbp(\bba,\bbb)\! = \!\textstyle \diag\Big[\!\ \big|\sum_{l=0}^{L-1}b_l\bbLam^l\big|^2\
\big|1-\sum_{m=1}^{M}a_m\bbLam^m\big|^{-2}  \Big].
\end{equation}
As in the AR and MA models, we can identify the model coefficients by minimizing the covariance distortion $D_\bbC(\hbC_x, \bbC_x(\bba,\bbb))$ or the PSD distortion $D_\bbp (\hbp_{\pg},\bbp(\bba,\bbb))$ [cf. \eqref{E:PSD_via_filter_coeff} and \eqref{E:opt_hFIR_coef_space}]. These optimization problems are computationally difficult.

Alternative estimation schemes can be obtained by reordering \eqref{E:cov_ARMA_param} into $(\bbI-\bbA)\bbC_x(\bbI-\bbA)^H=\bbB\bbB^H$ and solving for either the graph domain distortion
\begin{align}\label{E:ARMA_covariance_A_B}
(\hat{\bba},\hat{\bbb})
:= \argmin_{\bba,\bbb}\;D_\bbC\big((\bbI-\bbA)\hbC_x(\bbI-\bbA)^H,\bbB\bbB^H\big).
\end{align}
or the graph frequency domain distortion
%
%
\begin{align}\label{E:ARMA_psd_A_B}
(\hat{\bba},\hat{\bbb}) &
:= \argmin_{\bba,\bbb} \\ \nonumber &
\textstyle D_\bbp \Big[ \big|1-\sum_{m=1}^{M}a_m\bbLam^m\big|^{2}
\hbp_{\pg},\,
\diag\Big[\! \big|\sum_{l=0}^{L-1}b_l\bbLam^l\big|^2\Big]
\Big].
\end{align}
The formulations in \eqref{E:ARMA_covariance_A_B} and \eqref{E:ARMA_psd_A_B} can still be intractable but we observe that the problems have the same structure as the ones considered in \Sec \ref{Ss:parametric_PSD_est}. The tractable relaxation that we discussed for symmetric shifts and the tractable restriction to nonnegative filter coefficients for positive semidefinite shifts can be then used here as well. 

%
\begin{remark}\label{rmk_parametric_estimation_diffusion_processes}\normalfont
	
	The parametric methods in this section are well tailored to PSD estimation of diffusion processes -- see \Sec \ref{S:DiffusionProcStationarity}. When $L\ll N$ the dynamics in \eqref{E:local_diffusion_singlenode} are accurately represented by a low-order FIR model. Parametric estimation with a MA model as in \Sec  \ref{Ss:parametric_PSD_est} is recommendable. Single pole AR models arise when $\gamma^{(l)} = \gamma$ for all $l$ and $L=\infty$ as we already explained in \Sec \ref{Ss:parametric_PSD_est_AR}. An AR model of order $M$ arises when $M$ single pole models are applied sequentially. The latter implies that AR models are applicable when the diffusion constants $\gamma^{(l)}$ are constant during stretches of time or vary slowly with time. If we consider now $M$ diffusion dynamics with constant coefficients $\gamma^{(l)} = \gamma_m$ for all $l$ running in {\it parallel} and we produce an output as the sum of the $M$ outcomes we obtain an ARMA system with $M$ poles and $M-1$ zeros \cite{loukas2015distributed}.   
\end{remark}

%
\begin{remark}
	\normalfont The methods for PSD estimation that we presented in Secs. \ref{S:nonparametric_PSD_est} and \ref{S:parametric_PSD_est} can be used for covariance estimation as well. This follows directly for some of the parametric formulations in which we estimate filter coefficients that minimize a graph domain distortion -- these include \eqref{E:opt_hFIR_coef_space}, its relaxation in \eqref{E:opt_hFIR_coef_space_reformulated}, and the analogous formulations in Secs. \ref{Ss:parametric_PSD_est_AR} and \ref{Ss:parametric_PSD_est_ARMA}. When this is not done, an estimate for $\bbC_x$ can be computed from a PSD estimate as $\hat{\bbC}_x=\bbV \diag(\hbp_{\pg})\bbV^H$. \black{A further step will be to use the notion of stationarity and the models in this section to estimate the shift $\bbS$ itself (hence, the topology of the network) from a set of signal realizations. See, e.g., \cite{mei2015signal,segarra2016network} for two recent examples along these lines.}
\end{remark}


\section{Numerical experiments}\label{S:NumExper}

\begin{figure*}
	\centering
	
	\begin{subfigure}{.295\textwidth}
		\centering
		\includegraphics[width=\textwidth]{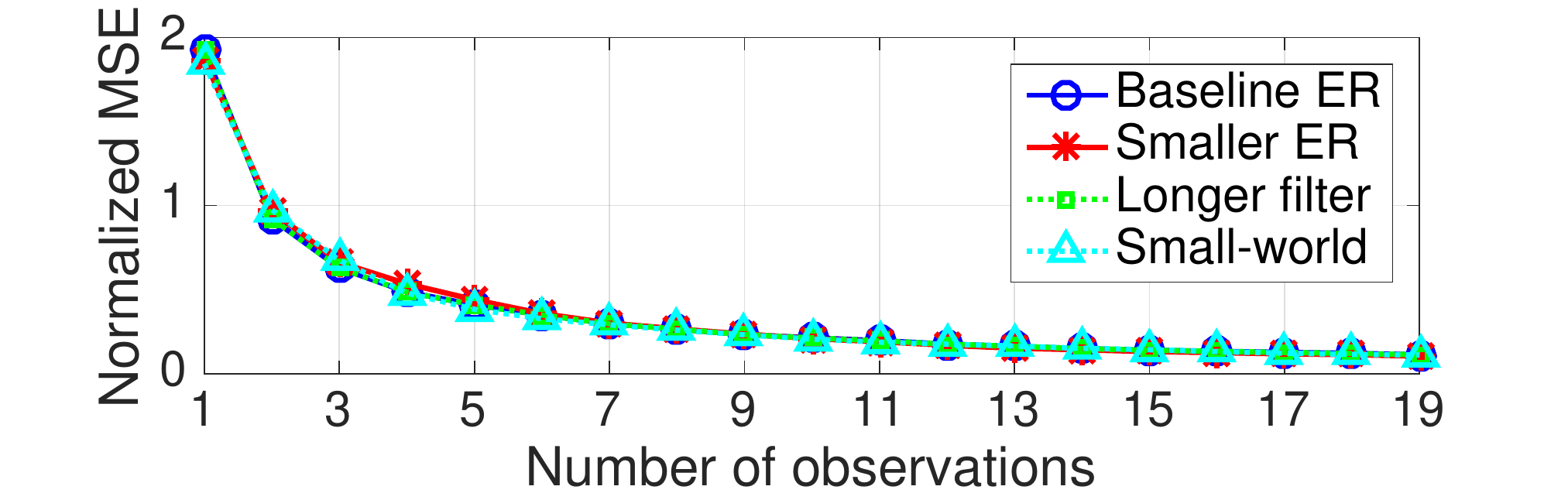}
		\includegraphics[width=\textwidth]{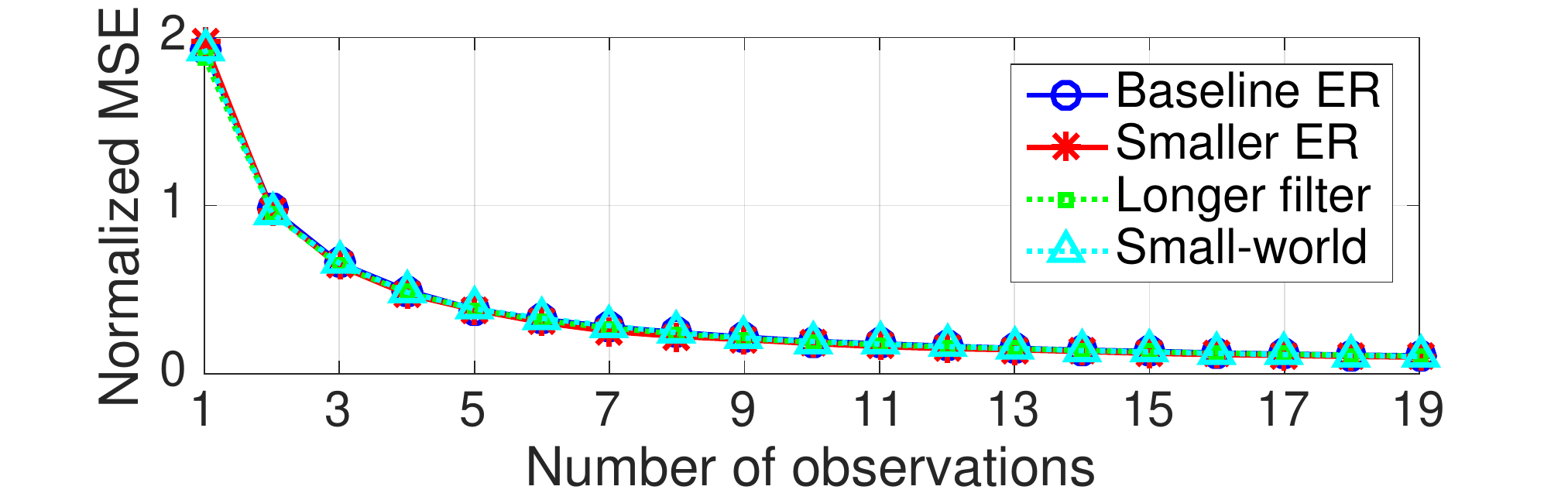}
		\caption{}
		\label{fig:sub1}
	\end{subfigure}%
	\begin{subfigure}{.295\textwidth}
		\centering
		\includegraphics[width=\textwidth, height =0.66\textwidth]{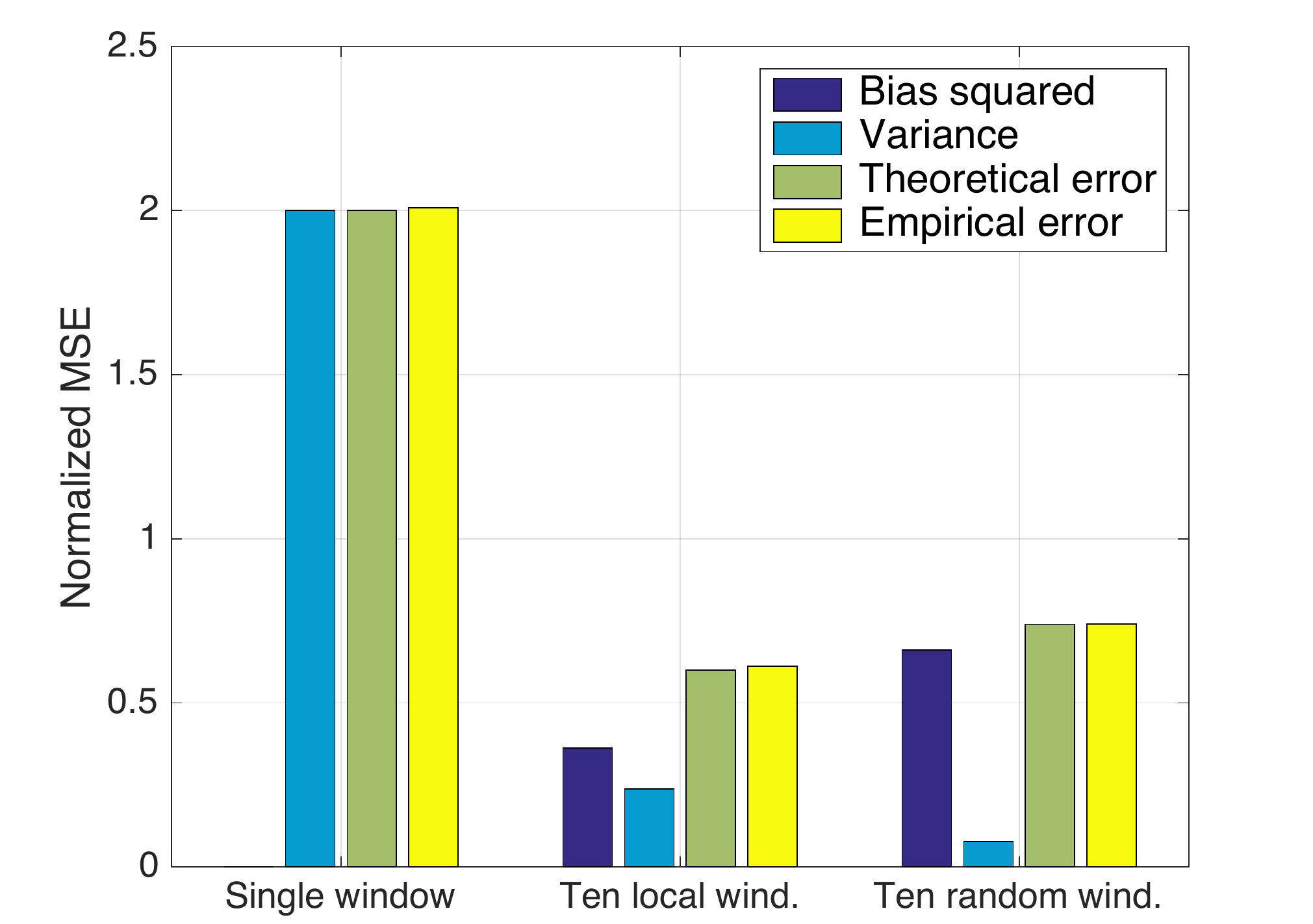}
		\caption{}
		\label{fig:sub2}
	\end{subfigure}%
	\begin{subfigure}{.31\textwidth}
		\centering
		\includegraphics[width=\textwidth]{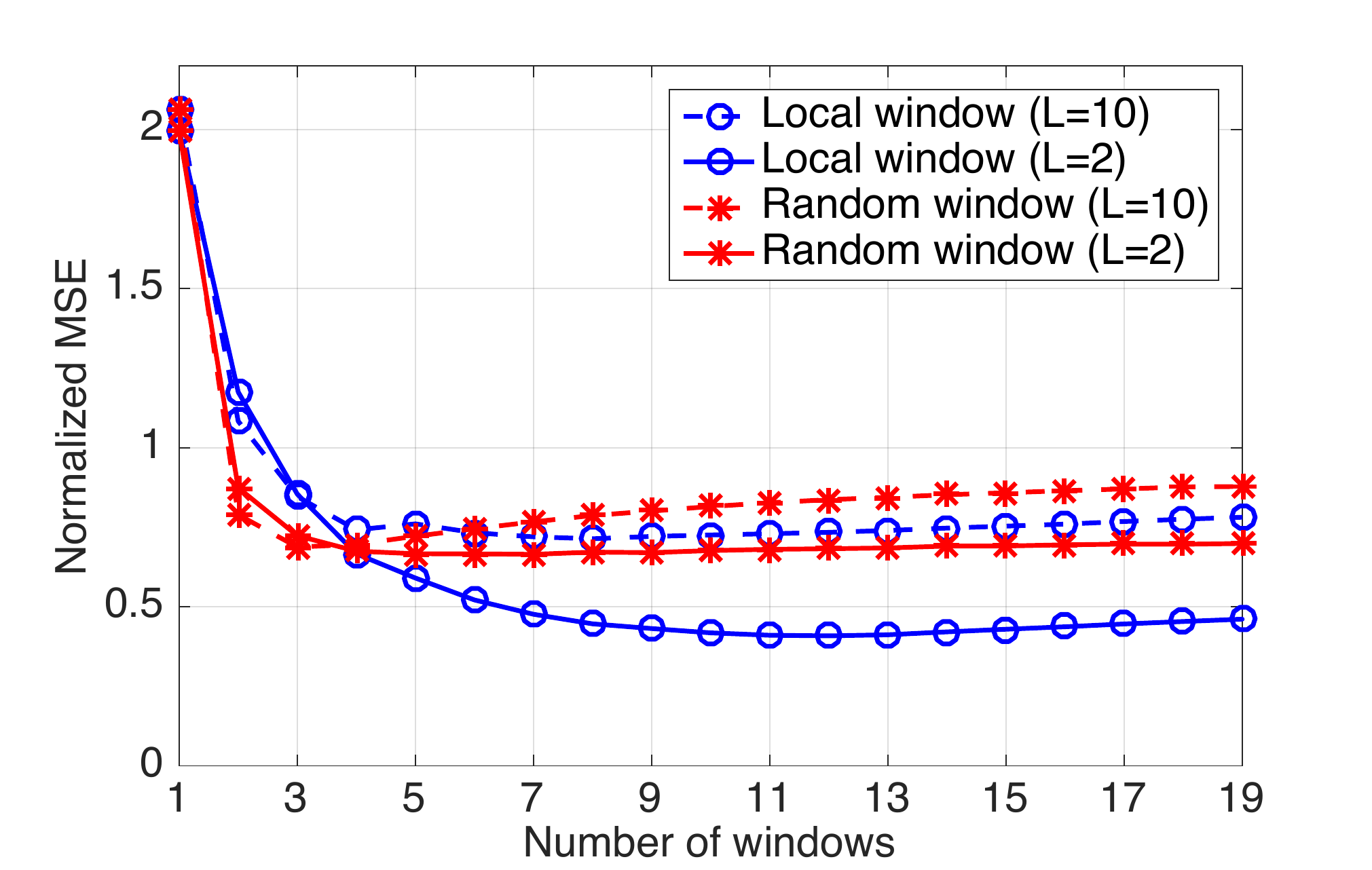}
		\caption{}
		\label{fig:sub3}
	\end{subfigure}
	
	\begin{subfigure}{.30\textwidth}
		\centering
		\includegraphics[width=\textwidth]{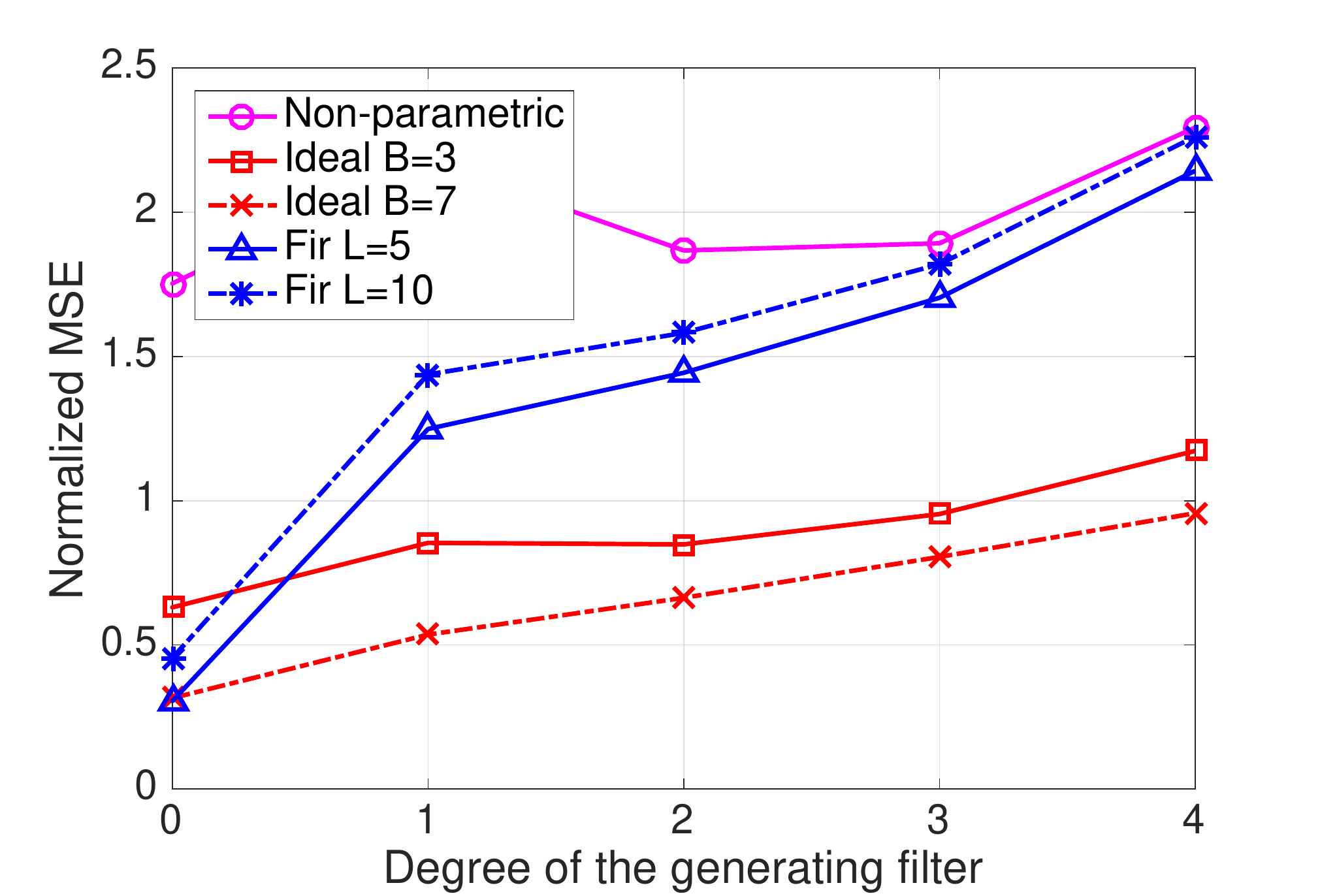}
		\caption{}
		\label{fig:sub4}
	\end{subfigure}%
	\begin{subfigure}{.30\textwidth}
		\centering
		\includegraphics[width=\textwidth]{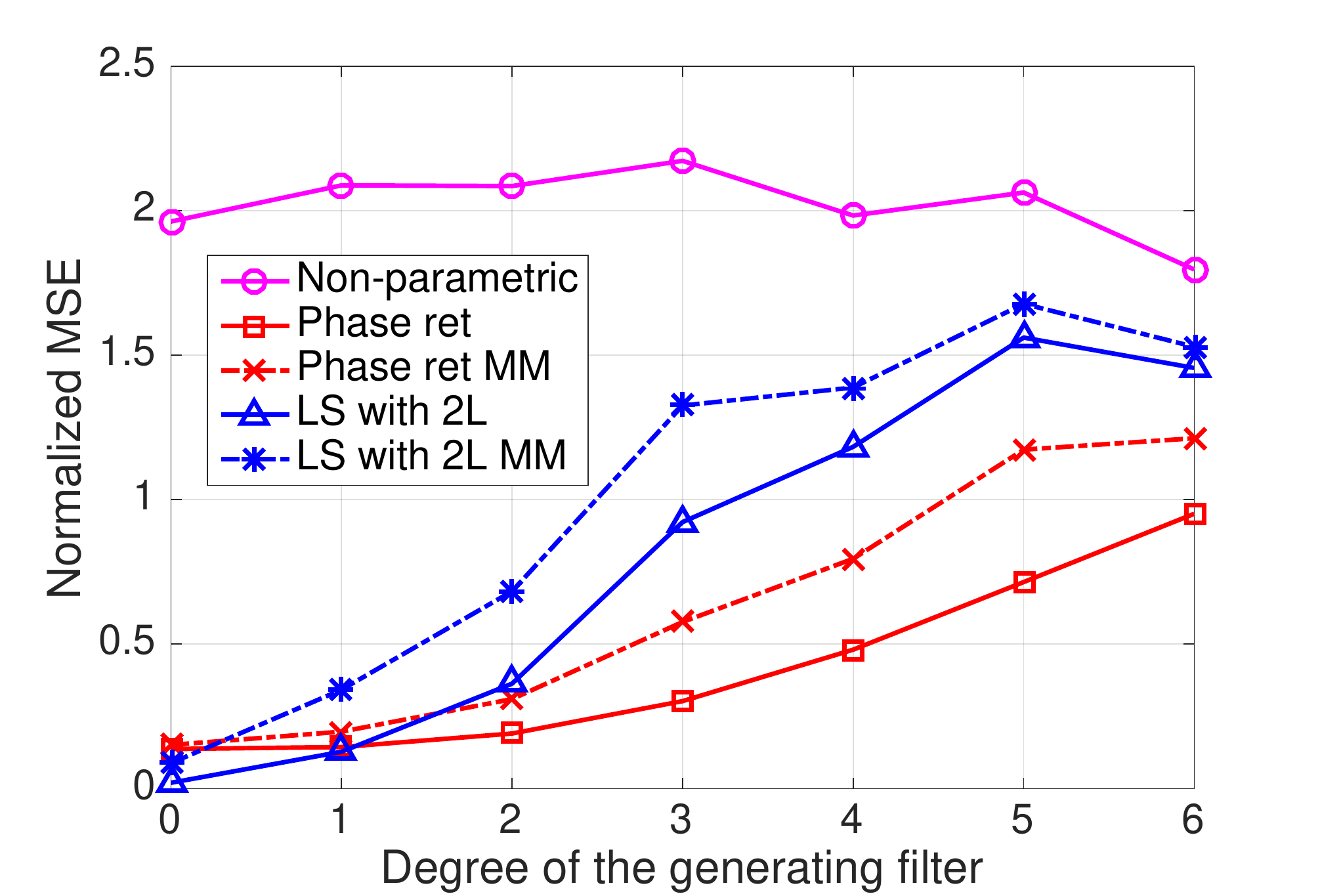}
		\caption{}
		\label{fig:sub5}
	\end{subfigure}%
	\begin{subfigure}{.30\textwidth}
		\centering
		\includegraphics[width=\textwidth]{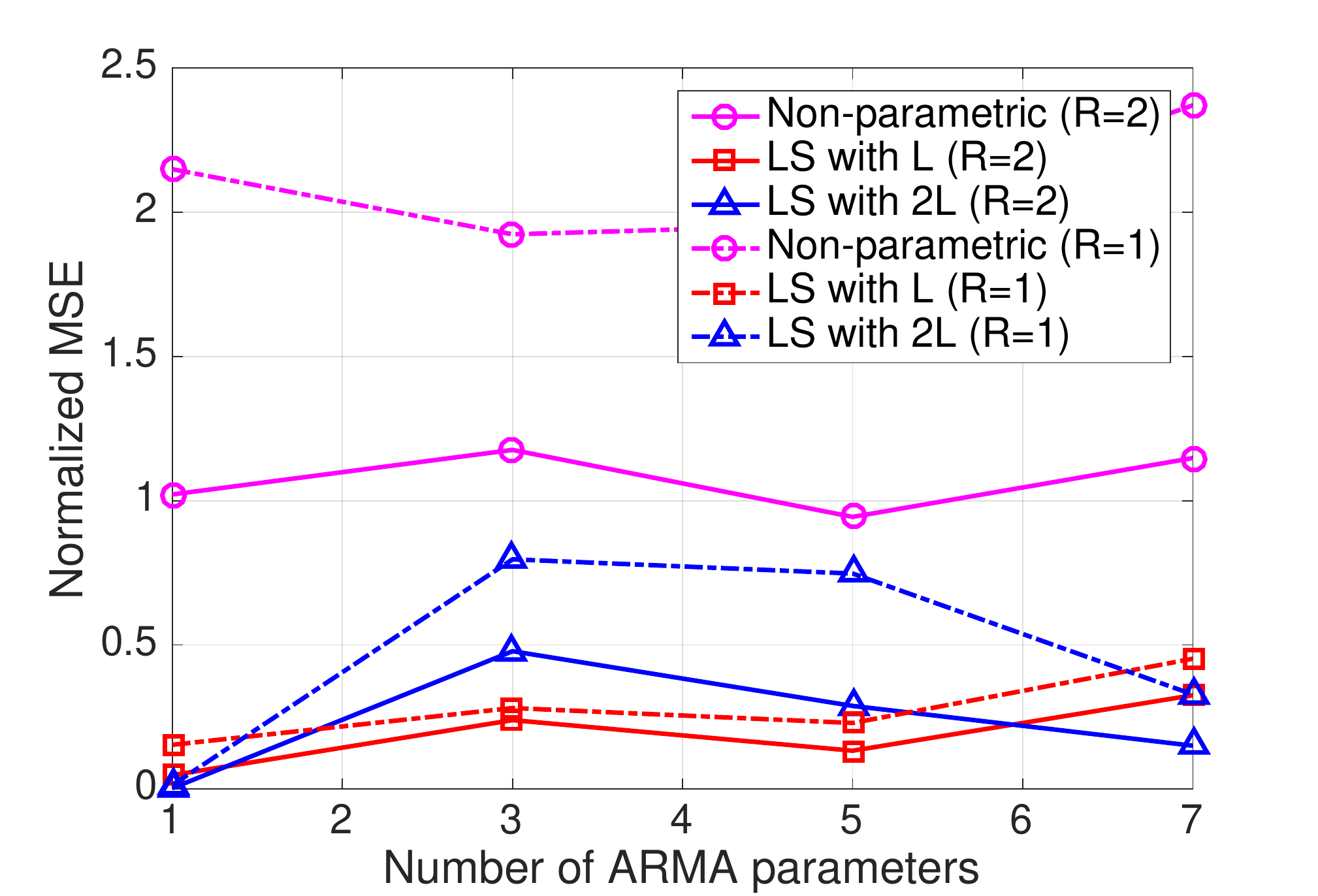}
		\caption{}
		\label{fig:sub6}
	\end{subfigure}
	
	\vspace{-0.05in}
	
	\caption{Normalized MSE (NMSE) for different PSD estimation schemes. (a) Top:  NMSE for the periodogram using Gaussian inputs. Bottom: NMSE for non-Gaussian inputs. (b) Theoretical and empirical NMSE for different window strategies. (c) NMSE as a function of the number of windows for local and random windows. (d) NMSE for  filter banks as a function of the degree of the generating filter. (e) NMSE for MA parametric estimation as a function of the degree of the generating filter. (f) NMSE for different ARMA parametric estimators based on $R=1$ and $R=2$ signal realizations.}
	\vspace{-0.15in}
	\label{F:PSD_est}
\end{figure*}

The implementation and associated benefits of the proposed schemes are illustrated through four test cases (TCs). TC1 and TC2 rely on synthetic graphs to evaluate the performance of nonparametric and parametric PSD estimation methods. TC3 and TC4 illustrate how the concepts and tools of stationary graph processes can be leveraged in practical applications involving real-world signals and graphs. Unless otherwise stated, the results shown are averages across 100 realizations of the particular experiment.

\medskip
\noindent {\bf TC1. Nonparametric methods:} 
We first evaluate the estimation performance of the average periodogram [cf. \eqref{E:nonpar_PSD_periodogram} and \eqref{E:mse_period}] as a function of $R$, the number of realizations observed.
Consider a baseline Erd\H{o}s-R\'{e}nyi (ER) graph with $N\!=\!100$ nodes and edge probability $p=0.05$ \cite{bollobas1998random}. We define its adjacency matrix as the shift and generate signals by filtering white Gaussian noise with a filter of degree 3. In this case, the normalized MSE equals $2/R$ [cf.~\eqref{E:mse_period}] as can be corroborated in Fig.~\ref{fig:sub1} (top). To further confirm this result, we consider three variations of the baseline setting: i) a smaller ER graph with $N=10$ nodes and $p=0.3$, ii) a small-world graph \cite{kolaczyk2009book} obtained by rewiring with probability $q=0.1$ the edges in a regular graph of the same size as the baseline ER, and iii) filtering the noise with a longer filter of degree 6. As expected, Fig.~\ref{fig:sub1} (top) indicates that the normalized MSE is independent of these variations. We then repeat the above setting but for signals generated as filtered versions of \emph{non-Gaussian} white noise drawn from a uniform distribution of unit variance. Even though the MSE expression in \eqref{E:mse_period} was shown for Gaussian signals, we observe that in the tested non-Gaussian setup the evolution of the MSE with $R$ is the same; see Fig.~\ref{fig:sub1} (bottom). \black{This similarity can be explained by the fact that a graph filter is a linear combination of shifted signals and shifting is a linear operation. Hence, invoking the central limit theorem, the larger the filter degree the closer the signals at hand are from being normal.}

The second experiment evaluates the performance of window-based estimators. To assess the role of locality in the window design, we consider graphs generated via a stochastic block model \cite{holland1983stochastic} with $N=100$ nodes and 10 communities with 10 nodes each. The edge probability within each community is $p=0.9$, while the probability for edges across communities is $q=0.1$. We design rectangular non-overlapping windows where the nodes are chosen following two strategies: i) $M=10$ local windows corresponding to the 10 communities, and ii) $M=10$ windows of equal size with randomly chosen nodes.
We use the Laplacian as shift and generate the graph process using a filter with $L=2$ coefficients. Fig.~\ref{fig:sub2} shows the theoretical and empirical normalized MSE for the two designs as well as that of the periodogram for a single window. We first observe that the periodogram has no bias and that the theoretical and empirical errors coincide for the three cases, validating the results in Propositions~\ref{P:CovariancePeriodogram} and \ref{P:BiasCovMSE_Av_W_Period}, respectively. Moreover, we corroborate that windowing contributes to reduce the variance of the estimator. Fig.~\ref{fig:sub2} also illustrates that windows that leverage the community structure of the graph yield a better estimation performance.
To gain insights on the latter observation, we now consider a small-world graph of size $N\!=\!100$ obtained by rewiring with probability $q=0.05$ a regular graph where each node has 10 neighbors.
Both local and random windows are considered, where the local windows are obtained by applying complete linkage clustering \cite{clusteringref} to a metric space given by the shortest path distances between nodes. In order to obtain increasing number of windows $M$, we cut the output dendrogram at smaller resolutions \cite{Carlsson2014}. The random windows are designed to have the same sizes as the local ones. 
The windows are tested for graph processes generated by two filters of different degrees: i) $L=2$, so that nodes that are more than 2 hops away are not correlated (cf. Property \ref{P:locality_stat_MA_process}); and ii) $L=10$, which is greater than the graph diameter, inducing correlations between every pair of nodes. In Fig.~\ref{fig:sub3} we illustrate the performance of local and random windows in these two settings as a function of $M$. We first observe that as $M$ increases, the error first decreases until it reaches an optimal point and then starts to increase. Intuitively, this indicates that at first the reduction in variance outweighs the increase in bias but, after some point, the marginal variance reduction when adding one extra window does not compensate the detrimental effect on the bias.
Moreover, it can be seen that local windows outperform the random ones, especially for localized graph processes ($L=2$). These findings are consistent for other types of graphs, although for graphs with a weaker clustered structure the benefits of local windows are less conspicuous.

\black{The last experiment evaluates the performance of filter-bank estimators. Two types of bandpass filters are considered.} The first type designs the $k$-th filter as an ideal bandpass filter with unit response for the $k$-th frequency and the $B$ frequencies closest to it, and zero otherwise [cf.~\eqref{E:filter_bank_window_freq}]. 
More precisely, being $\lambda_k$ the eigenvalue associated to the $k$-th frequency, we consider the closest frequencies as those with eigenvalues $\lambda_{k'}$ minimizing $|\lambda_k - \lambda_{k'}|$.
The second filter bank type designs the filters using the FIR approach in \eqref{E:filter_bank_fir_sol}. To run the experiments we consider the adjacency matrix of an ER graph with $N\!=\!100$ and $p\!=\!0.05$, and generate signals by filtering white noise.
Fig.~\ref{fig:sub4} shows the MSE performance of both approaches as a function of the degree of the filter that generates the process as well as the nonparametric periodogram estimation. 
We consider two ideal bandpass filters with $B\!=\!3$ and $B\!=\!7$ and two FIR bandpass filters with $L\!=\!5$ and $L\!=\!10$. Fig.~\ref{fig:sub4} indicates that filter banks contribute to reduce the MSE compared to the periodogram.
Moreover, the ideal bandpass filter outperforms the FIR design and, within each type, filters with larger bandwidth ($B\!=\!7$ and $L\!=\!5$) tend to perform better. The reason being that the periodogram-based estimation for a single observation is very noisy, thus, it benefits from the averaging effect of larger bandwidths. 

\begin{figure*}
	\centering
	
	\begin{subfigure}{.28\textwidth}
		\centering
		\includegraphics[width=\textwidth]{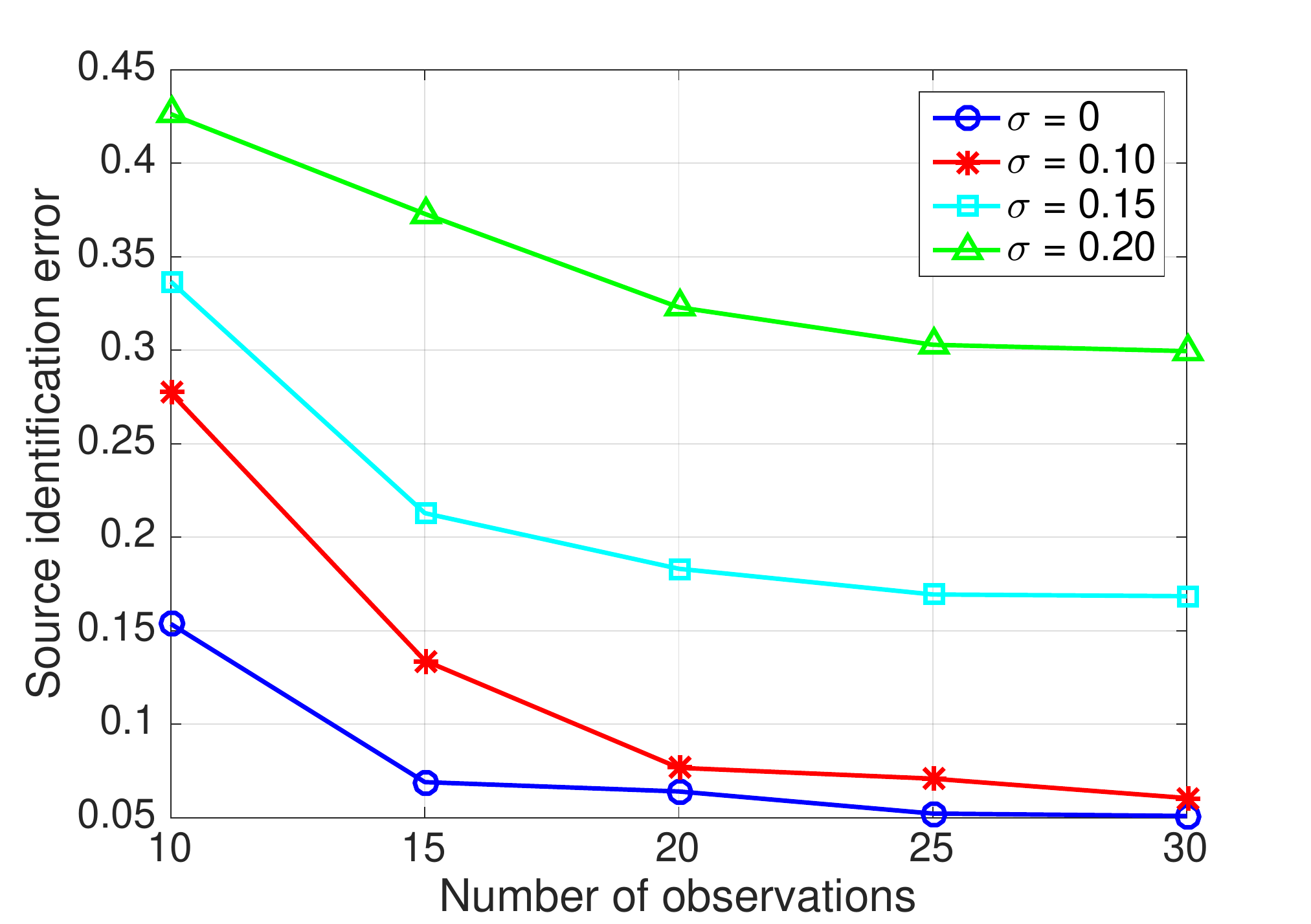}
		\caption{}
		\label{fig:sub10}
	\end{subfigure}%
	\begin{subfigure}{.12\textwidth}
		\centering
		\includegraphics[width=\textwidth]{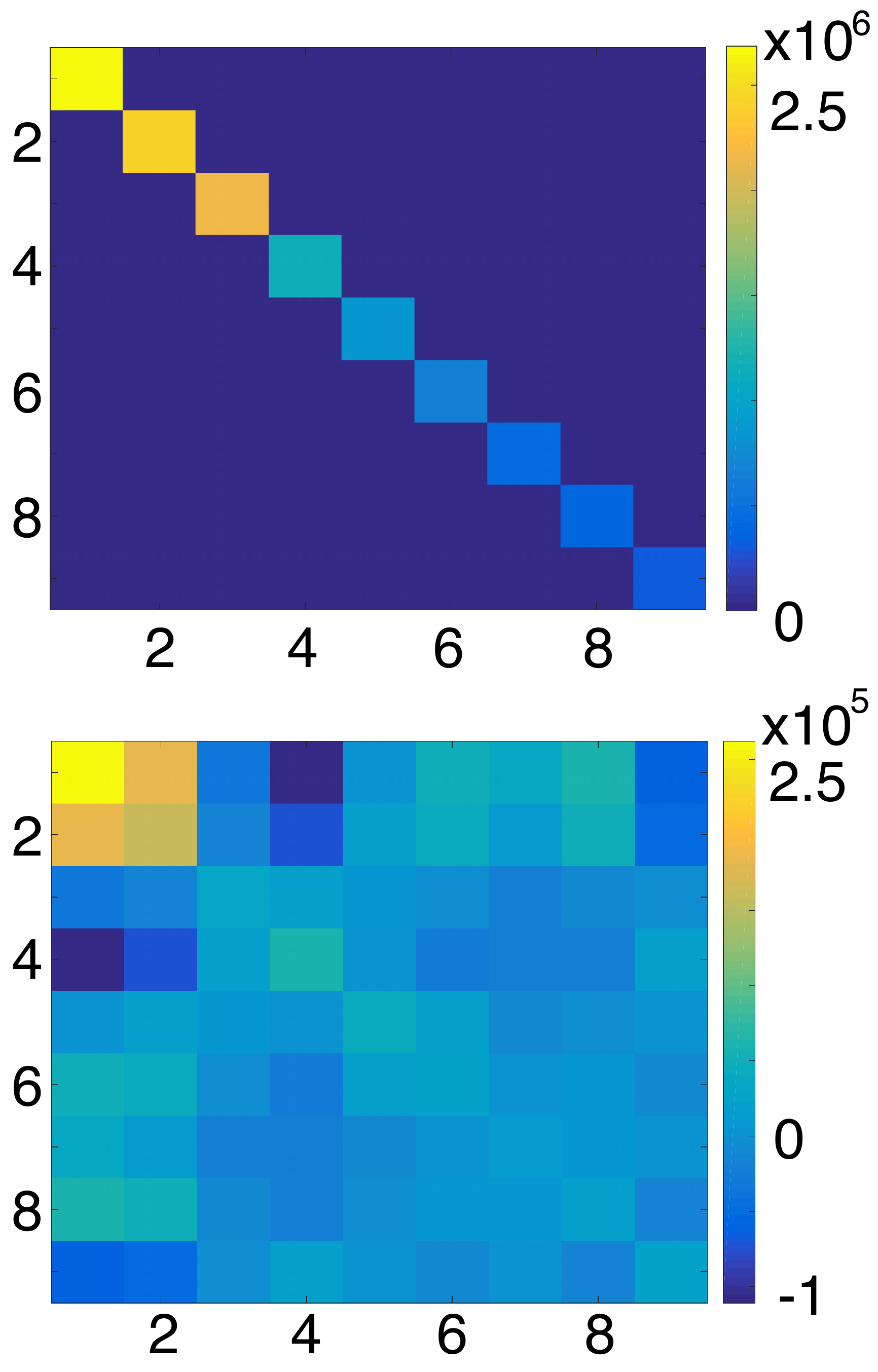}
		\caption{}
		\label{fig:sub101}
	\end{subfigure}%
	\hspace{0.01in} 
	\begin{subfigure}{.26\textwidth}
		\centering
		\includegraphics[width=\textwidth, height = 0.77\textwidth]{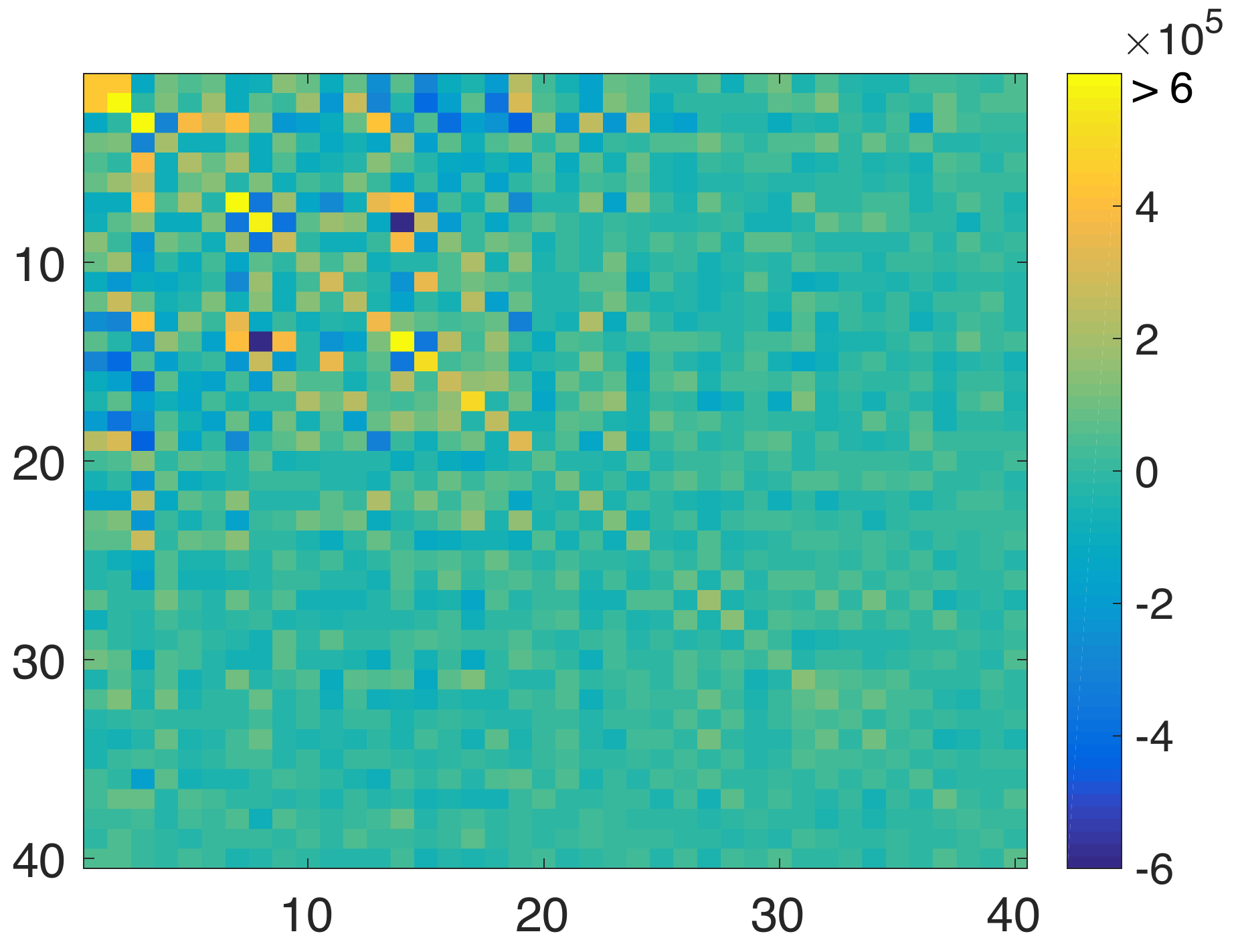}
		\caption{}
		\label{fig:sub11}
	\end{subfigure}%
	\begin{subfigure}{.26\textwidth}
		\centering
		\includegraphics[width=\textwidth]{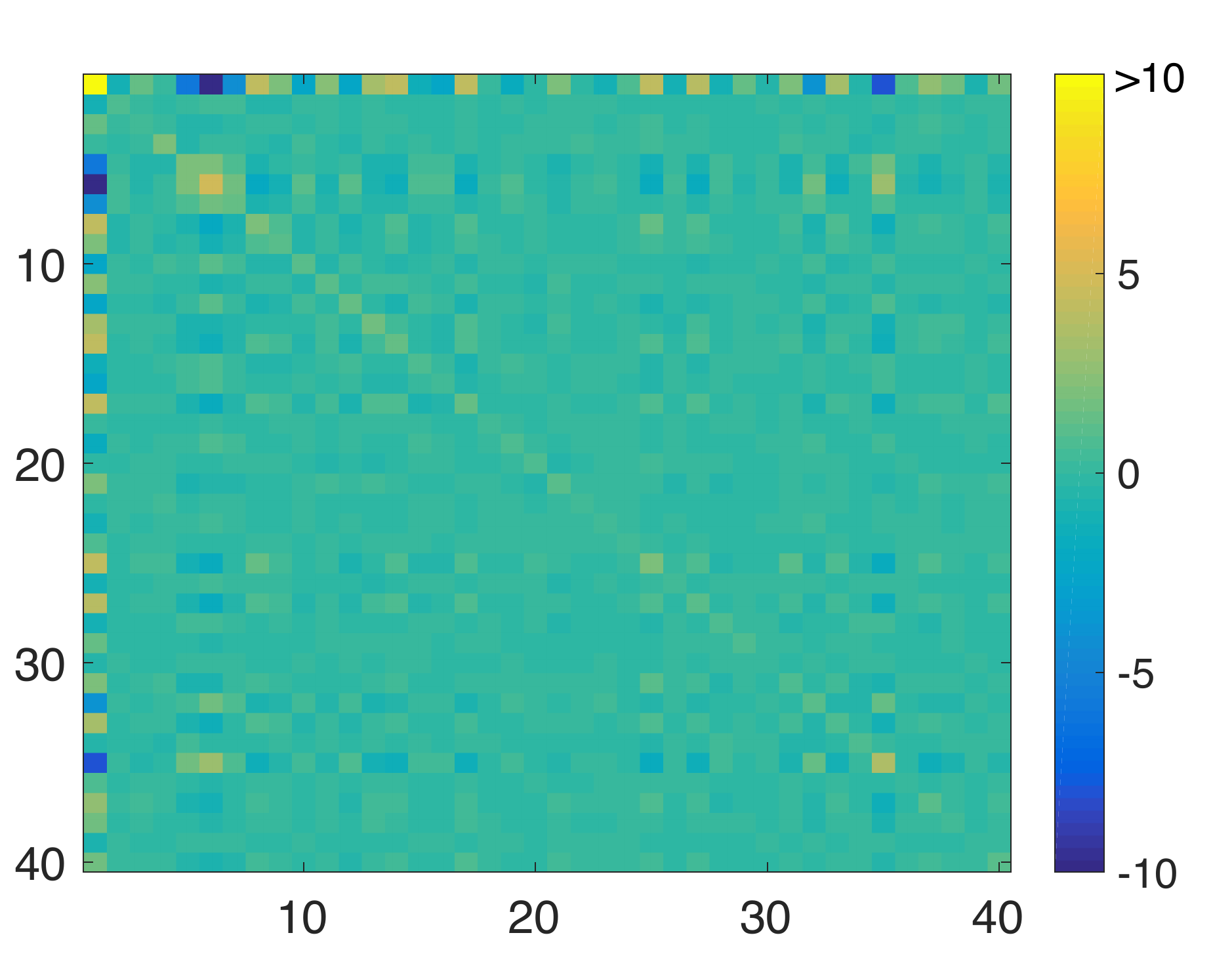}
		\caption{}
		\label{fig:sub12}
	\end{subfigure}
	
	\vspace{-0.05in}
	\caption{(a) Error in the identification of sources in opinion formation dynamics as a function of the number of observed opinions and parametrized by the noise level $\sigma$. \black{(b)-(c) Diagonalization of the sample covariance of the images of an individual with the basis of that same individual (b-top), a different one (b-bottom), and the ensemble sample covariance (c).} \black{(d) Diagonalization of the sample covariance of the functional brain signals in \cite{medaglia2016} with the basis corresponding to the structural brain network.} }
	\vspace{-0.015in}
	\label{F:multipleFigs_stationarity}
\end{figure*}

\medskip
\noindent {\bf TC2. Parametric methods:} We first illustrate the parametric estimation of a MA process. Consider the Laplacian of an ER graph with $N=100$ and $p=0.2$ and processes generated by an FIR filter of length $L$ whose coefficients $\bbbeta$ are selected randomly. The performance of a periodogram is contrasted with that of two parametric approaches: i) an algorithm that estimates the $L$ values in $\bbbeta$ by minimizing \eqref{E:opt_hFIR_coef_freq} via phase-retrieval \cite{candes2015wirtingerflow}; and ii) a least squares algorithm that estimates the $2L-1$ values in $\bbgamma$ by minimizing \eqref{E:cov_output_filter_FIR_eq3}.
The results are shown in Fig.~\ref{fig:sub5} (solid lines). It can be observed that both parametric methods outperform the periodogram since they leverage the FIR structure of the generating filter. Moreover, this difference is largest for smaller values of the degree, since in these cases a few parameters are sufficient to completely characterize the filter of interest. 
Furthermore, we also test our schemes for a model mismatch (MM) scenario where the MA schemes assume that the order of the process is $L+2$ instead of $L$ (dashed lines in Fig.~\ref{fig:sub5}). The results show that, although the model mismatch degrades the performance, the parametric estimates are still superior to the periodogram. 

The second experiment considers ARMA processes with $L$ poles and $L$ zeros. The coefficients are drawn randomly from a uniform distribution with support $[0,1]$ and the shift is selected as in the previous experiment. We compare the periodogram estimation with two schemes: i) a least squares (LS) algorithm that estimates $2L$ coefficients, i.e., the counterpart of \eqref{E:cov_output_filter_FIR_eq3} for the problem in \eqref{E:ARMA_psd_A_B}; and ii) a LS algorithm that estimates $L$ nonnegative coefficients, i.e., the counterpart of \eqref{E:par_PSD_est_phase_retr_positive} for \eqref{E:ARMA_psd_A_B}. Note that the latter is computationally tractable because both the eigenvalues of the shift and the coefficients of the filters are nonnegative. The algorithms are tested in two scenarios, with one and two signal realizations available, respectively. Fig.~\ref{fig:sub6} shows that the parametric methods attain smaller MSEs compared to the periodogram. Moreover, note that while increasing the number of observations reduces the MSE for all tested schemes, the reduction is more pronounced for nonparametric schemes. This is a manifestation of the fact that parametric approaches tend to be more robust to noisy or imperfect observations.

\medskip
\noindent {\bf \black{TC3. Real-world graphs with synthetic signals:}}  We now demonstrate how the tools developed in this paper can be useful in practice through a few real-world experiments.
The first one deals with source identification in opinion formation dynamics. We consider the social network of Zachary's karate club \cite{Zachary1977} represented by a graph $\ccalG$ consisting of 34 nodes or members of the club and 78 undirected edges symbolizing friendships among members. Denoting by $\bbL$ the Laplacian of $\ccalG$, we define the GSO $\bbS \!=\! \bbI \!-\! \alpha \bbL$ with $\alpha \!=\! 1/\lambda_{\max}(\bbL)$, modeling the diffusion of opinions between the members of the club. A signal $\bbx$ can be regarded as a unidimensional opinion of each club member regarding a specific topic, and each application of $\bbS$ can be seen as an opinion update. We assume that an opinion profile $\bbx$ is generated by the diffusion through the network of an initially sparse (rumor) signal $\bbw$. 
More precisely, we model $\bbw$ as a white process such that $w_i\!=\!1$ with probability $0.05$, $w_i\!=\!-1$ with probability $0.05$, and $w_i\!=\!0$ otherwise. We are given a set $\{\bbx_r\}_{r=1}^R$ of opinion profiles generated from different sources $\{\bbw_r\}_{r=1}^R$ diffused through a filter of unknown nonnegative coefficients $\bbbeta$. \black{Observe that the opinion profiles $\bbx_r$ are typically dense since the degree of the filters considered is in the order of the diameter of the graph.} Our goal is to identify the sources of the different opinions, i.e., the nonzero entries of $\bbw_r$ for every $r$.
Our approach proceeds in two phases. First, we use $\{\bbx_r\}_{r=1}^R$ to identify the parameters $\bbbeta$ of the generating filter. We do this by solving \eqref{E:par_PSD_est_phase_retr_positive} via least squares. 
Second, given the set of coefficients $\bbbeta$, we have that $\bbx_{r} = \sum_{l=0}^{L-1} \beta_l \bbS^l \bbw_{r}$. Thus, we estimate the sources $\bbw_{r}$ by solving a $\ell_1$-regularized least squares problem to promote sparsity in the input. In Fig.~\ref{fig:sub10} (blue) we show the proportion of sources misidentified as a function of the number of observations $R$. As $R$ increases, the estimates of the parameters $\bbbeta$ become more reliable, thus leading to a higher success rate. Finally, we consider cases where the observations are noisy. Formally, we define noisy observations $\hat{\bbx}_{r}$ by perturbing the original ones  $\hat{\bbx}_{r} = \bbx_{r} + \sigma \bbz \circ \bbx_{r}$ where $\sigma$ denotes the magnitude of the perturbation and $\bbz$ is a vector with elements drawn from a standard normal distribution. \black{Note that the elements $[\bbx_r]_i = 0$ remain unperturbed, which is equivalent to assuming that we can easily spot people that have not heard about the rumor.} As expected, higher levels of noise have detrimental effects on the recovery of sources. Nevertheless, for moderate noise levels ($\sigma = 0.1$) a performance comparable to the noiseless case can be achieved when observing 20 signals or more.

\begin{figure*}
	\centering
	
	\begin{subfigure}{.110\textwidth}
		\centering
		\includegraphics[width=0.9\textwidth, trim={2cm 1.4cm 1.5cm 1cm},clip]{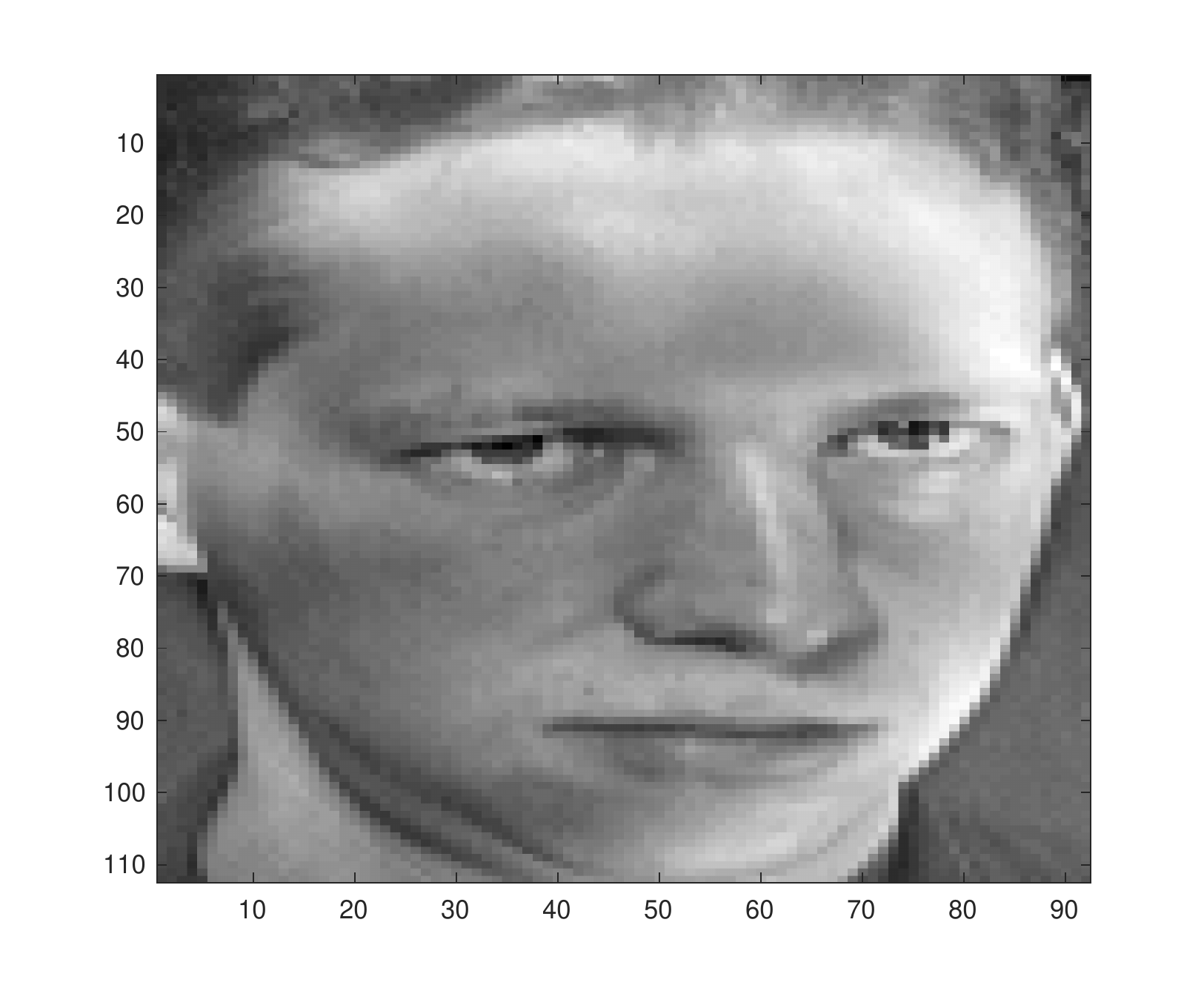}
		\caption{{\footnotesize Ground truth}}
		\label{fig:face_truth}
	\end{subfigure}%
	\begin{subfigure}{.110\textwidth}
		\centering
		\includegraphics[width=0.9\textwidth, trim={2cm 1.4cm 1.5cm 1cm},clip]{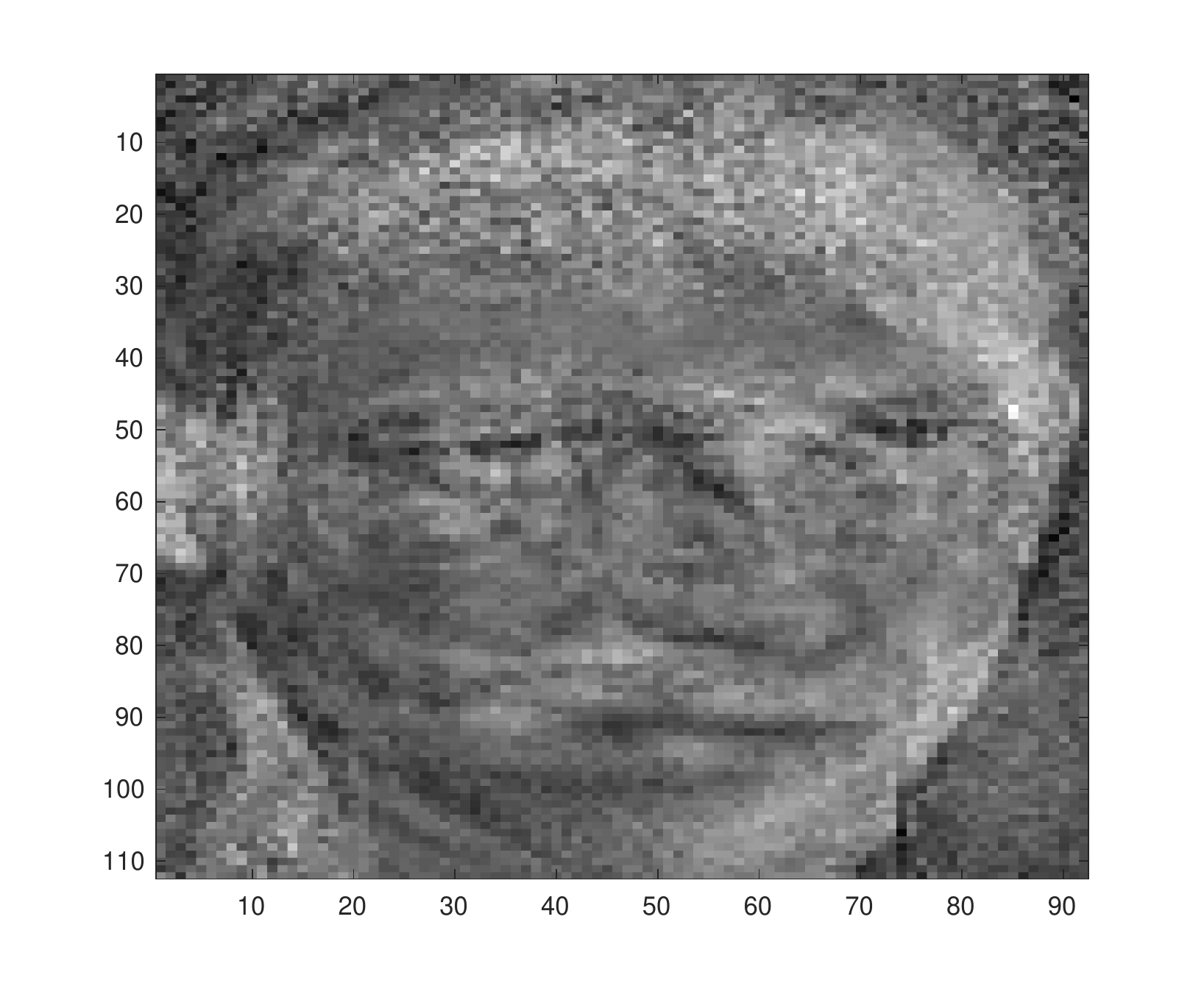}
		\caption{{\footnotesize Noisy}}
		\label{fig:face_noisy}
	\end{subfigure}%
	\begin{subfigure}{.110\textwidth}
		\centering
		\includegraphics[width=0.9\textwidth, trim={7.5cm 11.1cm 7.0cm 10.8cm},clip]{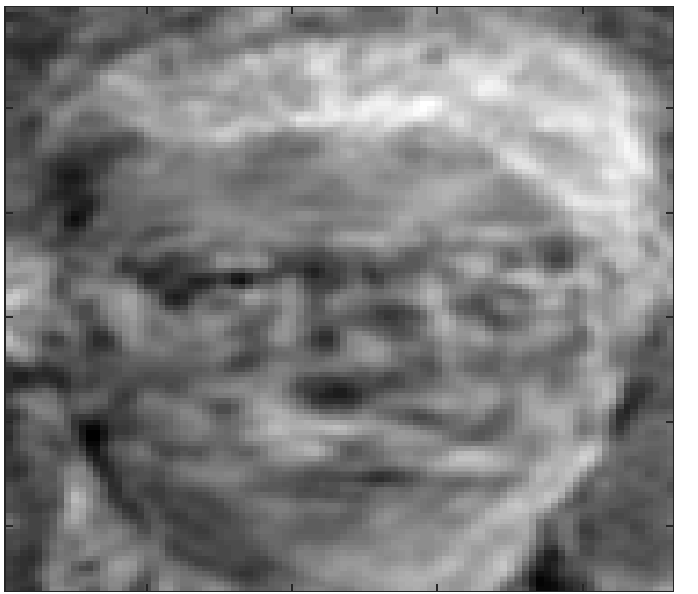}	
		\caption{{\footnotesize Gauss 2D}}
		\label{fig:face_gaussian_filter}
	\end{subfigure}%
	\begin{subfigure}{.110\textwidth}
		\centering
		\includegraphics[width=0.89\textwidth, trim={2cm 1.4cm 1.5cm 1cm},clip]{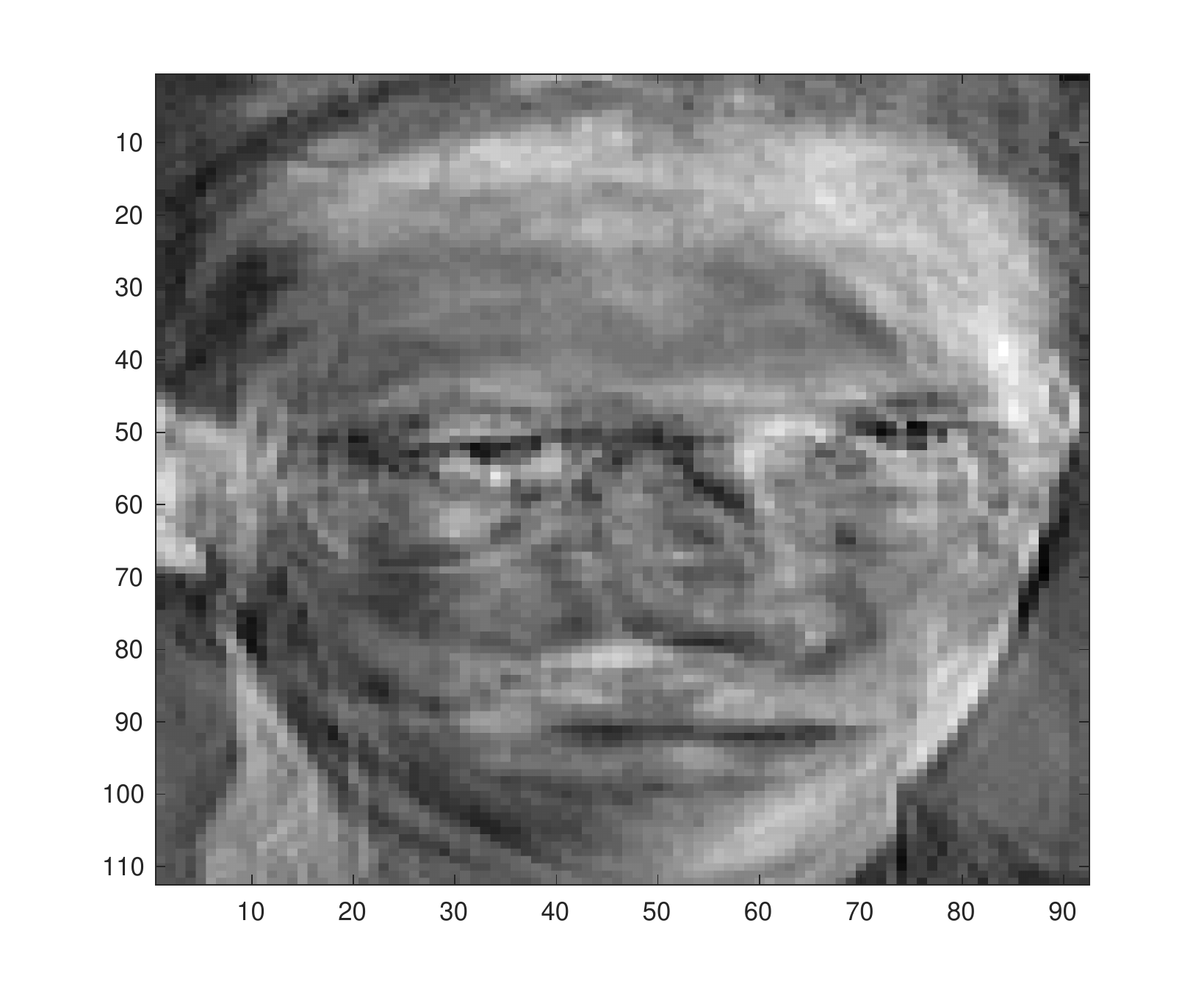}
		\caption{{\footnotesize Low-pass GF}}
		\label{fig:face_low_pass}
	\end{subfigure}%
	\begin{subfigure}{.110\textwidth}
		\centering
		\includegraphics[width=0.9\textwidth, trim={2cm 1.4cm 1.5cm 1cm},clip]{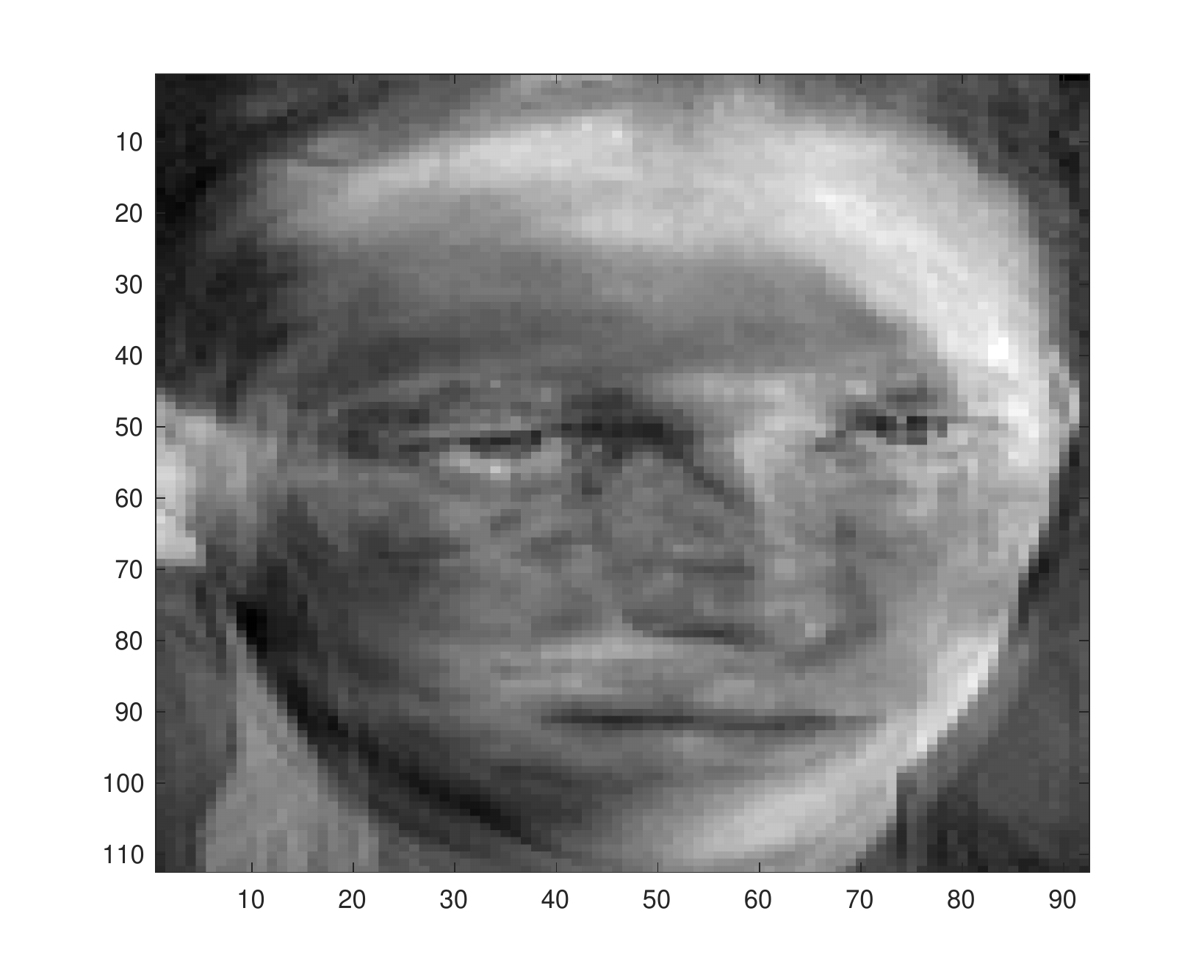}
		\caption{{\footnotesize Wiener GF}}
		\label{fig:face_smoother}
	\end{subfigure}
		\begin{subfigure}{.110\textwidth}
			\centering
			\includegraphics[width=0.9\textwidth, trim={2cm 1.4cm 1.5cm 1cm},clip]{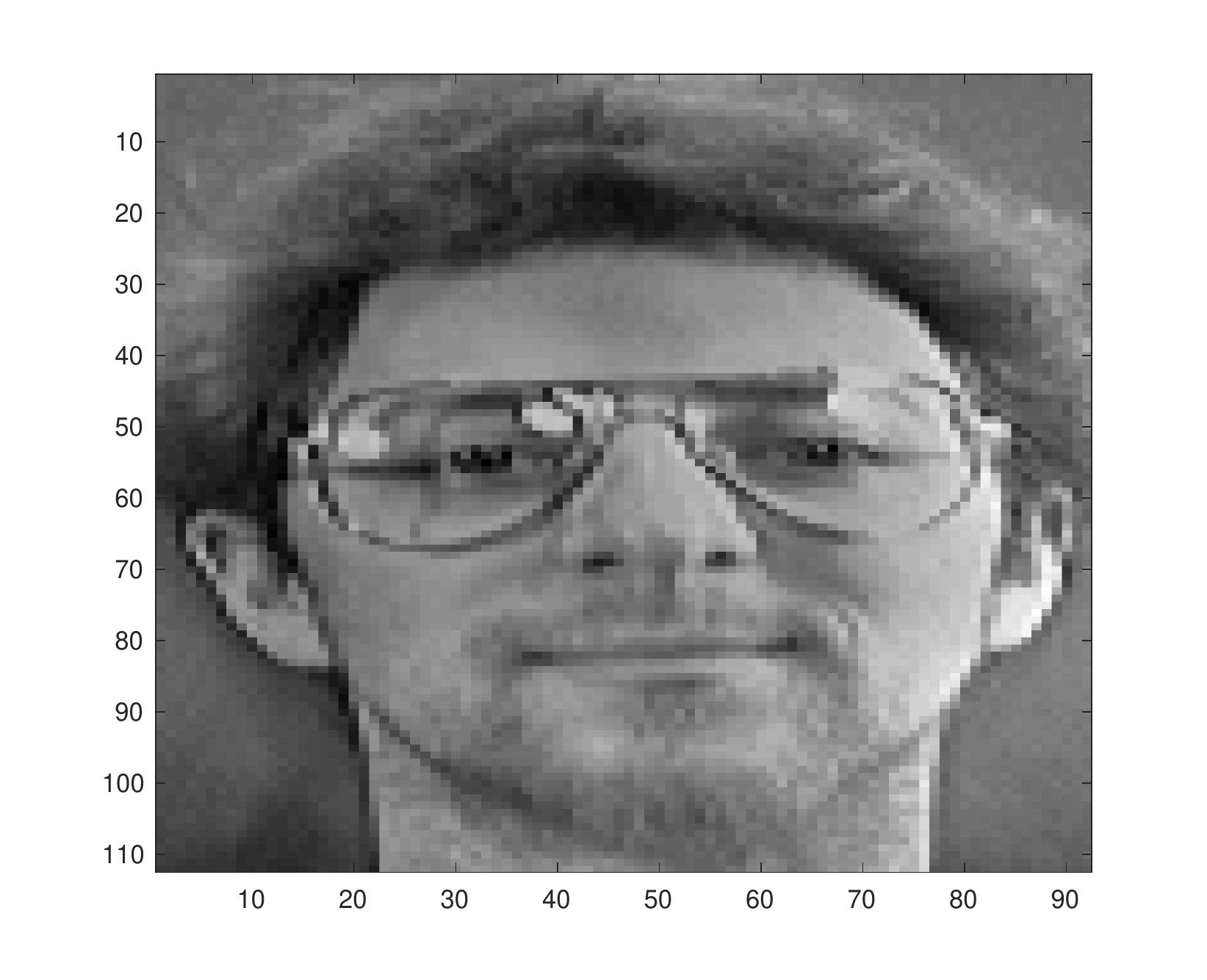}
			\caption{{\footnotesize Ground truth}}
			\label{fig:face_glasses_truth}
		\end{subfigure}%
		\begin{subfigure}{.110\textwidth}
			\centering
			\includegraphics[width=0.9\textwidth, trim={2cm 1.4cm 1.5cm 1cm},clip]{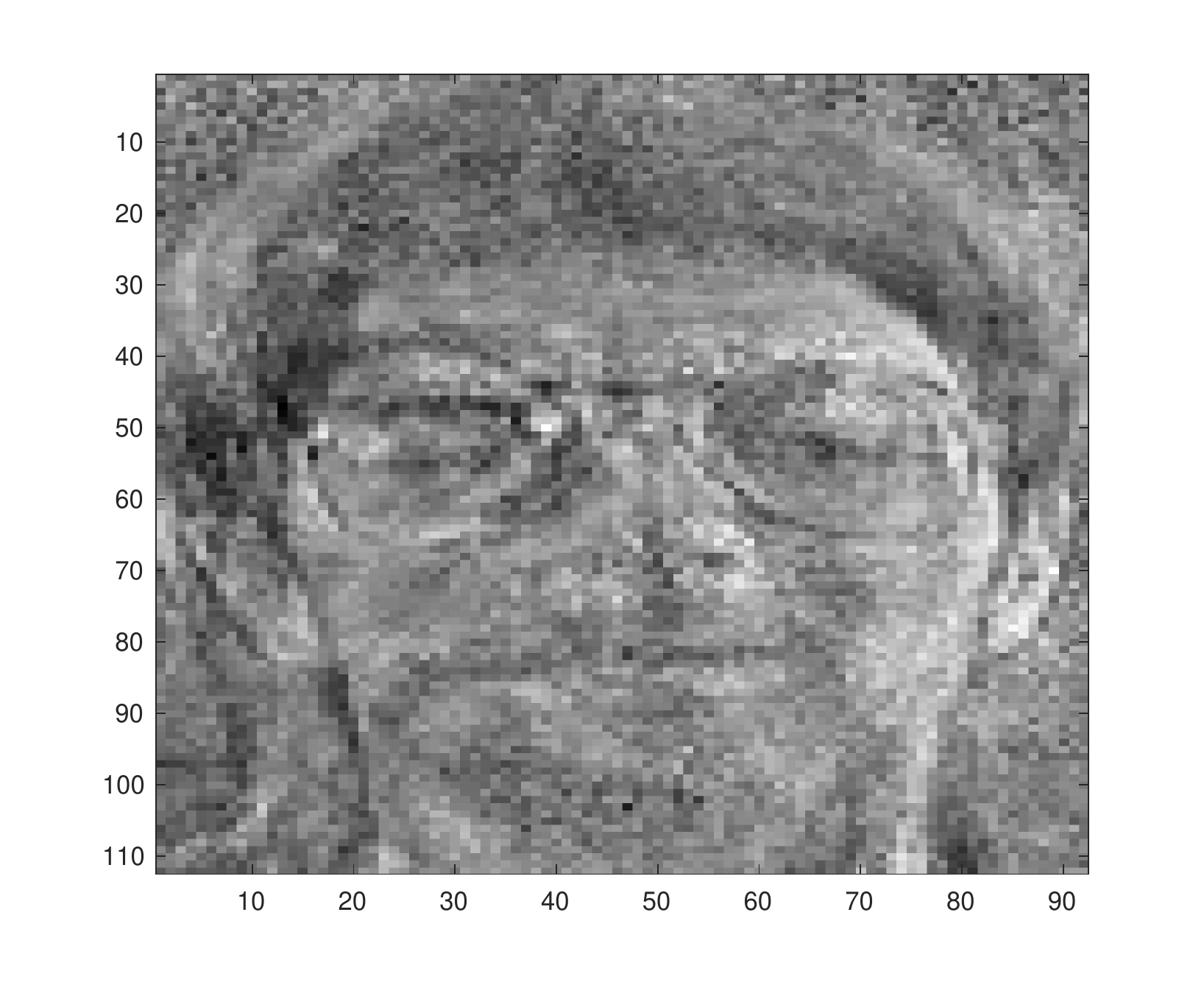}
			\caption{{\footnotesize Noisy}}
			\label{fig:face_glasses_noisy}
		\end{subfigure}%
		\begin{subfigure}{.12\textwidth}
			\centering
			\includegraphics[width=0.8256\textwidth, trim={2cm 1.4cm 1.5cm 1cm},clip]{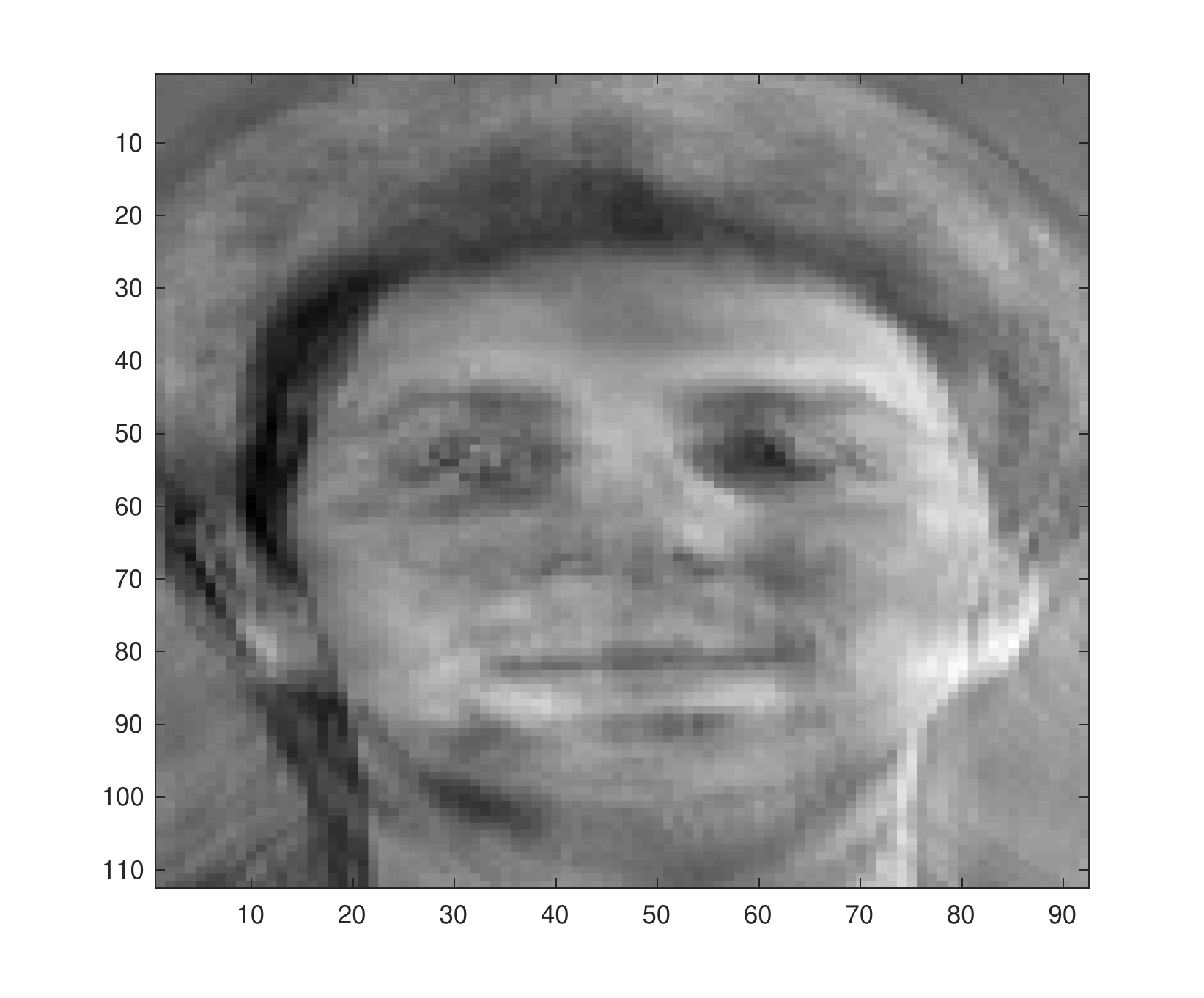}
			\caption{\hspace{-.05cm}{\footnotesize No-glasses \hspace{-.05cm}GF}}
			\label{fig:face_glasses_no_glasses}
		\end{subfigure}%
		\begin{subfigure}{.110\textwidth}
			\centering
			\includegraphics[width=0.9\textwidth, trim={2cm 1.4cm 1.5cm 1cm},clip]{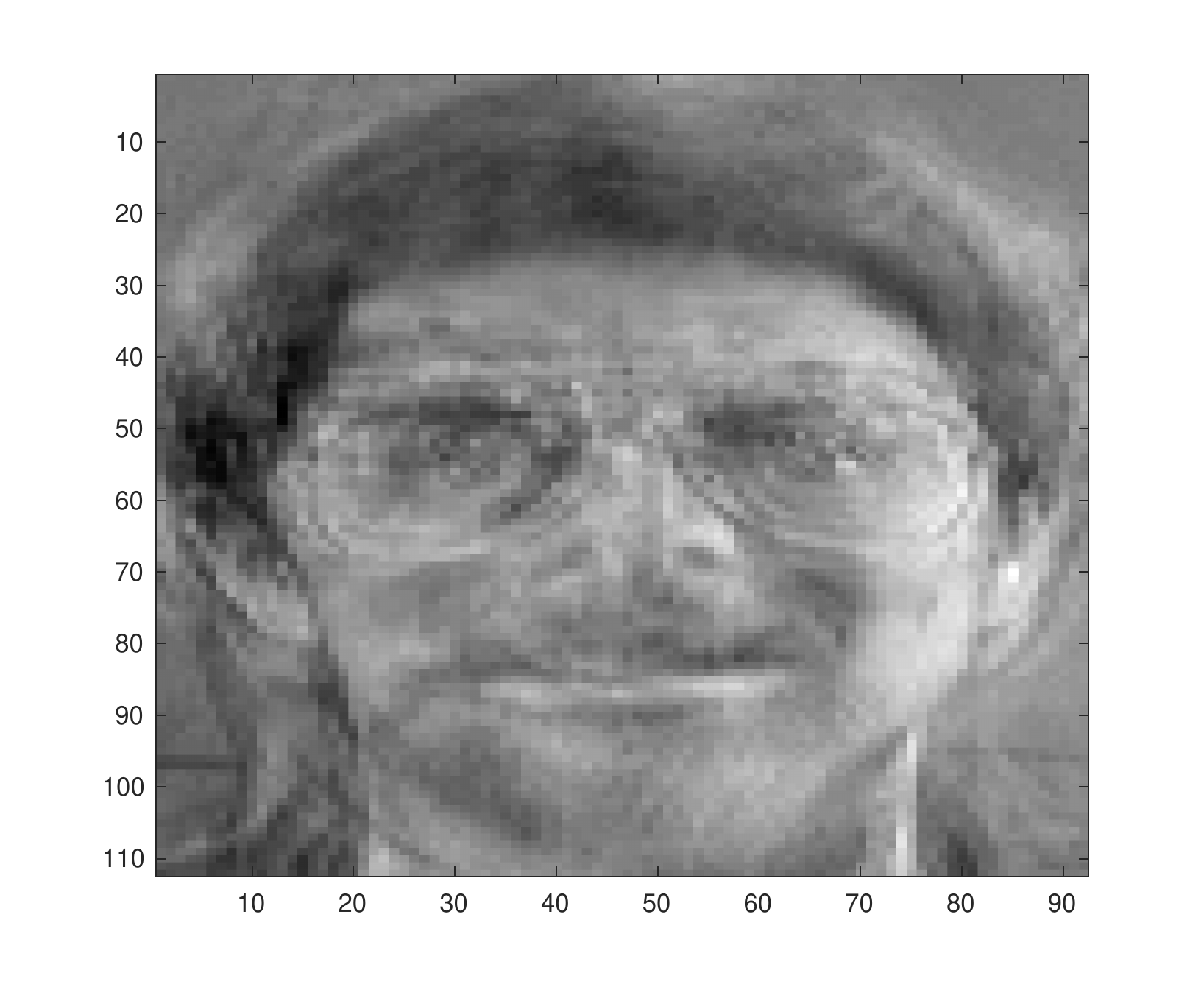}
			\caption{{\footnotesize Glasses GF}}
			\label{fig:face_glasses_glasses}
		\end{subfigure}
		
	\vspace{-0.05in}
	\caption{\black{The images (a-e) illustrate how, by leveraging the concept of graph stationarity, the Wiener graph filter (GF) in (e) achieves a better performance than a regular low-pass GF in (d) and both of them outperform the classical 2D Gaussian filter. The images (f-i) illustrate the difference in recovering the original image (f) from the noisy version (g) when assuming that the image is stationary in a shift build with faces without glasses (h) and with glasses (i). }}
	\vspace{-0.15in}
	\label{F:face_smoothing}
\end{figure*}

\medskip
\noindent {\bf \black{TC4. Real-world signals:}} \black{Three experiments with real-data are presented. The first one considers grayscale images of different individuals, the second one human brain signals measuring activity at different regions of the brain, and the last one cellular flow cytometry data.

\medskip
\noindent {\textit{Face images:}}}
\black{We consider a set of 100 grayscale images $\{\bbx_r\}_{r=1}^{100}$ corresponding to 10 face pictures of 10 different people\footnote{http://www.cl.cam.ac.uk/research/dtg/attarchive/facedatabase.html}. Formally, every image is represented by a vector $\bbx_r \in \reals^{10304}$ where the entries correspond to grayscale values of pixels, normalized to have zero mean. We consider the images $\bbx_r$ to be realizations of a graph process which, by definition, is stationary in the shift given by its covariance, here approximated by the sample covariance $\bbS = \hbC_x = \bbV \bbLambda_c \bbV^H$. Moreover, denote by $\ccalI_j$ the set of images corresponding to faces of person $j$ and consider the sample covariance $\hbC^{(j)}_x=\bbV^{(j)}_x\bbLambda_c^{(j)}\bbV^{(j)H}_x$ of the subset $\{\bbx_r\}_{r \in \ccalI_j}$. While matrix $\hbC^{(j)}_x$ will be perfectly diagonalized by $\bbV^{(j)}_x$, the question is whether it will be (approximately) diagonalized by $\bbV$ or $\bbV^{(k)}_x$, where $k \neq j$. After left and right multiplying the covariance $\hbC^{(j)}_x$ with the particular GFT, the resultant off-diagonal elements account for the cross-correlation among frequency components. Hence, invoking  Property \ref{P:uncorrelated_freq_components}, we know that if the process is actually stationary in the shift that generated the GFT, those off-diagonal elements must be zero. 
	To assess this, Figs.~\ref{fig:sub101}-(top), \ref{fig:sub101}-(bottom), and \ref{fig:sub11} plot blocks of the matrices $(\bbV^{(1)H} \hbC^{(1)}_x \bbV^{(1)})$, $(\bbV^{(2)H} \hbC^{(1)}_x \bbV^{(2)})$, and $(\bbV^H \hbC^{(1)}_x \bbV)$, respectively. The pictures reveal that while $\hbC^{(1)}_x$ is \emph{not} diagonalized by $\bbV^{(2)}_x$ (many of the off-diagonal elements are non-zero), it is \emph{approximately} diagonalized by the principal component basis $\bbV$. \black{To quantify this more rigorously, we consider the metric $\theta(\bbV, \hbC):=\|\diag(\diag(\bbV^H \hbC \bbV))\|_F/\| \bbV^H \hbC \bbV\|_F$, where only the $N$ elements of the diagonal are considered in the numerator. Then, for the example in Fig. \ref{fig:sub11}, we find that $\theta(\bbV, \hbC^{(1)}_x)=50\%$, confirming that the relative weight of the elements in the diagonal is high and, hence,} that the process generating face images of subject $j$ is approximately stationary in $\bbS=\hbC_x$. To illustrate why this is useful, we describe next a couple of experiments where traditional tools for processing stationary time-signals are applied to the graph signals at hand. Our goal is to explore potential application domains for the definitions and tools introduced in this paper. In particular, we use the graph counterpart of the Wiener filter \cite{hayes2009statistical} to denoise face images. More precisely, given a noisy version $\bby$ of an image of individual $j$, we model $\bby$ as stationary in $\bbS=\hbC_x$ and obtain the filtered version $\bby^{\mathrm{Wie}}$ where $\tilde{y}^{\mathrm{Wie}}_k  =  p_k/(p_k + \omega_k^2) \, \tilde{y}_k,$
	%
	%
	with $\omega_k^2$ denoting the noise power at frequency $k$. In the experiments we set the PSD as $\bbp=\hbp_{\cg}$, i.e., the eigenvalues of the sample covariance $\hbC_x$. For comparison purposes, we also consider a Gaussian 2D low-pass filter with unitary variance and a `low-pass' graph filter, where we keep unchanged the frequency components $\tilde{y}_k$ that correspond to active frequencies, i.e., all $k$ such that $p_k > 0$, and discard the rest.
	The results are illustrated in Fig.~\ref{F:face_smoothing} where inspection reveals that the Gaussian filter (which exploits neither $\bbV$ nor $\bbp$) yields the worst reconstruction performance, while the Wiener filter (which exploits both) achieves a better reconstruction than the low-pass graph filter (which exploits only $\bbV$). The second example also deals with image smoothing. In particular, we use the face with glasses in Fig.~\ref{fig:face_glasses_truth} and consider two different GSOs $\bbS_1$ and $\bbS_2$ corresponding to two different sample covariances that use the same number of images.  The shift $\bbS_1$ is constructed based on images of people not wearing glasses whereas most of the images used in generating $\bbS_2$ contained glasses. If we implement the Wiener filter to denoise Fig.~\ref{fig:face_glasses_noisy} using as shifts the sample covariances corresponding to $\bbS_1$ and $\bbS_2$ and as PSD their corresponding correlograms, we observe that the second filter is able to recover the glasses (Fig.~\ref{fig:face_glasses_glasses}) while the first one fails (Fig.~\ref{fig:face_glasses_no_glasses}), entailing a poorer recovery. This can be interpreted as a manifestation of the original image being closer to stationary in $\bbS_2$ than in $\bbS_1$.}

\medskip
\noindent \black{\textit{Brain signals:}} The \black{ensuing} experiment deals with brain signals. While in the setup with face images the supporting graph was unknown and estimated as the sample covariance, in this case the graph is given. In particular, the goal is to analyze brain functional signals using as support the so-called \textit{functional} and \textit{structural} brain networks \cite{huang2016,medaglia2016}. To that end, we use the 120 brain functional signals $\{\bbx_r\}_{r=1}^{120}$ and shifts $\bbS_{\text{fun}}$ and $\bbS_{\text{str}}$ provided in \cite{medaglia2016}. As done before, the goal is to \black{assess} if the signals are stationary in the provided graphs. The results indicate that $\{\bbx_r\}_{r=1}^{120}$ are approximately stationary in $\bbS_{\text{fun}}$. \black{In particular, we have that $\theta(\bbV_{\text{fun}},\hbC_x)=99\%$, so that the contribution to the norm of the off-diagonal elements of $(\bbV^H_{\text{fun}}\hbC_x\bbV_{\text{fun}})$ is negligible}. More importantly and interestingly, they also indicate that $\{\bbx_r\}_{r=1}^{120}$ are approximately stationary in $\bbS_{\text{str}}$, with the matrix $(\bbV^H_{\text{str}}\hbC_x\bbV_{\text{str}})$ being shown in Fig. \ref{fig:sub12}. \black{In fact, we have that $\theta(\bbV_{\text{str}}, \hbC_x)=67\%$}. This is more surprising since the construction of $\bbS_{\text{str}}$ is agnostic to $\{\bbx_r\}_{r=1}^{120}$. It also points out that functional and structural networks, often viewed as separate entities, are clearly related. Such an observation can be exploited in multiple tasks. One example is to improve the algorithms to identify the underlying networks. Our results for that particular application, which build on network topology identification algorithms that view the set of available observations as stationary in the network to be identified \cite{segarra2016networktopologyID}, are promising and will soon be reported in a separate contribution.

\begin{figure}
	\centering
		\begin{subfigure}{.2\textwidth}
		\centering
		\includegraphics[width=\textwidth]{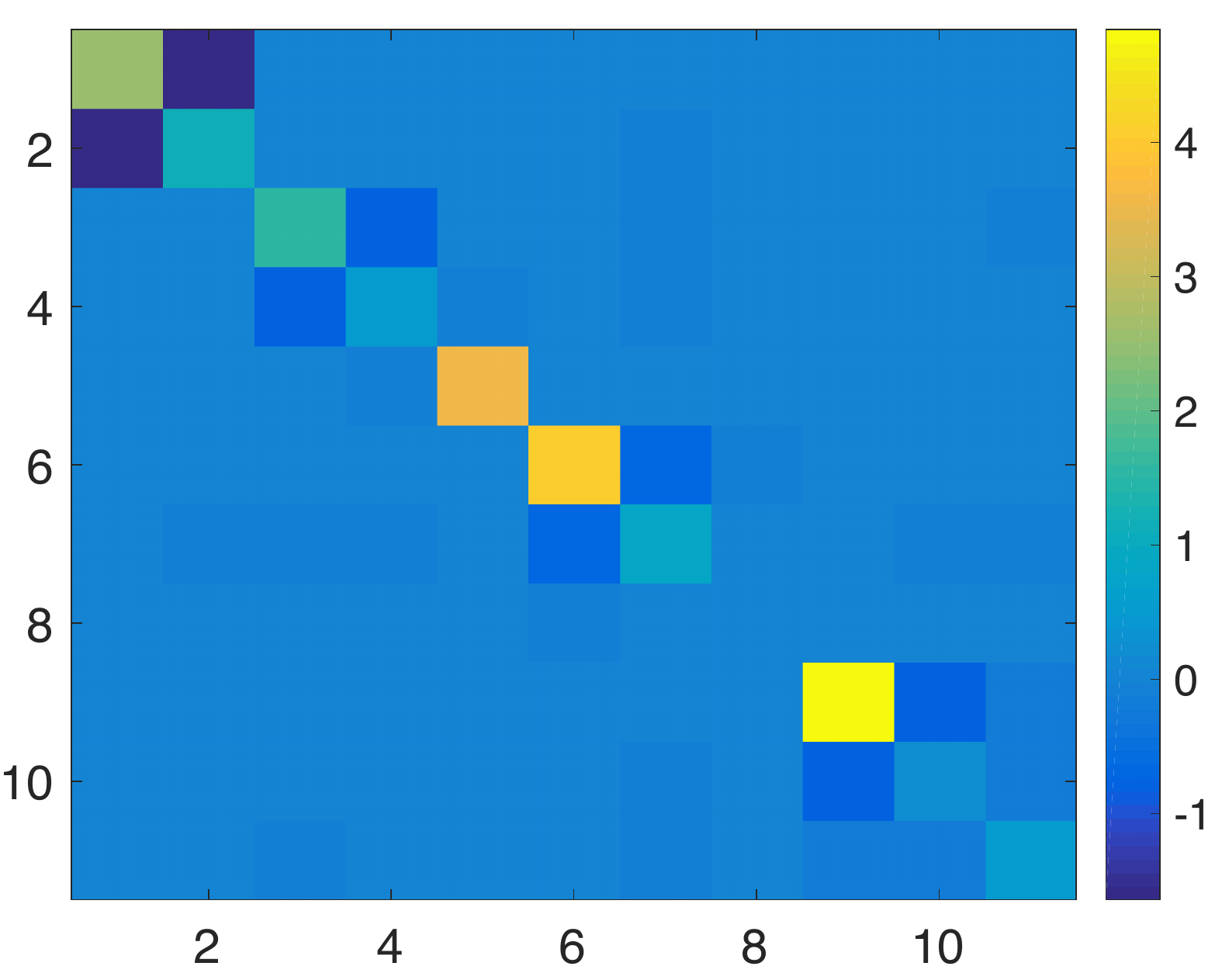}
		\caption{}
		\end{subfigure}%
		\begin{subfigure}{.25\textwidth}
		\hspace{0.2cm}
		\centering
		\includegraphics[width=0.75\textwidth, height = 0.655\textwidth]{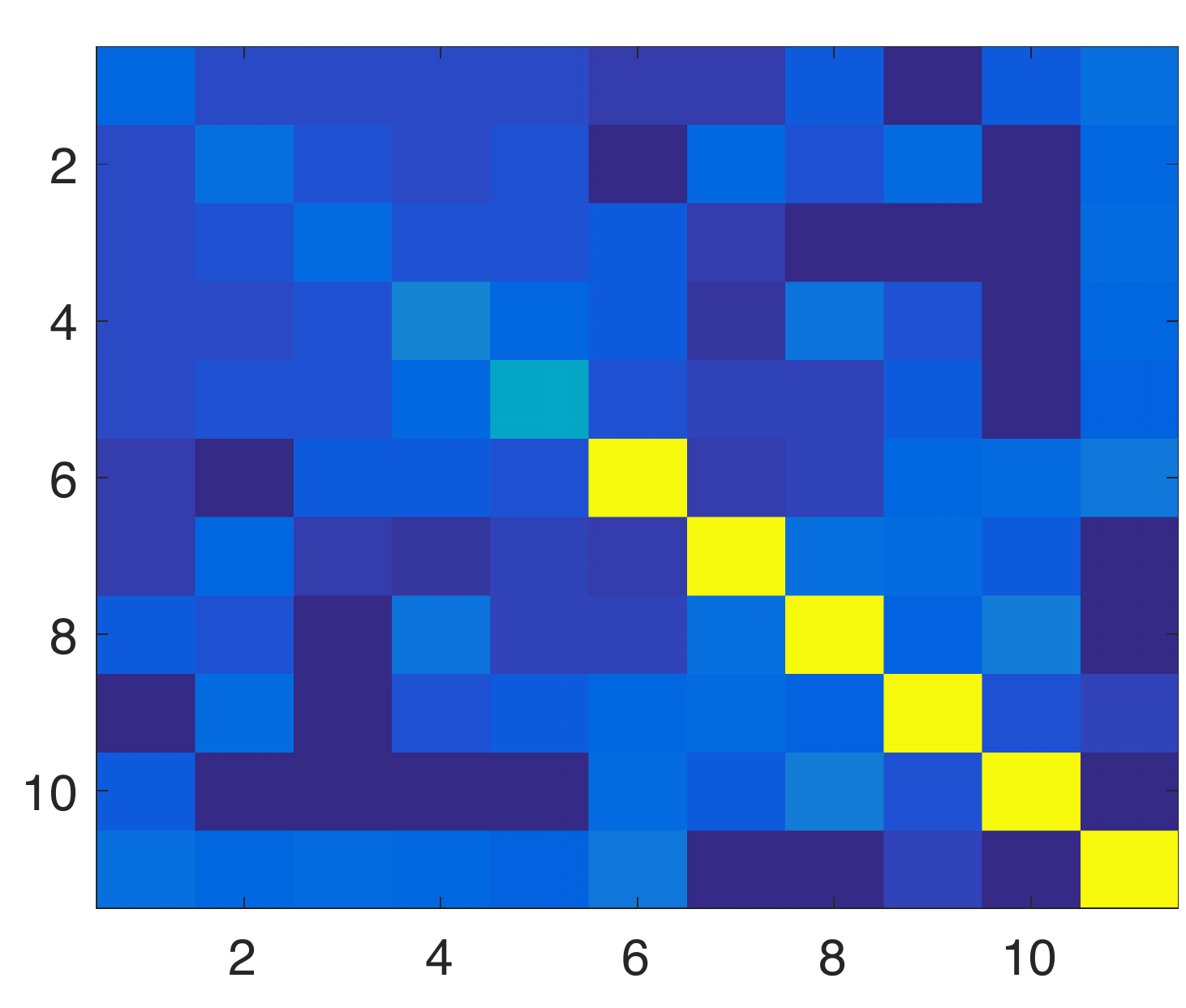}
		\hspace{-0.2cm}
		\includegraphics[width=0.152\textwidth]{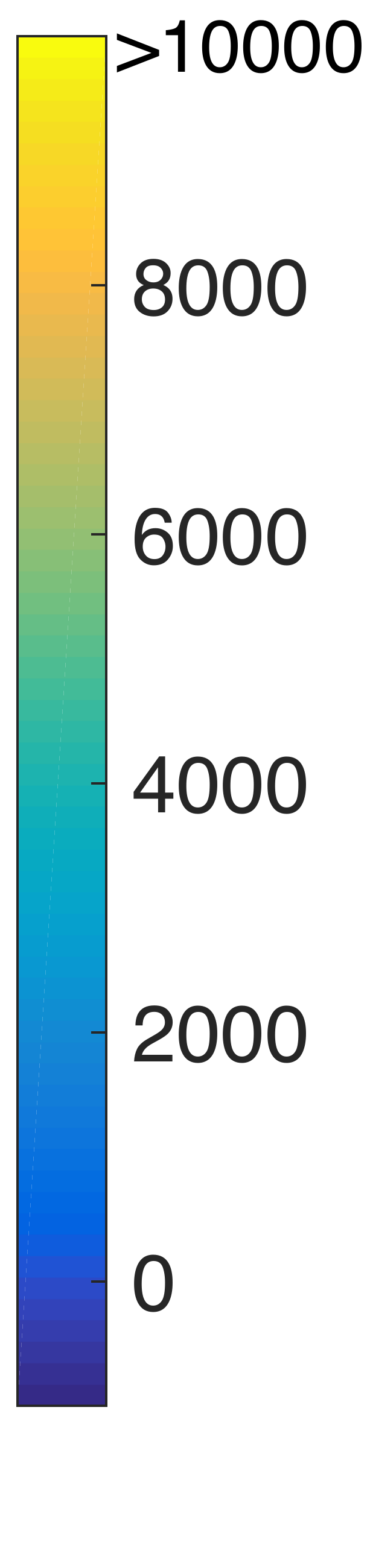}
		\caption{}
		\end{subfigure}%
	\caption{\black{(a) Shift operator $\bbS_{10^3}$ recovered from the implementation of graphical lasso. (b) Diagonalization of $\hat{\bbC}_x$ with $\bbV_{10^3}$. The graph process is approximately stationary in~$\bbS_{10^3}$.}}
	\vspace{-0.015in}
	\label{F:flow_cytometry}
\end{figure}

\black{
\medskip
\noindent \textit{Flow cytometry:} A common approach to identify the graph structure associated with an observed dataset is to construct a \emph{regularized} estimator of the precision matrix $\bbTheta$; see, e.g., graphical lasso \cite{GLasso2008}. Given that, by definition, a process is stationary in the graph given by its precision matrix (Example~3 in Section~\ref{S:DiffusionProcStationarity}), this same process is expected to be \emph{approximately} stationary in the graph obtained via graphical lasso. In order to illustrate this property, we consider a flow cytometry dataset on $N=11$ proteins and $R = 7466$ observations corresponding to different cells \cite{Sachs523} (\black{dataset also used in \cite{GLasso2008}}). More precisely, we are given a set $\{ \bbx_r \}_{r=1}^R$ of observations where $x_r \in \reals^{11}$ represents the levels of $11$ different proteins in an individual cell $r$. Denoting by $\hat{\bbC}_x$ the sample covariance of the observations, graphical lasso estimates the associated graph shift $\bbS_{\eta}$ as
\begin{equation}
\bbS_{\eta} = \argmin_{\bbS \in \ccalS^N_+} \,\,\, - \log | \bbS | +  \Tr(\hat{\bbC}_x \bbS) + \eta \| \bbS \|_1,
\end{equation}
where $\ccalS^N_+$ denotes the cone of $N \times N$ positive semi-definite matrices and $\eta \geq 0$ parametrizes the level of regularization in estimating $\bbS_{\eta}$. In Fig.~\ref{F:flow_cytometry}a we portray the shift operator recovered when $\eta = 10^3$. 
Denoting the eigendecomposition of $\bbS_{\eta} = \bbV_\eta \bbLambda_\eta \bbV_\eta^H$, the form of $\bbV_\eta^H \hat{\bbC}_x \bbV_\eta$ determines if the process is stationary on $\bbS_{\eta}$. One can first check that $\bbV_{0}^H \hat{\bbC}_x \bbV_{0}$ is a diagonal matrix, implying that the observed network process is stationary in $\bbS_{0}$. This is not surprising since $\bbS_{0} = \hat{\bbC}_x^{-1}$, i.e., graphical lasso with no regularization returns the inverse sample covariance. Interestingly, for $\eta = 10^3$ we have that $\theta(\bbV_{10^3}^H,\hat{\bbC}_x)=99\%$, demonstrating 
that the observed graph process is approximately stationary in $\bbS_{10^3}$, \black{a graph that resembles the known dependencies between proteins} \cite{GLasso2008, Sachs523}; see Fig.~\ref{F:flow_cytometry}b. Given that graphical lasso and related methods are widely used to unveil graph structures, the framework of stationary graph processes here developed is relevant to a broad range of applications.
}

\section{Conclusion}\label{S:Concl}
\black{Different equivalent ways of generalizing the notion of stationarity to graph processes associated with a normal shift operator were studied.
Given that graph stationary processes were shown to be diagonalized by the graph Fourier basis, the corresponding power spectral density was introduced and several estimation methods were studied in detail.} We first generalized nonparametric methods including periodograms, window-based average periodograms, and filter banks. Their performance was analyzed, comparisons with the traditional time domain schemes were established and open issues, such as how to group nodes and frequencies in a graph, were identified. We then focused on parametric estimation, where we examined MA, AR, and ARMA processes. We not only showed that those processes are useful to model linear diffusion dynamics, but also identified particular scenarios where the optimal parameter estimation problem is tractable.  


\section*{Appendices}\label{S:Appendices}
\subsection{Proof of Prop. \ref{P:CovariancePeriodogram}}\label{Ss:App_Proof_CovariancePeriodogram}

To prove that the estimate is unbiased, we use the fact that $\hbp_{\pg}=\hbp_{\cg}$ combined with \eqref{E:nonpar_PSD_correlogram} to conclude that $\EE[\hbp_{\pg}]= \diag[\bbV^HR^{-1}(\sum_{r=1}^R \EE[\bbx_r\bbx_r^H])\bbV] = \diag[\bbV^HR^{-1}R\bbC_x\bbV] = \bbp$. Hence, $\bbb_{\pg}=\EE[\hbp_{\pg}]-\bbp=\bb0$. 

To compute the covariance, start by writing $\bbSigma_{\pg} =\E{\hat{\bbp}_{\pg} \hat{\bbp}_{\pg}^H}- \bbp \bbp^H$. We may expand the leftmost term as
\begin{align}
	\!\!&\!\E{\hat{\bbp}_{\pg} \hat{\bbp}_{\pg}^H}= \E{ \diag(\bbV^H \hbC_x \bbV) \diag(\bbV^H \hbC_x \bbV)^H} \label{E:cov_diffk_eq1}\\
	\!\!&\!\hspace{-0.15cm}=\frac{1}{R^2}\!\!\sum_{r=1,r'=1}^R \!\!\! \!\!\!\E{ \diag(\bbV^H \bbx_r \bbx_r^H \bbV) \diag(\bbV^H \bbx_{r'} \bbx_{r'}^H \bbV)^H}\!. \label{E:cov_diffk_eq2}
\end{align}

We may split the above summation into the terms where $r \neq r'$ and those where $r = r'$. For the former case, since $\bbx_r$ is assumed to be independent from $\bbx_{r'}$, we have that
\begin{align}
\!\E{ \diag(\bbV^H \bbx_r \bbx_r^H \bbV) }\!  \E{ \diag(\bbV^H \bbx_{r'} \bbx_{r'}^H \bbV)^H} = \bbp \bbp^H\!.\! \label{E:cov_diffk_eq3}
\end{align}
By contrast, for the case where $r = r'$ we undertake an elementwise analysis of the elements in the expected value in \eqref{E:cov_diffk_eq2}. Notice that if we denote by $\bbv_i$ the $i$th column of $\bbV$ we have that $[\diag(\bbV^H\bbx_r \bbx_r^H \bbV)]_i = \bbv_i^H \bbx_r \bbx_r^H \bbv_i$. Replacing this expression in \eqref{E:cov_diffk_eq2}, setting $r=r'$, and using the formula for the quartic form of a Gaussian, see e.g. \cite[pp. 43]{matrixcookbook}, we obtain that
\begin{align}
	&\big[\E{ \diag(\bbV^H \bbx_r \bbx_r^H \bbV) \diag(\bbV^H \bbx_{r'} \bbx_{r'}^H \bbV)^H}\big]_{ij} \\
	&\,=\E{\bbv_i^H \bbx_r \bbx_r^H \bbv_i \bbv_j^H \bbx_r \bbx_r^H \bbv_j}\!=\!\bbv_i^H \E{\bbx_r \bbx_r^H \bbv_i \bbv_j^H \bbx_r \bbx_r^H} \bbv_j \nonumber\\
	&\,=p_i p_j + \bbv_i^H \bbC_x \bbv_j^* \bbv_i^T \bbC_x \bbv_j + \bbv_j^H \bbC_x \bbv_i \bbv_i^H \bbC_x \bbv_j. \nonumber
\end{align}
When $\bbS$ is symmetric, hence $\bbV$ is real, the above expression can be further simplified to obtain
\begin{align}
	&\E{ \diag(\bbV^H \bbx_r \bbx_r^H \bbV) \diag(\bbV^H \bbx_{r} \bbx_{r}^H \bbV)^H} \nonumber \\
	&\hspace{1cm}= 2 \, \diag^2(\bbp) + \bbp \bbp^H \! . \label{E:cov_diffk_eq4}
\end{align}
Upon substituting \eqref{E:cov_diffk_eq3} and \eqref{E:cov_diffk_eq4} into \eqref{E:cov_diffk_eq2}, the result follows.

\subsection{Proof of Prop. \ref{P:BiasCovMSE_Av_W_Period}}\label{Ss:App_Proof_BiasCovMSE_Av_W_Period}

From the definition of $\hat{\bbp}_\ccalW$ in \eqref{E:nonpar_PSD_avw_periodogram}, it follows that 
\begin{equation}\label{E:nonpar_PSD_avw_periodogram_proof_v0}
   \hbp_{\ccalW} \! = \! \frac{1}{M}\sum_{m=1}^M \diag( \bbV^H\diag(\bbw_m) \bbV \tilde{\bbx} \tilde{\bbx}^H \bbV^H\diag(\bbw_m^*) \bbV).
\end{equation}
Thus, using the definition of $\tilde{\bbW}_m$, we may write
\begin{equation}\label{E:nonpar_PSD_avw_periodogram_proof_v1}
   \E{\hbp_{\ccalW}} \! = \! \frac{1}{M}\sum_{m=1}^M \diag( \tilde{\bbW}_m \E{ \tilde{\bbx} \tilde{\bbx}^H} \tilde{\bbW}_m^H).
\end{equation}
Leveraging the fact that $\E{ \tilde{\bbx} \tilde{\bbx}^H}\! =\! \diag(\bbp)$ is diagonal and recalling that $\tilde{\bbW}_{mm}=\tbW_m\circ\tbW_m^*$, the result in \eqref{E:bias_avw_period_PSD} follows.
  
In order to show \eqref{E:cov_avw_period_PSD}, notice that only the diagonal elements of $\bbSigma_\ccalW$ are needed. Each of them can be found as 
\begin{align}\label{E:diag_elem_cov_avw}
[\bbSigma_\ccalW]_{k,k}=\E{[\hat{\bbp}_\ccalW]_k[\hat{\bbp}_\ccalW]_k}- \E{[\hat{{\bbp}}_\ccalW]_k}^2.
\end{align}
Rewriting $\hat{\bbp}_\ccalW$ as a sum across the $M$ windows, it holds that
\begin{align}
\nonumber\E{[\hat{\bbp}_\ccalW]_k [\hat{\bbp}_\ccalW]_k} & =\E{\frac{1}{M}\!\sum_{m=1}^M [ |\tilde{\bbW}_m \tilde{\bbx} |^2 ]_k
	\frac{1}{M}\!\sum_{m'=1}^M[ |\tilde{\bbW}_{m'} \tilde{\bbx} |^2  ]_k}\\
& \hspace{-1.45cm} =\frac{1}{M^2}\sum_{m=1,m'=1}^M\E{  [ |\tilde{\bbW}_m \tilde{\bbx} |^2 ]_k [|\tilde{\bbW}_{m'} \tilde{\bbx} |^2]_k}.\label{E:2nd_mom_psd_awp_v0}
\end{align}
Denoting the $k$th row of $\tilde{\bbW}_m$ by $\tbw_{k|m}^T$, the term $[ |\hat{\bbW}_m \tilde{\bbx} |^2 ]_k$ can be written as $\tbw_{k|m}^T \tbx \tbx^H \tbw_{k|m}^*$. Consequently, we have
\begin{align}
&\E{\!  [ |\hat{\bbW}_m \tilde{\bbx} |^2 ]_k [|\hat{\bbW}_{m'} \tilde{\bbx} |^2]_k \!} \!\!= \!\!\EE\! \left[\!\tbw_{k|m}^T\tbx \tbx^H \tbw_{k|m}^* \tbw_{k|m'}^T\tbx\tbx^H \tbw_{k|m'}^* \right] \nonumber\\
&\hspace{0.1cm} = \tbw_{k|m}^T \E{\tbx \tbx^H \tbw_{k|m}^* \tbw_{k|m'}^T\tbx \tbx^H }\tbw_{k|m'}^*,
\end{align}
where the middle factor is a quartic form of a Gaussian with covariance $\diag(\bbp)$  (cf.~Property~\ref{P:uncorrelated_freq_components}). Solving this fourth moment, see e.g. \cite[pp. 43]{matrixcookbook}, it follows that
\begin{align}\label{E:diag_elem_cov_avw_v2_2}
&\hspace{-0.015cm}\E{\!  [ |\hat{\bbW}_m \tilde{\bbx} |^2 ]_k [|\hat{\bbW}_{m'} \tilde{\bbx} |^2]_k \!} = \\
&\hspace{0.8cm} | \tbw_{k|m}^T |^2 \bbp | \tbw_{k|m'}^T |^2 \bbp \,\,+ | \tbw_{k|m}^T \diag(\bbp) \tbw_{k|m'}^* |^2 \nonumber \\
 &  \hspace{0.8cm} + \tbw_{k|m}^T \diag(\bbp) \bbV^H \bbV^* \tbw_{k|m'} \tbw_{k|m}^H \bbV^T \bbV \diag(\bbp) \tbw_{k|m'}^* \nonumber
\end{align}
When $\bbS$ is symmetric, hence $\bbV$ is real, the third summand in \eqref{E:diag_elem_cov_avw_v2_2} is equal to the second one. Thus, substituting first \eqref{E:diag_elem_cov_avw_v2_2} into \eqref{E:2nd_mom_psd_awp_v0} and then, \eqref{E:2nd_mom_psd_awp_v0} into \eqref{E:diag_elem_cov_avw} yields
\begin{align}\label{E:diag_elem_cov_avw_v2}
\textstyle [\bbSigma_\ccalW]_{k,k} \! = \frac{2}{M^2}\!\!\sum_{m=1,m'=1}^M\!\! |\tbw_{k|m}^T\diag(\bbp)\tbw_{k|m'}^*|^2.
\end{align}
Since $\tbW_{mm'}=\tbW_m\circ\tbW_{m'}^*$, it follows that $|\tbw_{k|m}^T\diag(\bbp)\tbw_{k|m'}^*| $ $= [\tilde{\bbW}_{mm'} \bbp]_k$, thus
\begin{align}\label{E:diag_elem_cov_avw_v3}
\textstyle [\bbSigma_\ccalW]_{k,k} \! = \frac{2}{M^2}\!\!\sum_{m=1,m'=1}^M\!\! |[\tilde{\bbW}_{mm'} \bbp]_k|^2.
\end{align}
Finally, using \eqref{E:diag_elem_cov_avw_v3} to write $\Tr[\bbSigma_\ccalW]=\sum_{k=1}^{N}[\bbSigma_\ccalW]_{k,k}$, we obtain \eqref{E:cov_avw_period_PSD} and the proof concludes.

\subsection{Proof of Prop. \ref{P:BiasCovMSE_FilterBank}}\label{Ss:App_Proof_BiasCovMSE_FilterBank}

Rewriting the norm in \eqref{E:filter_bank_window_freq} as the trace of the corresponding outer product we get that
\begin{equation}
\E{\hat{p}_{\tilde{\bbq}_k}} = \Tr[\diag(\tilde{\bbq}_k) \E{\tilde{\bbx} \tilde{\bbx}^H} \diag( \tilde{\bbq}_k^*)].
\end{equation}
Expression \eqref{E:mean_filter_bank_period_PSD} follows from replacing $\E{\tilde{\bbx} \tilde{\bbx}^H}$ by $\diag(\bbp)$ and noting that the trace of the product of diagonal matrices equals the sum of the entrywise products of the diagonals.

To prove \eqref{E:cov_filter_bank_PSD}, we first find $\EE{[\hat{p}_{\tilde{\bbq}_k} \hat{p}_{\tilde{\bbq}_k}]}$. Since $\hat{p}_{\tilde{\bbq}_k}  = \Tr[\diag(\tilde{\bbq}_k)\tilde{\bbx}\tilde{\bbx}^H\diag(\tilde{\bbq}_k)^H]=\Tr[\tilde{\bbx}^H\diag(|\tilde{\bbq}_k|^2)\tilde{\bbx}]$ then, since the argument of the trace is a scalar, we can write 
\begin{align}\label{E:second_order_moment_fb_v1}
&\E{\hat{p}_{\tilde{\bbq}_k} \hat{p}_{\tilde{\bbq}_k}}
=\E{\Tr[\tilde{\bbx}^H\diag(|\tilde{\bbq}_k|^2)\tilde{\bbx}\tilde{\bbx}^H\diag(|\tilde{\bbq}_k|^2)\tilde{\bbx}]}\nonumber\\
&=\Tr[\E{\tilde{\bbx}\tilde{\bbx}^H\diag(|\tilde{\bbq}_k|^2)\tilde{\bbx}\tilde{\bbx}^H}\diag(|\tilde{\bbq}_k|^2)],
\end{align}
where we have a quartic form of a Gaussian. From the expression of this quartic form, see e.g. \cite[pp. 43]{matrixcookbook}, we obtain
\begin{align}\label{E:second_order_moment_fb_v2}
&\E{\hat{p}_{\tilde{\bbq}_k} \hat{p}_{\tilde{\bbq}_k}}
= \left\|\diag\left(|\tbq_k|^2\right) \bbp \right\|_2^2 + ((|\tbq_k|^2)^T \bbp)^2 \\
& \hspace{0.5cm}+ \Tr[\diag(\bbp) \bbV^H \bbV^*\diag(|\tilde{\bbq}_k|^2) \bbV^T \bbV \diag(\bbp) \diag(|\tilde{\bbq}_k|^2)]. \nonumber
\end{align}
When $\bbS$ is symmetric, hence $\bbV$ is real, the trace in \black{\eqref{E:second_order_moment_fb_v2} is equal to} $\left\|\diag\left(|\tbq_k|^2\right) \bbp \right\|_2^2$ and, since $\var{\hat{p}_{\tilde{\bbq}_k}} = \E{\hat{p}_{\tilde{\bbq}_k} \hat{p}_{\tilde{\bbq}_k}} - \E{\hat{p}_{\tilde{\bbq}_k}}^2$, \black{using \eqref{E:second_order_moment_fb_v2} and} \eqref{E:mean_filter_bank_period_PSD}, the expression \eqref{E:cov_filter_bank_PSD} follows.

\bibliographystyle{IEEEtran}
%
\bibliography{citations}

\end{document}